# How to set up a psychedelic study: Unique considerations for research involving human participants

Marcus J. Glennon[2], Catherine I. V. Bird[3], Prateek Yadav[1], Patrick Kleine[1,8], Shayam Suseelan[1], Christina Boman-Markaki[1], Vasileia Kotoula[1], Matt Butler[1], Robert Leech[1], Leor Roseman[7,8], David Erritzoe[8], Deepak P. Srivastava[4,5], Celia Morgan[7], Christopher Timmermann[6,8], Greg Cooper[2], Jeremy I. Skipper[6], James Rucker[3,9,10], Sunjeev K. Kamboj[2], Mitul A. Mehta[1], Ravi K. Das[2], Anjali Bhat[1]

**Affiliations**

1 Centre for Neuroimaging Sciences, King's College London

2 Clinical Psychopharmacology Unit, Research Department of Clinical, Educational and Health Psychology, University College London

3 Psychoactive Trials Group, Institute of Psychiatry, Psychology and Neuroscience, King's College London

4 Department of Basic and Clinical Neuroscience, Institute of Psychiatry Psychology and Neuroscience, King's College London

5 MRC Centre for Neurodevelopmental Disorders, King's College London

6 Experimental Psychology, Psychology and Language Sciences, University College London

7 Department of Psychology, University of Exeter, Exeter, United Kingdom

8 Centre for Psychedelic Research, Department of Brain Sciences, Faculty of Medicine, Imperial College London, London.

9 The Department of Psychological Medicine, Institute of Psychiatry, Psychology and Neuroscience, King's College London, London, UK

10 South London and Maudsley NHS Foundation Trust, London, UK

**Corresponding author:**

Marcus Glennon (m.glennon@ucl.ac.uk)

# Contents

## Abstract

Setting up a psychedelic study can be a long, arduous, and kafkaesque process. Researchers are faced with a host of challenges in this rapidly-evolving field, necessitating a range of questions that remain largely unstandardised. Many of the complexities inherent to psychedelic research also challenge existing assumptions around, for example, approaches to psychiatric prescribing, the conceptual framing of the placebo effect, and definitions of selfhood. This review paper aims to formalise these unique hurdles by addressing the sociocultural, political, legal, pharmacological, safety, study-design and experiential facets inherent to a psychedelic study. We bring together several of the major psychedelic research teams across the United Kingdom, identify continuing areas of debate, and provide a practical, comprehensive, experience-based guide, with recommendations for policymakers and future researchers intending to set up a psychedelic research study or clinical trial.

## Highlights

- We call for a re-evaluation of inclusion and exclusion criteria in psychedelic studies, particularly regarding suicidality, psychosis risk, and prior psychedelic use. We also emphasise the importance of recruiting more representative samples and proactively addressing barriers to participation for marginalised groups.
- We argue that conventional randomised controlled trial designs are often a poor fit for psychedelic research, and propose alternative frameworks, including open-label extensions, active comparators, and other designs aimed at mitigating unblinding and placebo disappointment effects.
- We recommend a rethinking of how outcomes are measured in psychedelic research, highlighting the limitations of current questionnaires and advocating for approaches that better capture the richness of the psychedelic experience, such as microphenomenology, experience sampling, and the integration of subjective reports with neuroimaging data.
- We offer 39 practical recommendations covering the unique methodological, ethical, clinical, and regulatory considerations involved in psychedelic research, with the aim of supporting researchers, clinicians, and policymakers in designing safe, inclusive, and scientifically robust studies.

## 1. Introduction

In recent years, there has been a proliferation of research into – and widespread public interest in – psychedelic compounds (‘psychedelics’) such as lysergic acid diethylamide (LSD), psilocybin, and *N,N*-Dimethyltrypamine (DMT). Psychedelics elicit profound alterations in conscious experience (hence the term ‘psychedelic’, meaning ‘mind manifesting’) and have been shown in early phase clinical trials to significantly improve outcomes for patients with a range of psychiatric conditions, including treatment resistant and major depression, anxiety, post-traumatic stress disorder (PTSD), obsessive-compulsive disorder (OCD), and addiction (Feulner et al., 2023; Bogenschutz et al., 2015; Carhart-Harris, Bolstridge, et al., 2016; Johnson et al., 2014; Ko et al., 2023; Krebs & Johansen, 2012; Varker et al., 2021; Yao et al., 2024).

However, setting up a psychedelic study[1] can be a long, arduous, and kafkaesque process. This rapidly-developing field poses several unique challenges for researchers, necessitating a range of considerations that have not yet been standardised. Many of the complexities inherent to psychedelic research also challenge existing assumptions around, for example, approaches to psychiatric prescribing, the conceptual framing of the placebo effect, and definitions of selfhood. This review paper brings together several of the major psychedelic research teams across the United Kingdom[2] to formalise these unique considerations, identify continuing areas of debate, and provide a practical, experience-based guide, with recommendations for policymakers and future researchers intending to set up a psychedelic research study or clinical trial. Where there is debate between authors of this paper, we outline arguments for and against each position. We approach this such that the paper can either be read end to end, or treated as a manual: readers can dip into relevant sections as needed. A compiled table of recommendations is provided in the conclusion.

## 2. Sociocultural, political, and legal considerations

Psychedelic research is inseparable from its cultural and political context. In this section, we outline the current socio-political landscape of psychedelic research, its implications for dissemination, and the practicalities of setting up a study in this field.

### 2.1. Cultural context and implications for dissemination of findings

Before beginning to set up a psychedelic study, it is important to understand the cultural context of this field. Psychedelic research has been (and continues to be) influenced by cultural, political and legal factors to a unique degree. It is therefore important for researchers to consider the impact of their work on patients, public opinion, corporate agendas, and Indigenous and mixed-race (‘mestizo’) populations to whom psychedelics are culturally significant. In this subsection, we summarise some of these considerations, make suggestions about how findings should be disseminated given these cultural factors, and acknowledge areas of legislation that may be disproportionately influenced by political history.

#### *2.1.1. Reasons for Schedule I / 1 licensing and its impact on research*

Psychedelics were classified in the United States as Schedule I substances under the Controlled Substances Act in 1971, with countries around the world (including the UK, with Schedule 1) following suit. This designation, asserting *no medicinal value, high potential for abuse, and high risk of dependence* (United States Drug Enforcement Administration, 2024), effectively ended nearly all scientific research into psychedelics for decades. Before this, however, psychedelic science was already burgeoning: before its prohibition, LSD had appeared in over 1,000 scientific papers and had been administered to approximately 40,000 individuals (Dyck, 2005), and evidence was already emerging for psychedelics as treatments for depression, alcoholism, and other addictions (see Nichols & Walter, 2020, for a review).

[1] We focus here on studies that involve administration of psychedelics to human subjects. Several psychedelic studies use observational approaches such as recruiting participants after ayahuasca ceremonies which they have attended independently (e.g., Doss et al., 2024), but these are beyond the scope of this paper.

[2] The King’s College London Psychedelic Experience Neuroimaging Group (PENG) led by Robert Leech and Anjali Bhat; the KCL Psychoactive Trials Group led by James Rucker; the King’s Neuropharmacology Group and Centre for Innovative Therapeutics led by Mitul Mehta; the KCL Neuronal Circuitry and Neurodevelopmental Disorders Group led by Deepak Srivastava; the Imperial College London Centre for Psychedelic Research led by David Erritzoe; the Exeter University Psychopharmacology and Addiction Research Centre psychedelic research group led by Celia Morgan and Leor Roseman; the University College London (UCL) Clinical Psychopharmacology Unit led by Sunjeev Kamboj and Ravi Das; the UCL Centre for Consciousness Research co-directed by Jeremy I Skipper and Christopher Timmermann, and the UNITy (Understanding Neuroplasticity Induced by Tryptamines) Group at UCL led by Jeremy I Skipper and Ravi Das.

The reasons for Schedule I licensing of psychedelics were more political than scientific. The counterculture movement of the 1960s, and its connection with the civil rights movement, marked a dramatic shift in the public perception and use of psychedelics. Figures such as Timothy Leary, a former Harvard professor turned counterculture icon, were central to this transition. Leary's enthusiastic promotion of LSD as a tool for expanding consciousness and societal liberation (encapsulated in his famous phrase "turn on, tune in, drop out") diverted attention away from its therapeutic and scientific potential, steering it toward widespread recreational use (Lee & Shlain, 1985; Barber et al., 2024). LSD also became a symbol of resistance within activist movements, such as the Vietnam War protests, further entangling its image with political unrest (Hartogsohn, 2020; Dyck, 2008). By 1971, President Richard Nixon, declaring psychedelics a threat to public order, launched the War on Drugs. This period saw a surge in public health campaigns and anti-psychedelic propaganda, perpetuating fears about their dangers (Hartogsohn, 2017; 2020) . In 1966, amid growing public concern, the US Food and Drug Administration (FDA) classified LSD as an illegal narcotic. By 1968, research into psychedelics in North America was effectively criminalised, with government approval for clinical studies becoming nearly impossible to obtain (Carhart-Harris & Goodwin, 2017; Barber et al., 2024). This left the field of psychedelic research at an awkward, pubescent stage of development, wherein several studies had demonstrated widespread applications for psychiatry, but many had done so using scientific methods that would not meet modern standards of research (see Rucker, Iliff & Nutt, 2018), which was inevitably also a contributing factor to the widespread banning of psychedelic research (Hall, 2022; Oram et al., 2014).

Since the late 1990s, there has been a revival of scientific, cultural, and institutional focus on psychedelics, driven by advancements in research – in particular, with the emergence of brain imaging technologies such as positron emission tomography (PET), magneto-/electro- encephalography (M/EEG), and functional magnetic resonance imaging (fMRI), which have provided groundbreaking insights into how psychedelics affect the brain (Carhart-Harris et al., 2014; Carhart-Harris, Muthukumaraswamy, et al., 2016; Timmermann et al., 2019). However, the Schedule I classification of psychedelic substances remains in place, significantly prolonging the process (and substantially increasing the cost) of every step of the process of psychedelic drug research, from the manufacture of the drug through to its administration, as well as the entire process of authorising, funding and executing research. In Section 2.3, we share experiences and recommendations to help future researchers to streamline this process.

Given the accumulation of two decades of scientific research with psychedelics, now seems a pertinent time to consider reclassification. Despite the Schedule I designation, there are an increasing number of studies investigating the clinical utility of psychedelics that are funded by government bodies. That is to say, government agencies are funding research that directly challenges their own classification of these drugs as having no accepted medical benefit (Home Office, 2022). Rescheduling to Schedule II, alongside a restriction limiting use to Medicines and Healthcare Research Authority (MHRA) or National Health Service Research Ethics Committee (NHS REC) authorised research, would remove unnecessary licensing barriers to psychedelic drug production and research, while still maintaining adequate security and safe custody arrangements: the same that also apply to much more dangerous drugs such as diamorphine. This position is echoed in a recent UK Parliamentary Office of Science and Technology (POST) briefing, which highlights how current Schedule I restrictions create substantial research barriers and notes growing policy discussions around rescheduling to facilitate clinical trials (POST, 2025)

#### *2.1.2. Corporate interests and public perceptions*

Political shifts, such as the FDA granting 'breakthrough therapy' status to psychedelic-assisted treatments (Heal et al., 2023), have followed the resurgence of psychedelic research. Academic institutions, once hesitant to support this field, are now increasingly sponsors of psychedelic studies. Public perceptions of psychedelics shifted dramatically with the publication of Michael Pollan's influential book *How to Change Your Mind* (Noorani, 2020; Noorani et al., 2023).

There is a risk of psychedelics being over-hyped as a result of this resurgence (as they were in the 1960s). A further liability is posed by unregulated corporate interests that capitalise on the growing popular interest in psychedelic experiences. For example, unlicensed ayahuasca 'retreats' charging large sums of money and offering transformative experiences have begun to appear in areas of the world where Indigenous psychedelic rituals are traditionally practised, as well as in several Western countries. Some of these retreats have been seen to shirk appropriate safety precautions, with dangerous consequences for attendees (dos Santos, 2013; Evans & Adams, 2025). Indeed, the US government recently issued a warning advising travellers to Peru, a common destination for these retreats, not to ingest ayahuasca while travelling, stating, "U.S. citizens in Peru have… reported being sexually assaulted, injured, or robbed… at "healing" or "retreat" centers. Facilities or groups offering ayahuasca/kambo are not regulated by the Peruvian government and may not follow health and safety laws or practices." (U.S. Embassy in Peru, 2025). Furthermore, the surge in Western interest into psychedelic rituals has been economically and ecologically damaging to Indigenous communities (Bathje et al., 2022; Gonzalez et al., 2021; Tupper, 2009).

Given the lingering stigma from pre-prohibition propaganda, as well as the growing hype, building a robust evidence base is key to inform future mental health treatments and a balanced public perception. It is essential that researchers take a balanced approach to dissemination of results, being careful not to portray psychedelics as a panacea, while still acknowledging the profound potential for psychedelic medicine to provide options for patients who previously may not have had any treatment options. The research and policy determined at this time will be foundational in the way we approach psychedelic medicine in the future. A careful and patient approach to phasing these compounds into mainstream medicine is key. Balancing the inevitable advancement of therapeutic applications of psychedelics with rigorous scientific standards will be crucial in ensuring the safety of patients and preventing history from repeating itself.

#### *2.1.3. Cultural variation in medical practice: implications for psychedelic medicine*

Psychedelics have been used for medicinal purposes by Indigenous cultures for centuries (De la Garza, 2012; Herrera, 1992; Munn, 1973). There are fundamental differences in the way that this psychedelic medicine is conducted and underlying assumptions in modern medicine that are important to consider as we introduce these compounds into modern clinical practice. Indigenous societies such as the Shipibo-Konibo of Peru (who have been particularly engaged in discourse with foreign parties) view individual wellbeing as inherently connected to the broader ecosystem, emphasising that true healing requires an assessment of the individual's relationship with their environment (Allen, 2019). There is a fundamental contrast between the Western dualist views of selfhood, and the Indigenous and Eastern perspectives of selfhood and consciousness as fluid and relational (Mori, 2017). Influences on the medical profession of dualist ideas about the mind (e.g., the separation of psychiatry, psychotherapy, and neurology as clinical disciplines), and Western essentialist ideas about the self as a static entity (Mosig, 2006; Barker & Iantaffi, 2019; Percival, 2015; Bathje et al., 2022), constrain the scientific understanding of the psychedelic experience, which involves significant changes in the sense of self (Nour et al., 2017; Härter, 2021; Hayes et al., 2020).

A central concept that embodies this limitation is 'ego dissolution' (Carhart-Harris et al., 2014; Grof, 1980), which is assessed on many of the major psychedelic experience scales outlined in this paper. The term has been criticised for its nebulousness, and scales such as the 5-Dimensional Altered States of Consciousness (5D-ASC) include leading items that ask, for example, 'To what extent did you experience ego-dissolution?' (Dittrich et al., 2006). What is particularly revealing about this term is that the alterations of the sense of self under psychedelics are characterised as a *dissolution*, implying loss, which calls for conceptual clarifications of how this experience differs from the one associated with disordered states characterised by fragmented selfhood (as in dissociative disorders). Indeed, terms used synonymously in the literature include "ego death" and "ego loss" (Johnson et al., 2008) , while some scales include items that ask participants about their 'dread of ego-dissolution' (Dittrich et al., 2006). While some individuals may experience dread and other negative emotions during these alterations in selfhood, many subjects' reports involve descriptors such as a sense of 'oneness', 'bliss', and 'connectedness' with all beings (Carhart-Harris et al., 2018; Petranker et al., 2022) – seemingly better suited to the descriptor 'self-expansion' (Aron & Aron, 1986). We propose that it would be of value to update the notion of ego-dissolution with rigorous empirical study of the influence of psychedelics on selfhood, using emerging methods of precisely mapping subjective experience (see Section 5) and taking an unbiased exploratory approach that avoids leading questions (c.f. Timmermann, Bauer, et al., 2023). More broadly, we also suggest that valuable insights may be gained from existing medical practices using psychedelics (e.g., by Indigenous communities).

| **Recommendations** | |
|---|---|
| 1 | Psychedelics should be reclassified as Schedule II drugs, with ongoing consideration of how regulatory frameworks can stymie research. |
| 2 | Researchers should be aware of the impact of their findings on patients, public opinion, corporate agendas, and Indigenous or mestizo populations to whom psychedelics are culturally significant. Findings should be disseminated in a careful and balanced manner to avoid stoking unwarranted hype or stigma. |
| 3 | Valuable insights may be gained from existing traditional practices using psychedelics (e.g., those used by Indigenous communities), such as the possible limitations of dualist assumptions that underlie some elements of Western medical practice and psychometric measurements of selfhood. |

### 2.2. Funding for psychedelic research

One of the first challenges faced when setting up a psychedelic trial is obtaining funding. Costs vary, but it is not uncommon for clinical trial costs to range from one hundred thousand to several million pounds. Costs are dependent on factors such as sample size, manufacturing costs, costs of moving the drug between the manufacturer and dispensing pharmacy, as well as other study-related expenditures (e.g., personnel costs or scanning costs if using neuroimaging). In our experience, most public funding agencies may be apprehensive to fund psychedelic research given the historical stigma ascribed to these compounds.

Despite growing public interest in psychedelics, much of the published research from the UK so far has depended on philanthropy and private donations (Carhart-Harris, Erritzoe, et al., 2012; Carhart-Harris, Leech, et al., 2012; Carhart-Harris et al., 2013; Carhart-Harris, Muthukumaraswamy, et al., 2016); see also Marks & Cohen (2021). We identified eleven psychedelic studies in the UK for the past 5 years (Jan 2020 - Jan 2025).While most studies reported a single funder, some reported two sources;. five reported public funding, five reported industry funding, two reported charitable funding, one reported crowd-funding, and one appeared to be funded using 'soft money.' There are also a number of recently funded studies which are yet to report their findings. These include National Institute for Health Research (NIHR) UK grant for a clinical trial exploring psychedelics for opioid addictions; Wellcome Trust Fellowships for an open-label study of psilocybin in function neurological disorders (Butler et al, 2024), an experimental investigation into psilocybin's effects on human learning, and a neuroimaging study investigating the effects of DMT on hazardous drinking behaviour; as well as a Beckley Foundation philanthropic grant for a precision neuroimaging study of ongoing subjective experience under LSD.

Despite promising research results, we note that large pharmaceutical companies have been hesitant to invest in psychedelic therapy development. Nonetheless, the continuation of mixed funding sources, including public and charitable grants, is an encouraging step towards more accessible funding for psychedelic research. Researchers may consider unconventional sources of funding such as crowd-funding or donations from non-profit organisations such as The Centre for MINDS, as well as in kind donations (e.g., some drug companies may provide drugs for research free of cost).

| **Recommendations** | |
|---|---|
| 4 | Researchers may benefit from considering unconventional sources of funding (e.g., from crowd-funding or private donations) as well as in kind donations (e.g., study drug donations for research purposes) in order to finance their psychedelic studies. |
| 5 | Public investments into psychedelic research should be encouraged, given the promise of medical benefits for patients who currently lack treatment options, as well as the possible economic benefits. |

### 2.3. Regulatory and ethical approval

It can take six months to four years to obtain all the necessary approvals to begin conducting a new study involving psychedelics. As with all clinical and psychopharmacological research, there are multiple approval processes involving numerous regulatory bodies that need to be completed before participants can be enrolled into a psychedelic study. Below, we overview some of the key steps (finding funding, sponsorship review, ethical approval, Home Office licensing, manufacture and transport/import of the drug) in the study approval process, highlighting processes and considerations that are specific to psychedelic research. Each of these stages will implicitly include substages (occasionally multiple cycles thereof) of application, review, requests for amendments or further information, responses to these requests by the researchers, and final review and approval (see Figure 1 for summary).

#### *2.3.1. Study sponsorship/Capacity and Capability (C&C)*

The first domain of approval needed for a psychedelic study to begin is the researcher's own institution or organisation. The organisation will have to assess the proposed project for 'capacity and capability' for the project to be completed and to ensure the institution has the resources to support the project. All trials and basic research studies involving drugs require a 'sponsor,' which is the institution or organisation that is responsible for the set up and conduct of the study/clinical trial. Crucially, it is the sponsor's responsibility to hold indemnity for the study. In some cases, the sponsor is also the funder, but not always. For example, a non-commercially funded study (public grant, charity, philanthropic donation) will most likely have the academic institution at which the study is taking place as the Sponsor (i.e. sponsor and funder are different), but a commercially funded trial might have the same industry partner as the sponsor (i.e. sponsor and funder are the same). Private organisations and academic institutions are often better placed than public organisations to act as a sponsor (e.g., larger universities often have greater scope for indemnity than the National Health Service).

At an academic institution, practically, the first stage of approval for a study involving a psychedelic drug administration conducted is internal sponsorship review, in which the study protocols are reviewed by the host institution (usually the main affiliation of the chief investigator) before confirming their commitment to sponsoring indemnity for the study. The process usually involves a risk assessment by the institution, followed by an extended review of contractual obligations of all parties involved. Unlike ethical and regulatory review, this process does not require procedure timelines, but is vital to the management of costs and liabilities of the host institution. Thus, in practice, this stage can be one of the most time-consuming and variable across institutions. In the experience of some of the authors, this process can take up to two years, although it is possible that with multiple studies the process can reduce over time. The inclusion of psychedelic drug administration can significantly lengthen this process due to the complexities of drug procurement, importing and accountability under a Home Office license.

Once the study protocol, information sheets and consent forms have been designed by researchers, the drug manufacturer and their protocols for handling the study drug are approved by the pharmacy responsible for storing and dispensing it. The study will then undergo an internal risk assessment. Once internal formalities (which may vary between institutions) are completed, the research governance office will issue a confirmation of sponsorship, which typically opens the door for submission to an ethics committee (although in some cases these processes may happen in parallel).

#### *2.3.3. MHRA, HRA and REC approval*

In the UK, psychedelic studies require approval from Research Ethics Committees (RECs). Depending on the institution, psychedelic studies may also require approval – via the online Integrated Research Application System (IRAS) – from The Health Research Authority (HRA). Clinical Trials of Investigational Medicinal Products (CTIMPS) additionally require approval from the Medicines and Healthcare products Regulatory Agency (MHRA). See Appendix 1 for a sample IRAS form submitted and recently approved by an NHS REC and the HRA.

Following successful submission of an application to a REC, either internally or via IRAS, researchers will attend a REC meeting where a panel will assess the project's safety and ethical implications. In our experience, the main focus of these meetings for psychedelic studies tends to be ensuring that there are distress protocols in place for adverse events (see Section 4.2.1 – 'Distress protocols'). Other possible questions that could be applicable across psychedelic studies include: 1. How will the drug be transported between the dispensing pharmacy and administration site?; 2. What will the participant's environment look like during drug dosing? (see Section 6 – 'Experiential Considerations' for recommendations). 3. Who will be in the room with the participant? 4. How will the participant get home after dosing? 5. How will you screen participants?

While this ethical oversight is essential, each stage of bureaucracy has not necessarily been introduced with the overall process in mind (see Husain, 2024, 2025), resulting in lengthy proceedings involving significant duplication of effort (many requirements overlap among the HRA, MHRA, and RECs as well as the sponsorship review process). The arduous nature of the approval process can discourage smaller research teams, which may lack the resources or institutional backing to navigate such a demanding system, and exploratory studies, even though these smaller-scale projects often provide valuable insights into how psychedelics function and their therapeutic potential. The obstacle in larger institutions is often a lack of coherence between administrative departments, making it unclear for a researcher or trial manager which bureaucrat is expected to sign off on each form at each step of the process: independent researchers are often expected to go back and forth between various university departments and external regulators. In some instances, funding may run out by the time all approvals are in place, or one of the many necessary approvals may expire by the time another one has been obtained.

Streamlining approval systems, reducing redundancies, and enhancing flexibility could encourage more inclusive research while maintaining the high standards required for participant safety and scientific rigor. Collaborative efforts between regulators, academic institutions, and private entities are essential to advancing psychedelic research, reducing stigma, and ensuring equitable access to emerging treatments (Silva et al., 2025). A promising advancement that exemplifies this is a novel integrated review process recently announced by the MHRA, which may improve the inclusivity and feasibility of psychedelic research. As research teams and review committees develop productive relationships, regulatory hurdles may become less restrictive. With each successfully completed psychedelic study, trust grows, leading to shorter approval processes and a more efficient pathway for future trials. This evolving collaboration suggests that, despite initial challenges, regulatory frameworks can adapt, allowing for a more integrative approach to psychedelic research that acknowledges both drug effects and contextual factors which evidence shows shape these effects (Hartogsohn, 2017).

| **Recommendations** | |
|---|---|
| 6 | Regulators should look to streamline approval processes across bodies in order to help facilitate research for patient benefit. |
| 7 | Researchers should be aware of and budget the time required to undergo the process of obtaining all necessary approvals. |

#### *2.3.4. Controlled Drug Licencing*

Presently, classic psychedelics are still classified in the US and the UK as Schedule I / 1 substances (see Section 2.1.1). They are also Class A drugs in the UK, meaning they warrant the highest penalty for possession, use and/or distribution. As a result, studies performed in the UK involving administration of psychedelics to human subjects require approval from the Home Office Drug and Firearms Licensing Unit. Separate Home Office licenses are required for every part of the process, from the manufacture, certification and encapsulation, to the storage, and dispensing and, finally, administration of the drug. It is a tortuous, arcane, long-winded and expensive process that, in our experience, offers no additional security or safe custody benefit beyond that already afforded by Schedule 2/II, whilst acting to significantly inhibit the process of legitimate research.

The license for administration must name the lead researcher and the prescribing clinicians (all of whom must have an active registration with the General Medical Council), who will have to write a prescription for the study drug every time it is to be administered to a participant. To obtain the license for administration, the lead researcher will need to obtain an enhanced Disclosure and Barring Service (DBS) check of their criminal record, in order for the Home Office licensing to proceed.

#### *2.3.5. Manufacture, transport and dispensing of the study drug*

The Schedule 1 designation of psychedelic substances mandates that not only the sites and pharmacies dispensing and administering the substances possess a licence, but also that there is a chain of Schedule 1 approved links from manufacture through to packaging through to transporting to the research site. Pharmaceutical companies supplying the psychedelic study drug are required to manufacture the drug to Good Manufacturing Practice (GMP) standards – which include detailed records of reagents used in manufacture as well as stability and purity tests – but may not themselves be equipped to provide the logistical options for moving the drugs to a research site.

In practice, this means that anyone wishing to set up a psychedelic study will also be required to seek third party input to receive the drug. These third parties must be in possession of a Schedule 1 license and, if the drug is being manufactured in another country, an import license. They are responsible for any of: vetting the manufacturing drug company, shipping, storing the drug prior to transfer, quality control, labelling and bottling drugs (dependent on study arms and blinding), and distribution of the drug to site.

Due to a combination of the associated costs of these services, and the relative lack of market competition, the costs to research groups for their services can be significant (in the order of tens or even hundreds of thousands of pounds), with additional charges for bureaucratic hurdles (e.g., if drugs need to be relabelled on entering the UK, which they may if being imported from another country). It is inevitable that such costs are broadly inflated due to the added burdens of Schedule 1, and that removing these would render research considerably more affordable.

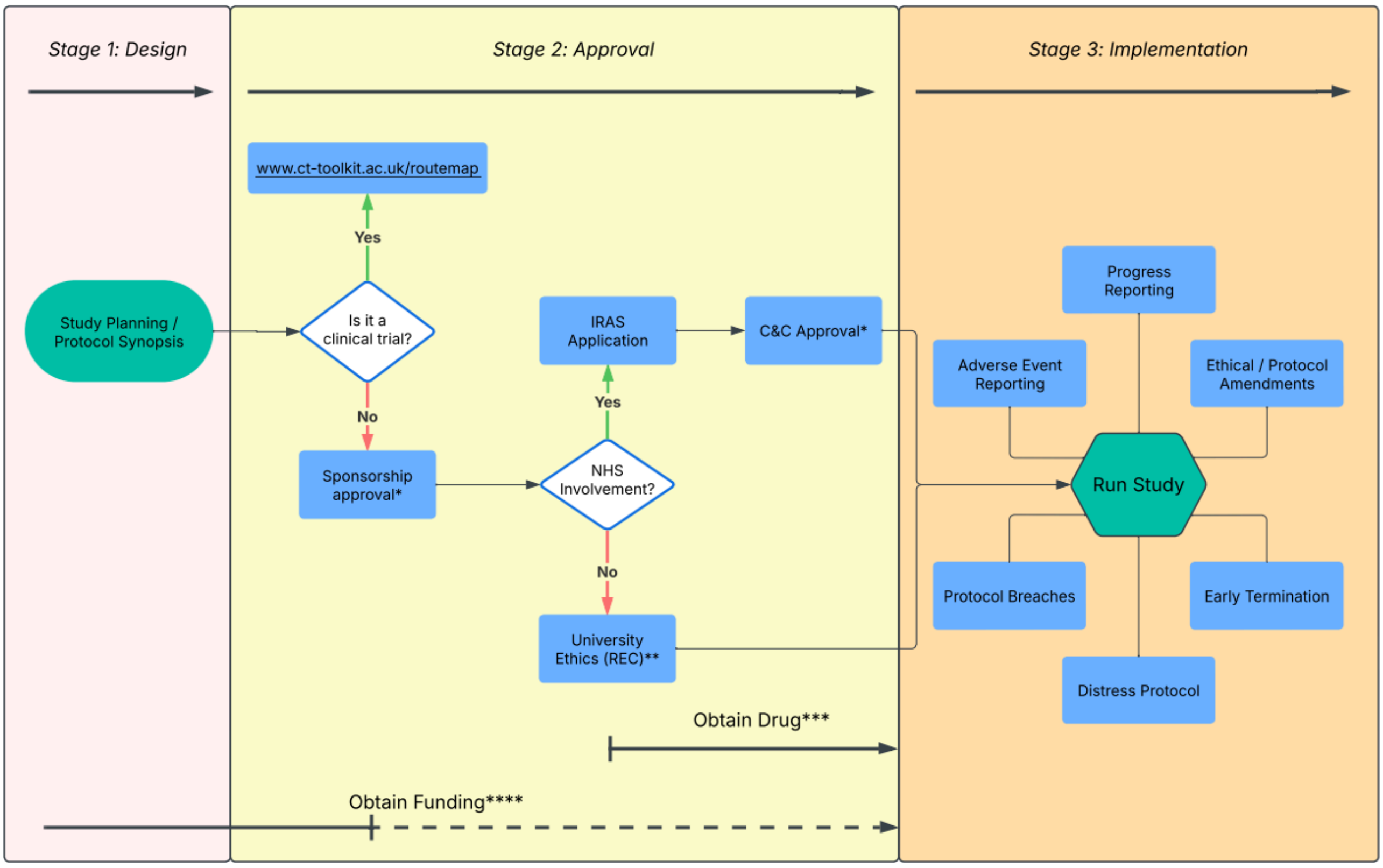

**Figure 1. Flowchart summarising the process of setting up a psychedelic study in the UK.** *Sponsorship/Capacity and Capability (C&C) approval provide insurance/indemnity for the study. **Health Research Authority (HRA) approval may be necessary if National Health Service governance is deemed to be required – e.g., in the case of a clinical trial of an investigational medicinal product (CTIMP). ***Notes regarding the schedule 1 status of psychedelic drugs: 1. The drugs must be GMP- standard; 2. Sufficient funds must be factored in for their import and prescription licences; 3. Researchers must establish a clear adverse event (AE) and serious adverse event (SAE) reporting framework in accordance with regulatory requirements. ****The stage at which funding must be obtained can vary between institutions. At King's, for example, evidence of funding must be provided prior to sponsorship approval. At University College London, evidence of funding may be provided at any stage. *IRAS=Integrated Research Application System; REC=Research Ethics Committee; GMP=Good Manufacturing Practice.*

## 3. Pharmacological considerations

There are a number of different psychedelics currently under investigation for the treatment of neuropsychiatric conditions. Psilocybin is perhaps the most commonly used psychedelic in modern trials, and this has to do with its pharmacokinetics, safety profile, subjective duration of action, and the relative absence of attached cultural stigma to the name. Practically, the administration of oral psilocybin can occur as an outpatient day case, whereby a dosing visit can be completed within normal working hours. This is in contrast to compounds with a longer duration of action, such as LSD, which may require participants to stay overnight in research institutes to ensure psychedelic effects have ceased. The characteristics of psychedelic compounds also affect modes of delivery. Psilocybin is suitable for oral administration, given it is dephosphorylated to its active metabolite psilocin in the liver. LSD is also absorbed orally. Psilocybin, LSD, DMT and 5-MeO-DMT can be administered intravenously, with a more rapid onset of action and shorter duration in comparison to oral routes, but with requirements for peripheral venous access.

The speed of onset of intravenous psilocybin is dramatically faster, with subjective intensity peaking within 5 mins (Carhart-Harris et al., 2011), compared to intravenous LSD, where subjective drug effects appear 5-15 mins post-infusion, and peak intensity is reached within 60-120 mins (Carhart-Harris, Kaelen, et al., 2016; Carhart-Harris, Muthukumaraswamy, et al., 2016). DMT, 5-MeO-DMT, and related compounds are rendered inactive by monoamine oxidase degradation during first-pass metabolism, so are not suitable for oral dosing. These compounds may be administered via inhalation, insufflation, intravenous, intramuscular, sublingual, or buccal routes, usually resulting in a faster onset of effects compared to oral dosing (Barker, 2022; Reckweg et al., 2022; Rucker et al., 2024; Timmermann, Roseman, et al., 2023).

| Compound | Route of Administration | Dose Range | | | Onset | Peak | Duration |
|---|---|---|---|---|---|---|---|
| | | Low | Med | High | | | |
| Psilocybin | Oral | <10mg[a, b] | ~15mg[d,e] | >25mg[a, b] ≈ 20-30mg/70kg[a, c] | ~30-60 min [c, f] | ~90-180 min [c, f] | ~6-7h [c,f] |
| | Intravenous | ~1.5mg ≈8mg oral[a] | ~2mg ≈15mg oral[a] | - | ~60 sec[a] | 4-5min[a] | ~20 min*[a] |
| LSD | Oral | 25-50µg [a] | ~75-100 µg[a] | 100-200µg [a] | ~30-60 min[a, b, c] | 120 - 170 min[a, b, c] | ~7-11h[a, b, c] |
| | Intravenous | - | 75µg ≈ 100 µg oral[c] | - | ~10 min[a, b] | ~60-120 min[a, b] | ~7-8h[a, b] |
| DMT | Inhalation (Freebase) | 10-20mg[e] | 20-40m[e] | 40-60mg[a, b, c, e] | ~60 sec[d] | ~2-3 min (similar to IV[d]) | ~20-30 min[a, c, d] |
| | Intravenous (Fumarate) | ~7mg[a, b] | ~15mg[a, b] | ~20mg[a, b] | ~60 sec[c] | ~2-3 min[a] | ~20-30 min[a,c] |
| | Intramuscular | - | 0.7mg/kg[a] | - | ~2-5 min[a, b] | ~10-15 min[b] | ~30-90 min[a, b] |
| 5-MeO-DMT | Inhalation | 2-4mg[a,b, c,e] | 4-8mg[a,b, c,e] | 8-20mg[a,b,c, e] | ~30-60 sec [a,c,d] | 2-5 min [a,b,c] | ~20-30 min[b,c] |
| | Intravenous | - | 0.5-2mg[a] | >2mg[a] | | | |
| | Intranasal | 5-10mg[a] | 8-15mg[a] | 10-25mg[a,c] | ~4-7 min[a,b] | ~10-30 min[a] | 30-45 min[a,b] |

**Psilocybin IV**: [a](Carhart-Harris et al., 2011) * 1.5mg ~ 20 min duration, 2mg total duration not reported, **Psilocybin Oral**: .[a](Roseman, Demetriou, et al., 2018), [b] (Carhart-Harris, Bolstridge, et al., 2016), [c](Griffiths et al., 2016), [d](Preller et al., 2020) *0.2mg/kg [d] converted to= 14mg/70kg, [e] (Vollenweider and Kometer, 2010) <0.215mg/kg = low/moderate, [f] (Holze et al., 2022) | **LSD IV**: [a](Carhart-Harris, Kaelen, et al., 2016), [b](Carhart-Harris, Muthukumaraswamy, et al., 2016), [c] (Liechti, 2017), **LSD Oral**: [a](Holze et al., 2021), [b] (Dolder et al., 2017), [c] (Holze et al., 2022) | **DMT Inhalation**: [a]Barker, 2022, [b] (Falchi-Carvalho et al., 2024), [c] (Michael et al., 2021), [d](Barker, 2018), [e](“Erowid DMT Vault : Dosage,” 2025.), **DMT IV**: [a](Timmermann et al., 2019), [b] Timmermann, Roseman, et al., 2023, [c] (Strassman et al., 1994), **DMT IM**: [a] Szára, 1956, [b] (Gillin et al., 1976) | **5-MeO-DMT Inhalation**: [a] (“Erowid 5-MeO-DMT Vault : Dosage,” 2025.), [b](Ermakova et al., 2022), [c] Davis et al., 2018, [d](Uthaug et al., 2020), [e]“5-MeO-DMT Information & Vital Education,”2025), **5-MeO-DMT Intranasal**: [a] (“Erowid 5-MeO-DMT Vault : Dosage,” 2025.), [b](Ermakova et al., 2022, [c](Rucker et al., 2024), . **5-MeO-DMT IV** [a](Ermakova et al., 2022), (Rucker et al, 2024).

### 3.1. Psilocybin

Psilocybin's effects include altered states of consciousness, affective change, illusions, and hallucination-like phenomena (Studerus et al., 2011). A systematic review and meta analysis of 6 studies (n=528) found psilocybin administration to be associated with short-term side effects including headache, nausea, anxiety and dizziness, but no serious adverse events (Yerubandi et al., 2024).

Dosing and regimens in psilocybin trials have varied over its history (Carhart-Harris, Bolstridge, et al., 2016; Griffiths et al., 2006; Malitz et al., 1960; Pahnke, 1969). The use of higher doses for fewer sessions is more common in modern clinical trials (Passie et al., 2022). Doses can be fixed (Carhart-Harris, Bolstridge, et al., 2016; Goodwin et al., 2022; Raison et al., 2023), or individualised by weight, either mg per kg approaches (Bogenschutz et al., 2015; Ross et al., 2016; Stauffer et al., 2020) or dosing per unit weight (Doss et al., 2021; Griffiths et al., 2006). Post-hoc analysis comparing weight-based dosing to fixed dosing did not produce significant differences in subjective responses and therefore the convenience and lower cost of fixed dosing may be preferable (Garcia-Romeu et al., 2021; Spriggs et al. 2023). Pharmacokinetic study after psilocybin administration showed that body weight had no effect on either Area Under Curve (AUC) or maximum concentration of active metabolite psilocin (Dahmane et al., 2020). A 'standard' dose commonly used in trials is 25 mg, whereas 35mg may be considered a high dose (MacCallum et al., 2022). An oral dose of 25mg appears equivalent to 0.3mg/kg in studies; there is a linear relationship between oral psilocybin dose and psilocin exposure in the range of 0.3–0.6 mg/kg in healthy adults which covers most commonly used doses in clinical trials (Brown et al., 2017). Further studies on the minimum effective dose required to elicit therapeutic efficacy would allow a dose range to be defined using a principled approach. There is some evidence and anecdotal reports of tolerance to the effects of psilocybin, and cross-tolerance with LSD, on repeated dosing but further study is required in this area (Baumann et al. 2022; Isbell at al., 1961).

Intravenous psilocybin has a rapid onset of effects within minutes, while the oral bioavailability is estimated at 52.7% +/- 20% (Hasler et al., 1997). When given orally the first subjective changes are noticed at 20-40 minutes and peak effects experienced at 60 to 90 minutes, although there is a significant degree of variation in these estimates between individuals (Brown et al., 2017; Hasler et al., 1997). These findings match the timing of plasma concentration curves of active metabolite psilocin after oral psilocybin ingestion (Lindenblatt et al 1998). Measures of subjective effects appear to have a dose-response relationship (Hirschfeld & Schmidt, 2021). The plasma half-life is variable between individuals with a range of between 1 and 3 hours (Brown et al., 2017; Holze at al., 2022). The small amount of psilocin that is renally excreted means that psilocybin is likely safe in people with mild to moderate renal impairment (Brown et al., 2017).

Psilocybin at higher doses (>215 micrograms/kg) has been shown to produce a transient rise in mean arterial blood pressure within 2 hours of ingestion, while having no measurable effect on body temperature or heart rhythm on electrocardiogram (Hasler et al., 2004) – although anxiety is a potential confound to this observation. There is evidence that psilocin can mildly prolong the QTc interval on electrocardiogram, however this only reached the threshold of concern (>10 milliseconds increase in QTc) at psilocin concentrations equivalent to ingesting 75mg psilocybin (Dahmane et al., 2020). Caution is advised in patients with hypertension or cardiovascular risk factors and these are often exclusion criteria. Transient increases in hormones including thyroid stimulating hormone, cortisol, prolactin and adrenocorticotropic hormone have been recorded with high doses but these normalise within 300 minutes of onset, and do not correlate with subjective reports of anxiety symptoms (Hasler et al., 2004).

There have been few case reports of fatal intoxication of psilocybin or magic mushroom ingestion, with most reported being due to polysubstance use or acute behavioural effects of psilocybin resulting in self-harm or accident (van Amsterdam et al., 2011). Only two fatalities due to psilocybin overdose have been described in the literature (Gerault & Picart, 1996; Buck, 1961). A case report associated magic mushroom ingestion with increased measured seizure frequency in a patient with refractory epilepsy; caution may be required in patients with a diagnosis of epilepsy (Blond & Schindler, 2023). In an international survey, only 0.2% of those using psilocybin or magic mushrooms recreationally sought emergency medical attention, with commonest symptoms being anxiety and paranoia, and all respondents but one reported being asymptomatic by 24 hours (Kopra, Ferris, Winstock, et al., 2022). It is unlikely that psilocybin plays any significant role in the evolution of serotonin syndrome since it has limited activity at the serotonin reuptake transporter, and is not a full agonist at serotonin receptors; nonetheless, idiosyncratic cases may occur (Amarnani et al. 2024).

### 3.2. LSD

LSD produces changes in subjective experience in domains including time perception, affect, and visual and auditory perceptions (Katz et al., 1968; Passie et al., 2008). LSD appears to produce cognitive impairments including poorer performance in measures of attention, recall, memory, reaction time, estimates of the passage of time, and altered language including increasing verbosity and decreased range of vocabulary (Abramson et al., 1955; Aronson et al., 1959; Hintzen, 2006; Jarvik et al., 1955; Lienert, 1959, Sanz et al., 2021).

The minimum dose to experience subjective effects has been reported to be 25 micrograms (Holze et al., 2021). However, trials of 'microdoses' have found subjective effects can be detected at doses as low as 5 micrograms (de Wit et al., 2022; Hutten et al., 2020). A dose of 100 micrograms is a commonly used dose recreationally and is considered an intermediate dose, whereas 200 micrograms would be considered a high dose (Holze et al., 2019; Passie et al., 2008).

Anxiety and ego dissolution increase at doses higher than 100 micrograms, but other subjective effects appear to experience a ceiling effect beyond 100 micrograms (Holze et al., 2021). A meta-analysis of psychometric data found a sigmoid dose-response curve for subjective effects, with a similar plateau at 100 micrograms (Hirschfeld et al., 2023). Peak effects are experienced at 1.5 to 2.5 hours with a total effect duration 10 – 12 hours, but pre-dose food consumption can affect absorption and onset (Passie et al., 2008). Intravenous, intramuscular and oral LSD were not found to be qualitatively different, only differing in higher speed of onset for parenteral routes (Hoch, 1956; Passie et al., 2008). There is some evidence including animal model data and anecdotal reports indicating tolerance to repeated doses of LSD, and cross tolerance with psilocybin, but this needs to be explored further to inform dosing schedules (Baumann et al. 2022; Gresch et al., 2005; Isbell at al., 1961).

LSD produces moderate increases in blood pressure and heart rate at doses higher than 50 micrograms and so caution is advised in patients with cardiovascular risk factors (Holze et al., 2021; Schmid et al., 2015). LSD has been shown to produce a small decrease in creatinine clearance but no other changes in electrolytes and lipids (Hollister & Moore, 1965). It has been shown to acutely increase serum cortisol, prolactin, oxytocin and epinephrine (Schmid et al., 2015). It has also been shown to increase growth hormone levels (Meltzer et al., 1981).

Beyond the capacity for contextual psychological toxicity, LSD is not considered a physiologically dangerous drug in humans, even when taken in overdose (Nichols & Grob, 2018). One case series of massive accidental overdose of LSD in eight patients measured gastric levels of 1,000 to 7,000 micrograms per 100 ml, implying an estimated ingested dose in the milligrams range, induced varied effects among the 8 including coma, platelet dysfunction, agitation and psychosis (Klock et al., 1974). However, after supportive care in hospital, all 8 patients were asymptomatic at 12 hours, and were discharged within 48 hours. 5 patients available to follow-up over 1 year showed no overt residual effects (Klock et al., 1974).

### 3.3. Dimethyltryptamine (DMT)

Dimethyltryptamine (DMT) produces vivid visual hallucinations, a labile affective state, and alterations of somatic sensation. At higher doses the experience is described as completely replacing preceding mental and sensory experience during its effects (Strassman et al., 1994). Bolus and infusion regimes of intravenous DMT produce different trajectories of effects; boluses produce highly intense experiences peaking within two minutes, whereas infusions without bolus produce longer lasting, dose-dependent and gradual effects that increase with accumulation of central DMT over the course of the infusion (Luan et al., 2024; Vogt et al., 2023). The subjective experience of an intravenous bolus of DMT peaks at one to two minutes, and the overall experience lasts an average of 30 minutes (Strassman et al., 1994). The mean half life of DMT is 10 – 12 minutes, however there is significant variability between individuals (Good et al., 2023). The peak in both blood level and subjective effects is within two minutes of intravenous administration (Strassman & Qualls, 1994). 0.2mg/kg of intravenous DMT has been found to be the threshold for hallucinogenic effects (Strassman et al., 1994). Inhaled DMT has a similar profile to intravenous with rapid onset and a duration less than 30 minutes with doses used at 25-40mg (Pallavicini et al., 2021; Riba et al., 2015). Studies of DMT pharmacokinetics have previously used weight-based doses (Kaplan et al., 1974; Strassman & Qualls, 1994). However, BMI and weight are not predictive of peak levels, indicating that fixed dosing regimes may be sufficient (Good et al., 2023).

DMT is the main psychoactive component of ayahuasca, but is also found endogenously in many mammals including humans (Chaves et al. 2024). Ayahuasca also contains β-carboline alkaloids that render the DMT orally bioavailable via monoamine oxidase inhibition; DMT alone is not orally bioavailable (Good et al., 2023). As a result, studies have administered DMT by inhalation (Riba et al., 2015), intramuscular injection (Rosenberg et al., 1964) or intravenously (D'Souza et al., 2022). Parenteral DMT avoids the number of confounding psychoactive substances in ayahuasca affecting results of clinical trials (Vogt et al., 2023). This also avoids potentially unwanted effects of ayahuasca such as severe gastrointestinal symptoms, despite some communities that use ayahuasca considering these symptoms integral to an experience (Fotiou & Gearin, 2019). Combination intranasal DMT and bucco-oral harmine is a newer method, informally known as 'pharmahuasca' or ayahuasca-analogue, which is well tolerated and may be an alternative to oral and parenteral routes, and may allow for better predictability and fine tuning of the psychological experience via intermittent nasal doses (Barker, 2022; Dornbierer et al., 2023).

DMT produces rapid, dose-dependent rises in blood pressure and heart rate, pupil diameter and temperature (Strassman & Qualls, 1994). A tolerability study found DMT generally increases heart rate, systolic and diastolic blood pressure, however one participant experienced symptomatic bradycardia that resolved with IV fluids and re-positioning (D'Souza et al., 2022). Hence as with other psychedelics, caution is advised in those with hypertension and other cardiovascular risk factors. DMT increases prolactin and cortisol 5-15 minutes after subjective experience peaks, and growth hormone 15-30 minutes after peak experience (Strassman & Qualls, 1994). While earlier studies suggested that DMT does not produce tolerance to its psychedelic effects, and only showed reduced hormonal and physiological responses with repeated dosing (Strassman et al., 1996), more recent research using continuous intravenous infusion has demonstrated moderate acute tolerance to subjective effects over time (Erne et al., 2024).

A 2007 review of the risks of psychoactive recreational substances found that evidence on lethal DMT doses in humans is scarce, and failed to find any reports of human fatalities relating to DMT use (Gable, 2007). In rat models, the equivalent of 50 times the 'standard' oral dose (a maximum of 15mg/kg given) of DMT during rituals was tolerated acutely, producing behavioural changes but with no permanent damage found at 14 days (Pic-Taylor et al., 2015). Systematic reviews found no evidence ayahuasca or DMT given in controlled settings to be associated with prolonged psychotic reactions, with the most common adverse effects being vomiting, nausea and anxiety (dos Santos et al., 2016; dos Santos et al., 2017). However, it should be noted that participants with previous psychiatric history are often screened out in study settings (dos Santos et al., 2017). Case reports of ayahuasca or DMT related psychosis in those with previous diagnosis of a psychotic disorder, a family history of psychosis or concurrent polysubstance use indicate that caution is required in these cases and this should inform exclusion criteria for studies (Paterson et al., 2015; dos Santos et al., 2017; Szmulewicz et al., 2015; Warren et al., 2012).

### 3.4. 5-MeO-DMT

Although described as an ‘atypical’ psychedelic due to its uniquely distinctive pharmacology and phenomenology (Dourron et al., 2023), 5-MeO-DMT is worthy of mention in this review given the recent surge in clinical research with this drug (Ermakova et al., 2021; Reckweg et al., 2021, 2023; Rucker et al., 2024; Uthaug et al., 2020). Given the potent acute psychological effects of 5-MeO-DMT, as well as its Class A and Schedule I status in the UK, research with 5-MeO-DMT faces similar ethical, scientific, and logistical considerations as research with classical psychedelics, therefore necessitating similar approaches and precautions (Ermakova et al., 2021).

5-MeO-DMT is found in the toxins of the Colorado river toad, in a variety of plant and fungi species, has been suspected to be endogenously produced in humans, and induces a variety of effects including emotional changes, perceptual distortions, described ‘mystical’ experiences and ego dissolution (Dourron et al., 2023; Reckweg et al., 2022). It is metabolised by monoamine oxidases (MAO, predominantly MAO- A) and cytochrome p450 2D6 (CYP2D6) to produce inactive 5-methoxyindoleacetic acid and active metabolite bufotenine (Shen, Wu, et al., 2010; Yu et al., 2003). CYP2D6 status affects response to 5-MeO-DMT, and CYP2D6 status may in the future inform individualised prescribing, as is beginning to be the case with other psychotropics (Bousman et al., 2023; Shen, Wu, et al., 2010).

In non-medical settings, 5-MeO- DMT has been used alongside MAO inhibitors such as harmaline to extend and intensify the experience, consistent with studies demonstrating increased exposure when harmaline is concurrently given (Ott, 1999; Shen, Jiang, et al., 2010). Harmaline’s 5-HT agonism and MAO inhibition, slowing both 5-HT and 5-MeO-DMT metabolism, can lead to toxicity (Halman et al., 2024; Shen, Jiang, et al., 2010). Case reports describe lethal toxicity from this combination (Halman et al., 2024; Sklerov et al., 2005; Tanaka et al., 2006). The interaction between MAO inhibitors, CYP2D6 status, 5-MeO-DMT and bufotenine is complex and further study is required in humans; caution is advised with concurrent use of CYP2D6 inhibitors and MAO inhibitors (Shen, Jiang, et al., 2010; Shen, Wu, et al., 2010).

5-MeO-DMT can be inhaled (Reckweg et al., 2021; Uthaug et al., 2019) insufflated (Rucker et al., 2024), smoked as part of a preparation of extracts for non-medical use (Barsuglia et al., 2018), given intramuscularly (Uthaug et al., 2020), or given intravenously (Timmermann et al., 2018). 5-MeO-DMT is not active when consumed orally (Shulgin and Shulgin, 1997). Intravenous administration produces subjective effects almost immediately, peaking at 2-3 minutes and mostly subsiding by 20 minutes (Timmermann et al., 2018). Peak experience is described in non-medical use to occur in seconds to minutes when smoked, or at 30 – 40 minutes if insufflated (Davis et al., 2018; Shulgin and Shulgin, 1997). Clinical studies on insufflated 5-MeO-DMT found subjective experience peaked within 10 minutes and resolved by 45 – 90 minutes, and this correlates with peak plasma concentrations achieved at 8-10 minutes (Rucker et al., 2024). Peak subjective experience with intramuscular delivery is reported, in surveys and anecdotally, to be achieved at 1 – 6 minutes, with an experience that can last 40-60 minutes (Uthaug et al., 2020). Intensity of subjective experience appears to be significantly associated with dose (Reckweg et al., 2021; Rucker et al., 2024). Mean half-life measures of insufflated 5-MeO-DMT ranged from 15 – 27 minutes (Rucker et al., 2024). Plasma measures of 5-MeO-DMT are very low at 1 hour post-dose and nearly undetectable at 3- 4 hours post-dose (Reckweg et al., 2021; Rucker et al., 2024). Plasma levels of metabolite bufotenine appear not to be above detectable thresholds at any time post-dose in human studies (Reckweg et al., 2021; Rucker et al., 2024).

There is evidence to suggest that individual subjective responses to a dose vary (Reckweg et al., 2021; Reckweg et al., 2023; Rucker et al., 2024). A study of inhaled 5-MeO-DMT in healthy participants found all doses higher than 2mg (doses: 6mg, 12,mg ,18mg) elicited significantly higher MEQ-30 ratings than 2mg, however 8 of 22 participants reported peak experiences at differing doses (Reckweg et al., 2021). Similarly, Rucker et al. found that some participants taking insufflated 5-MeO-DMT achieved peak subjective effects at 8mg, with others not achieving it at 12mg (Rucker et al., 2024). To overcome this some studies have used multiple doses to facilitate achieving peak experience, and the short half-life allows for multiple doses on a single day (Reckweg et al., 2023).

The majority of adverse events in clinical studies are mild and self-limiting and none severe or requiring participants to withdraw (Reckweg et al., 2021, 2023; Rucker et al., 2024). In a survey of use in non-controlled settings, 37% of responders described having challenging experiences, both psychological (most commonly anxiety, sadness) and somatic (most commonly feelings of body shaking, heart racing) (Davis et al., 2018)  Post-dose transient rises in blood pressure and heart rate that normalise by 90 minutes have been reported, with no clinically significant changes on electrocardiogram, laboratory blood tests or other vital signs (Rucker et al., 2024). Inhaled 5-MeO-DMT has been found to increase salivary cortisol and decrease salivary interleukin-6, with neither correlating with ratings of psychedelic experience, and to have no significant effect on salivary CRP or interleukin- β (Uthaug et al., 2019).

Reactivation experiences appear to occur with higher frequency with 5-MeO-DMT than other psychedelics, with re-experiencing of aspects of the psychedelic state over the following days to weeks in manner similar to HPPD (Davis et al., 2018; Ortiz Bernal et al., 2022). The reactivation associated with 5-MeO-DMT was described as positive or neutral by the vast majority of survey respondents from both structured settings (96%) and the general population (93%) (Davis et al., 2018; Ortiz Bernal et al., 2022). The 5-MeO-DMT reactivation phenomenon appears to be short-lived when compared to the descriptions of long durations of HPPD, and is more similar to HPPD Type 1 benign short-lived experiences, than HPPD Type 2 which is more distressing (Martinotti et al., 2018; Ortiz Bernal et al., 2022). An online survey found that inhaled 5-MeO-DMT has a higher likelihood of reactivation than intramuscular injection; 69% versus 21% respectively in 27 participants, although re-dosing being more common when inhaled is a potential confounder (Uthaug et al., 2020). Increased odds of reactivation has been found to be associated with being female, older age at first dose, higher educational level and use in a structured group setting (Ortiz Bernal et al., 2022). Further study of this phenomenon is required to inform clinical use of 5-MeO-DMT.

## 4. Safety considerations

### 4.1. Inclusion and exclusion criteria

In all human research studies, carefully considered inclusion and exclusion criteria should be included to ensure participant suitability, representativeness of the target population, and safety. Strict criteria can result in higher internal validity at the expense of external validity, and vice versa. Given the potent acute psychological and physiological effects and unique cultural significance of psychedelics, there are some specific eligibility issues to consider. We outline these below along with recommendations for how to address them. These relate to mental health history, neurological conditions, and previous psychedelic experience. There is no universal consensus on exactly how these should be considered within study eligibility criteria, but they will have a substantial impact on findings and generalisation thereof. As such, this section provides a breakdown of some of the key criteria that we believe should be considered in all human psychedelic research. For an example of exclusion/inclusion criteria of an ongoing research study (a non-CTIMP investigating the neural correlates of subjective experience under LSD at King's College London), see Appendix 1.

#### *4.1.1. History of psychiatric illness*

Psychedelics can have psychologically destabilising effects in some individuals, so it is important that care is taken in the selection of participants (Johnson et al., 2008). Assessing the mental health history of participants is crucial, particularly in the context of research on novel or less well-established psychedelic compounds and interventions. However, if further reassuring data on the safety of the short- and long-term psychological effects of psychedelics continues to be obtained, mental health related exclusion criteria may gradually be relaxed in order to be as inclusive as possible, reaching groups that are currently disenfranchised from developments in psychedelic science and medicine.

Similarly, there has been a prevalent assumption that psychedelics have the potential to induce psychosis in predisposed individuals (Nichols, 2016). This has contributed to many studies screening out those with a personal or family history of psychosis or psychosis-related disorders. However, evidence for this assumption remains inconclusive, with mixed findings in the literature. Population-level analysis shows no strong evidence for a link between psychedelic use and psychosis (Dourron et al., 2025; Krebs & Johansen, 2013b; Sabé et al., 2025), though it remains possible that risk is dose or context-dependent (indeed, Simonsson et al., 2025, found evidence for a context-dependent link between psychedelics and psychosis, with those taking psychedelic drugs in illegal contexts having a higher risk of symptoms). A recent survey of 100 people who used psychedelics found that use was associated with a decrease in the number of psychotic symptoms in individuals with a personal history of psychotic disorders (but an increase in psychotic symptoms in individuals with a personal or family history of bipolar disorder; Honk et al., 2024).

Aligning with this, in a large sample of Swedish adolescent twins, psychedelic use was significantly associated with lower rates of psychosis (but higher rates of manic symptoms in those genetically predisposed to schizophrenia or bipolar disorder; Simonsson, Mosing, et al., 2024). On the other hand, a retrospective cohort study found that individuals with emergency department visits involving hallucinogens had a 21-fold higher risk of developing a schizophrenia spectrum disorder compared to the general population (Myran et al., 2024). While this finding suggests an association, the underlying causal mechanisms, potential mediating factors, and, crucially, differences between different hallucinogens (a term used by Myran et al. to encompass substances beyond psychedelics, including dissociative drugs such as ketamine), remain unclear. It is important to note that much of the evidence discussed above is based on population-level data, which reflects recreational psychedelic use rather than administration in controlled clinical or research settings, potentially limiting its direct applicability to these contexts.

In parallel to psychosis, history (or apparent risk) of suicidality has been listed as an exclusion criterion in psychedelic clinical trials for depression (e.g. Carhart-Harris et al., 2021; Goodwin et al., 2022; Rucker et al., 2021). This is due to the concern that psychedelics may exacerbate suicidal tendencies in those already vulnerable. Research has found that some individuals attempt suicide in the context of negative psychedelic experiences (Carbonaro et al., 2016). It is of note that this was three (0.15%) of the sample of 1,993 individuals assessed in this study. Firm evidence for an exacerbating effect of psychedelic use on suicidality is limited, with population-level analysis and a systematic review showing no strong evidence for a link between psychedelic use and increased suicidal ideation or behaviour (Johansen & Krebs, 2015; Zeifman et al., 2021).

Compiled evidence shows that psychedelic use is associated with reduced suicidality (Hendricks, Johnson, et al., 2015; Hendricks, Thorne, et al., 2015). In clinical settings, meta-analyses have revealed that psychedelic therapy is associated with significant reductions in suicidality (Zeifman et al., 2022), and that no suicides have been reported in psychedelic research studies in an aggregated sample of 3504 participants (Hinkle et al., 2024). One systematic review found consistent reductions in suicidal ideation following psilocybin-assisted therapy in randomised controlled trials, though results from non-randomised and observational studies were more mixed (Meshkat et al., 2025). Taken together, this evidence suggests that suicidal ideation or risk as an exclusion criterion may be overly stringent. In addition, given the estimates that 37.7% of patients with depression experience suicidal ideation and that suicide planning prevalence is 15.1% (Cai et al., 2021), it could be considered inequitable to exclude those with a history of suicidal ideation from these trials, as they constitute a large portion of patients with MDD – particularly treatment-resistant depression – and are potentially the ones who could benefit the most from these treatments. The evidence suggests that it may be safe to remove this exclusion criterion from late-phase psychedelic clinical trials for depression.

That said, it is important to distinguish between people experiencing suicidal ideation and those with a history of attempted suicide, especially if there have been multiple attempts. Repeated suicide attempts are often a feature of Cluster B personality disorders (Pompili et al., 2005; McClelland et al., 2023). This can make clinicians wary about administering a psychedelic, given that coping mechanisms or psychological robustness may already be limited in these patients. Indeed, Marrocu et al. (2024) found that negative responses in psychedelic trials were disproportionately associated with personality disorders, suggesting increased risk in this population.

Nevertheless, in a selection of prospective observational studies, whilst Gordon et al. (2024) found that psychedelic use amongst individuals with personality disorders led to several cases of increased anxiety and depression severity, they also found overall improvements in psychological functioning. Another study observed improvements in emotion dysregulation in individuals with personality disorders who underwent a single ayahuasca experience (Domínguez-Clavé et al., 2019), suggesting psychedelics may have treatment potential in personality disorders (Zeifman & Wagner, 2020). These mixed findings indicate the need for further research on the effects of psychedelics on those with personality disorders, but at this early stage, it may be wise to include personality disorders as an exclusion criterion for psychedelic research.

In summary, with the accumulation of more data on adverse events, it may become evident that psychedelics, when administered in a safe and controlled environment, do not exacerbate the risk of psychosis or suicidality. This would allow researchers to consider removing these conditions as exclusion criteria, thereby reaching a broader demographic. At this stage, special attention should be given to individuals with personality disorders, although this risk might be mitigated with careful psychological support and a strong therapeutic alliance (Marrocu et al., 2024). Additionally, there may be value in understanding which specific aspects of personality disorder phenotypes (e.g., interpersonal difficulties that may make establishing rapport with a therapist challenging, or need for control over self/other/environment that might make the experience more distressing; Wood et al., 2024) particularly contribute to the higher risk for negative immediate responses to psychedelics, and what psychological mechanisms nevertheless predict long-term positive outcomes for some of these patients.

| **Recommendations** | |
|---|---|
| 8 | Given that psychedelic administration in controlled settings increasing the risk of psychosis remains unsubstantiated, and participants with a family history of psychosis are not contraindicated from benefiting from psychedelic therapies, their exclusion from psychedelic trials must be carefully considered. |
| 9 | Exclusion criteria considering suicidality should be more specific to suicide planning or previous suicide attempts, as opposed to the much broader criterion of suicidal ideation. Even in cases of past planning or attempts, automatic exclusion may not be justified. |

#### *4.1.2. Neurological conditions*

It has been proposed that psychedelics may have the potential to treat a variety of neurological conditions, including stroke, traumatic brain injury, functional neurological disorder, and neurodegenerative disorders (Butler et al., 2020; Khan et al., 2021; Saeger & Olson, 2022). However, due to the largely unknown effects of psychedelics on individuals with neurological conditions, such individuals are typically screened out of psychedelic studies.

It is unclear whether psychedelic use is associated with changes in seizure frequency (Soto-Angona et al., 2024). Therefore, a cautious approach is warranted when considering individuals with epilepsy for psychedelic research participation. More research is needed on the effects of psychedelics in preclinical animal and organoid models of epilepsy (Freidel et al., 2023). If it appears to be the case that psychedelics do not worsen seizures, epilepsy could potentially be removed as an exclusion criterion. However, if the psychedelic trial is not aiming to target neurological conditions, it is wise to exclude them from the study to avoid unforeseen adverse effects. At least, until more is known on how such conditions respond to psychedelics.

#### *4.1.3. Physiological contraindications*

There are also a number of physiological factors that are considered as exclusion criteria in most psychedelic trials. Many of these are considered because they are standard precautionary exclusions in many clinical trials: for example, participants with hypo- and hypertension, brady- and tachycardia, people who are severely underweight or severely obese, disorders of hepatic metabolism, and ages outside of 21-65. There are also exclusion criteria based on potential drug interactions, such as prescriptions of monoamineoxidase-inhibitors (MAOIs) or other serotonergic drugs. However, while there may be reasons for excluding people with cardiac abnormalities given the effects of some psychedelics on the human heart (Neumann et al., 2024), there is little evidence that directly supports any of these physiological exclusion criteria – more research would be needed to confirm their relevance to psychedelic studies.

#### *4.1.4. Previous psychedelic use*

Previous psychedelic use is crucial to consider as a criterion for recruitment of participants. Some studies and trials recruit individuals with no psychedelic experience (e.g. Griffiths et al., 2006; Timmermann et al., 2019), some only recruit psychedelic-experienced individuals (e.g. Carbonaro et al., 2018; Carhart-Harris, Erritzoe, et al., 2012), and some recruit a mix of both (e.g. Carhart-Harris, Bolstridge, et al., 2016; Johnson et al., 2014). Some trials have also, in their healthy volunteers, used a criterion of a previous 'mind-altering' experience (with psychedelics, ketamine, or even high dose THC) in order to balance expectation and anxiety (Gabay et al., 2018).

Inclusion of psychedelic-experienced individuals depends on the study goals. For example, if the study aims to assess the safety and feasibility of a novel psychedelic-assisted intervention, or aims to administer psychedelics in an environment that may induce additional stress in participants (e.g., fMRI), it may be a safer option to recruit individuals more familiar with psychedelics (Johnson et al., 2008). However, if a study aims to assess the efficacy of a psychedelic-assisted intervention that may, in future, be implemented on a large scale, it is advisable that psychedelic-naïve participants are included in the study to better reflect the population that this intervention will target. Given that an estimated 17% of adults aged 21 to 64 in the United States have used a psychedelic drug (in this case, LSD, psilocybin, or mescaline) at least once in their lifetime (Krebs & Johansen, 2013a), the external validity of clinical trials only recruiting psychedelic-experienced individuals can be brought into question.

Incorporating this recommendation will help avoid selection bias, positive effect size inflation and provide additional data on how prior psychedelic experience might mediate study outcomes, particularly if extent of previous use is measured. Psychedelic-naïve participants may also have fewer preconceived expectations about the effects of the intervention (Aday et al., 2022), although this possibility may be diminished as public discourse about psychedelics grows.

Crucially, there is a safety consideration regarding previous psychedelic experience; if a prospective participant reports a high proportion of negative experiences with psychedelics, it may be unsafe or unethical to include them in the study, as they may have a higher likelihood of experiencing a negative reaction to the study drug. This determination should be made by qualified clinical psychologists, psychiatrists, or equivalent professionals.

| **Recommendations** | |
|---|---|
| 10 | Inclusion or exclusion of participants on grounds of previous psychedelic use are not universally delineable: they should be tailored to the goals of the given study. |

#### ***4.1.5. Screening***

Screening for psychedelic studies is often a multi-step process: e.g., an online questionnaire to make an initial assessment based on inclusion and exclusion criteria, followed by a phone call, after which another screening may be conducted by a clinical psychologist or psychiatrist. In clinical psychedelic trials and studies with drug-naive participants, expert psychological or psychiatric final assessment is advisable. This is particularly pertinent for some criteria relating to current and historical psychiatric symptoms and other issues that may prove to obstruct the establishment of rapport or safe psychedelic administration. Accredited mental health professionals can, if necessary, justify inclusion of participants who might go on to experience an adverse reaction. Psychedelic studies will often require a medical doctor or clinical psychologist to be involved in screening assessments and make a final judgement on eligibility. Most commercial clinical trials require a secondary, external 'medical monitor' and/or independent rater's approval on top of the study site's decision. Although these safeguards are important, we acknowledge that dedicated participants – motivated by curiosity about the psychedelic experience, rather than ill-intentions – can inappropriately gain admittance to psychedelic studies through deception, which even experienced mental health professionals might not detect. This underscores the particular importance of proper documentation of decision-making during screening and assessment for suitability.

Accordingly, some previous psychedelic trials have included subjective clinical judgement as a potential exclusion criterion (e.g. Rucker et al., 2021). In addition to the approved inclusion and exclusion criteria of a study, there are other elements of a potential participant's situation that need to be considered at the point of screening. These considerations relate to psychosocial elements of a participant's presentation. Although tending to be less well-defined than psychiatric diagnostic eligibility criteria, consideration of these in decision-making on inclusion is left to the discretion of the research team, and they ultimately contribute to the optimisation of the participant's safety. These include, but are not limited to: availability (e.g. do their work commitments permit participation; caring responsibilities?, support network (do they have an adequate support network given the potential for a destabilising experience?), previous experiences with drugs (particularly psychedelics), previous therapy experiences, and ability to build rapport with at least one member of the research team. There is some debate within our authorship about this: researchers focused on basic mechanisms of psychedelic action question the scientific rigour of studies with flexible inclusion and exclusion criteria; researchers conducting clinical trials argue that patient safety and wellbeing is the overarching ethical principle of clinical trials, and that there are considerations which are not always captured in the inclusion and exclusion criteria of a study.

We therefore recommend that, at least for clinical trials, there is a balance to be struck between gatekeeping access to a clinical trial (often related to the study team's anxieties around keeping the potential participant safe), and making sure the trial is a safe and realistic option for them at point of screening. This does not necessarily have to be an authoritative process, and can involve open conversations with the participant. However, some participants will inevitably be disappointed with the decision made by the team. Importantly, for some individuals, hopes of efficacy for long-lasting mental health conditions, or the incentive of potential receipt of pharmaceutical grade psychedelic drugs may cause them to downplay or exaggerate symptoms. A degree of trust is required on behalf of the experimental team, but a multi-stage and multi-agent recruitment process provides some safeguard against these cases.

Additionally, to avoid participant misrepresentation or misremembering regarding previous health conditions that may render them ineligible, we recommend, if resources and ethical permission allow, that participant medical records are obtained and assessed by the research team. This is particularly applicable to clinical trials, in which confirmation of an official diagnosis (or treatment status) is required, and in studies of psychiatric/mental health populations, where diagnosis cannot be verified by a biological diagnostic test. This is not to generate a culture of mistrust between patient and study team, but given the current hype around the potential of these drugs, potential participants may omit information in order to get accepted onto a trial. However, there is debate amongst the authorship regarding whether this approach is necessary for safety or undermines participant autonomy, with some suggesting that researchers should trust self-reported histories unless there is clear reason for doubt. Regardless, itt may often be the case that the sponsor for a trial requires this measure.

### 4.2. Recording and managing adverse events

Adverse events (AEs) in psychiatric research are often highly subjective, and within psychedelic research this becomes more nebulous still. Adverse events from psychedelics, as they are technically defined by medicine's regulators, are very common. However the regulatory definition is problematic from the get-go. Psychedelics likely amplify experiential intensity in a non-specific way, with the specific effects shaped by a range of non-pharmacological factors. From a regulatory perspective, an adverse event can be any experience whatsoever under a psychedelic that might not otherwise have happened if the user hadn't taken it. Taken together with regulatory hysteria around the risk of psychedelics inducing mania or psychosis, this has resulted in some bizarre absurdities. For example, recording 'positive mood' as an adverse event in a trial of psilocybin for treatment resistant depression.

Beyond the consistently-reported effects of psychedelics (e.g. visual distortions, altered time perception, emotional lability, etc.), it becomes challenging to categorise the wide range of experiences reported by psychedelic users in a clinically meaningful way. Such effects are often positively experienced and, while they may technically fall within regulatory definitions of adverse events, are difficult to clinically classify as such. Delineating what constitutes an adverse effect, an expected effect, or a component of the intended therapeutic process is especially challenging (Hinkle et al., 2024). Both what the participant chooses to report and how the researcher interprets and records it are likely to strongly influence what is ultimately documented. Moreover, asking participants about adverse events may itself introduce iatrogenic effects, particularly in the context of heightened suggestibility induced by psychedelics (Simonsson, Goldberg, & Hendricks, 2024). Inevitably this leads to varied, non-specific reporting of AE data. However, at present, there is no clear alternative to this approach. In general, we have taken a pragmatic approach to AE reporting, recording events that are clearly beyond the bounds of the expected, distressing to the participant, or otherwise objectively identifiable (high BP, for example). Serious adverse events should always be reported.

In general, severe, serious, or prolonged adverse events are unusual, even with recreational use of psychedelics in non-controlled settings. Fatalities are even rarer, and most psychedelic associated deaths are due to accidents or polysubstance use (Kopra et al., 2025). Most of the danger and challenges arise as a result of psychological effects, such as anxiety or psychotic-like symptoms, which might result in behavioural disturbances, residual mental health problems, self-injury, or accidents (Evans et al., 2023; Kopra, Ferris, Rucker, et al., 2022; Kopra, Ferris, Winstock, et al., 2022; Kramer et al., 2023).

Given that early psychedelic studies were often conducted by researchers with a favourable view of psychedelics – understandably, as few others were motivated to pursue such research during periods of stigma – it is plausible that the adverse event rate was (albeit unintentionally and unconsciously) under-reported (Bender & Hellerstein, 2022; Breeksema et al., 2022; Hinkle et al., 2024; de Laportalière et al., 2023). Regardless, it speaks to the need for pharmacovigilance to remain a priority in psychedelic research, and continue to be emphasised if and when they are approved for wider use. An agreed framework for reporting (S)AEs in psychedelic studies (similar to the one proposed by Palitsky et al., 2024, for psychedelic-assisted therapies) will allow synthesis of these data across study sites and an ability to detect less common adverse events.

| **Recommendations** | |
|---|---|
| 11 | Wider 'suitability for inclusion' factors, such as social support network, therapy and therapist suitability, and ability to commit to often lengthy trials, may be necessary to consider at screening, at least for clinical trials. |
| 12 | Researchers should record and report, as far as is reasonable, all adverse events (and particularly serious adverse events), and contribute this information to public registries that can be used by other researchers to prepare their risk assessments and ethical reviews. |

#### *4.2.1. Distress protocols*

In psychedelic studies, the phrase 'set and setting' is often employed to refer to the psychological mindset and physical environment (respectively) surrounding an acute psychedelic drug experience. Given that psychedelics can render the user more sensitive to their environment, psychedelic studies necessitate specific dosing environments (see Section 6 – Experiential considerations). In the experience of the authors, internal and external review boards are interested in reviewing safety procedures and distress protocols. These protocols refer to the ways in which distress elicited by the (effects of the) psychedelic experience is managed.

Distress protocols vary across studies, but generally, for psychedelic administrations, involve 'low-intensity' behavioural interventions (grounding techniques, breathing techniques, dialogue between participant and therapist) in the first instance, followed by rescue medication only if absolutely necessary (i.e. if participant is seriously distressed and all non-pharmacological interventions have failed, and/or at risk of harming self or others).

The listed examples are simple behavioural techniques, which do not require extensive training of researchers, and are readily learned and deployed by participants. By contrast, the use of rescue medications including oral and parenteral benzodiazepines and antipsychotics is invasive, and employed only in the very rare cases that non-pharmacological de-escalation has failed (see Section 4 – Safety considerations) (Romeo et al., 2024). However, use of rescue medication in such contexts is debated, and lacks evidence as an effective approach. In addition, usability is partially dependent on the specific drug, i.e. being ineffective for DMT or other short-acting psychedelics. If a participant becomes distressed during an acute dosing session, it will be the role of the professional in the room with them (often a psychotherapist or facilitator with substantial experience in supporting psychedelic experiences) to escalate this to the wider team if in need of further assistance. As with all delicate clinical decisions, practitioners must consider the immediate needs of the participant, their professional expertise, and the study protocol.

The standard use of both psychological and pharmacological rescue protocols highlights the need for an interdisciplinary approach, in which psychotherapists, psychiatrists, and scientists collaborate. Some practitioners might argue that there is no such thing as a ‘bad trip’ (a bad acute psychological experience), and that negative material that arises during an acute experience needs to be, and should be, worked through (Barrett et al., 2016; Carbonaro et al., 2016; Carhart-Harris, Kaelen, et al., 2016; Gashi et al., 2021).

Such a position appears, inherently, to lack clinical equipoise and, we would not, therefore, echo it. Whilst the vast majority of clinical scenarios with psychedelics might be managed non-pharmacologically, serious and prolonged distress, or imminent violence to self or others, may necessitate medical intervention.

#### *4.2.2. Long-term adverse outcomes*

Some users of psychedelics experience persisting negative effects after psychedelic use. These are significantly more common in situations where the psychedelic is used recreationally, outside of a controlled environment, and often may be due to polypharmacy or impurity of the substance (see Section 3 – ‘Pharmacological considerations’). Hallucinogen persisting perception disorder (HPPD), a syndrome characterised by chronic perceptual distortion lasting weeks, months, or even years after the cessation of psychedelic use, is one such example (Martinotti et al., 2018). The prevalence of HPPD is unclear, with estimates differing depending on the type of HPPD (type I or II) and frequency of hallucinogen use (Orsolini et al., 2017). Its aetiology is notably opaque, and under-ascertained. Research on HPPD should aim to elucidate its frequency, nature, functional impact, aetiology and evolution over time (Doyle et al., 2022).

Other long-term adverse effects which may arise from psychedelic use can include more severe forms of anxiety, social phobia, depersonalization, and derealization (Evans et al., 2023). Further research is needed on static and modifiable risk factors for such long-term adverse psychological responses following psychedelic use and how these may be mitigated by optimising ‘tripping’ environments or dosing practices.

| **Recommendations** | |
|---|---|
| 13 | Approval frameworks should focus on contextual safety factors that reduce the likelihood of distress occurring, rather than pharmacological interventions for distress. |
| 14 | There should be ongoing clinical and research consideration of the delicate distinction between allowing negative emotions that arise to be addressed in a safe and contained context, and permitting a distressing experience with potential long lasting psychological toxicity to continue. |

# 5. Study design considerations

## 5.1. Participant Recruitment

### *5.1.1. Representative samples*

Mirroring trends in clinical trials and drug development more generally, the lack of representative samples has been an ongoing issue in psychedelic research, both for basic science studies as well as clinical trials (Hartmann et al., 2013; Michaels et al., 2018). There have been minimal improvements over the past few years, with 85.6% of participants enrolled in psychedelic clinical studies between 2017-2024 being non-Hispanic White (Hughes & Garcia-Romeu, 2024) and predominantly wealthy – e.g., in Raison et al., 2023, ~50% of participants had an annual income of over $100,000, which is in the top 10% globally (www.wid.world/income-comparator). Recent findings suggest that the psychological effects of psychedelics differ by gender and education, reinforcing the need for intersectional recruitment strategies (Viña, 2025). Furthermore, there have been concerns that psychedelic trials are failing to adequately include disabled people, thereby perpetuating accessibility barriers and health inequities within healthcare (Mintz et al., 2022).

Additionally, in the experience of the authors, a larger proportion of male subjects often volunteer for non-clinical psychedelic studies, which is an important consideration given sex-dependent variations in psychedelic drug absorption and potential estrogen-serotonin interactions (Shadani et al., 2024). The reasons for these unrepresentative samples are unclear, but such observations are not restricted to studies using psychedelics, and tend to apply across the board. Factors may include a lack of trust in the medical system, medical establishments failing to proactively reach out to underrepresented groups, cultural and personal attitudes towards psychedelics varying by demography, and social barriers (e.g. lack of time to participate in an unpaid study). Strategies for increasing diversity include (but are not limited to), diversifying research teams, training to encourage cultural competency, and inclusive design of inclusion/exclusion criteria throughout screening and consent processes (Williams et al., 2020). In addition, to account for potential sex-based differences in response, we recommend stratified randomization by sex in study designs where feasible.

In the case of psychedelic clinical trials, an example of inclusive study criteria in a psychedelic clinical trial would be permitting diagnosis of a condition (e.g. PTSD) to be made at an initial screening visit by a clinician, rather than requiring evidence of diagnosis via general practitioner (GP) records. This is particularly relevant for groups that may not engage with their local GPs for mental health treatment but have preferred to seek private psychotherapy, where formal diagnoses may not be made. An additional consideration in this clinical setting is the non-prescriptive task of matching participants with a 'suitable' therapist dyad for the duration of their trial participation. This is more of 'art' than 'science', but in our experience matching participants with therapists purely based on demographic characteristics ('ethnomatching', for example) is to be avoided in favour of a more nuanced approach.

Finally, researchers face the challenge of recruiting a variety of participants not only in terms of demographics but also attitudes towards, and experience with, psychedelic drugs. To maintain generalisability of study results, prior to launching recruitment, it may be prudent to develop approximate recruitment targets based on recent census and survey data regarding participant demographics and the proportion of individuals who have used psychedelic drugs previously (see Section 4.1.5 – 'Previous psychedelic use'). It is, however, intuitively obvious that those who volunteer for a psychedelic study are unlikely to hold a heavy negative bias towards them. Ambivalence is, perhaps, the ideal position of a participant (but usually rather hard to actually find).

#### *5.1.2. Advertising*

For most scientific studies, enrolling sufficient participants can be challenging and time-consuming. However, psychedelic research sometimes – uniquely – faces the opposite challenge. This is particularly the case if the study population is large (e.g., healthy volunteers or patients with depression) – a niche study population and/or strict exclusion criteria that significantly limit eligibility may inherently mitigate this challenge. Due to widespread excitement around psychedelics (see Section 2 – Sociocultural, political, and legal considerations), the prospect of receiving a psychedelic can attract a great deal of enthusiasm. In our experience, we have had to find diplomatic ways to turn down hopeful subjects, some of whom may lobby forcefully for an opportunity to participate. Some ethics committees or sponsors may request advertising material as a part of the application process, and it may for this reason be necessary to explain that advertising is not necessary (or indeed may be a hindrance) due to this unusual level of enthusiasm for participation. Leaders of research teams may need to be mindful of discussing their research on mass-media outlets, lest it results in their teams being overwhelmed by a deluge of self-referrals.

#### *5.1.3. Participant Deceit*

A further challenge arising from participants' eagerness to take part is the potential for participant deceit: participants may provide false information to gain entry into the trial. For example, participants in some of the studies conducted by authors of this paper have attempted to complete initial screening questionnaires under different email accounts after being rejected the first time (as identified from submission metadata). While the issue of participant deceit is a recognized concern across many areas of human research (Resnik & McCann, 2015), it is likely more pronounced in psychedelic research.

To mitigate the risk of deceitful participants passing through pre-screening or screening processes, it is advisable to keep certain inclusion and exclusion criteria (particularly those that could be easily falsified, such as criteria relating to previous psychedelic experience or history of mental health conditions) hidden from prospective participants. In addition, measures such as reviewing medical records or conducting hair, urine, and blood tests can help minimise the risk of participant deceit regarding medical history and drug use. However, the potential benefits of these measures must be weighed against the logistical challenges in executing them, the expense, and the impact it may have on the participant.

| **Recommendations** | |
|---|---|
| 15 | Psychedelic trialists should proactively aim to be inclusive, for example via recruiting samples with representative proportions of protected characteristics (e.g. ethnicity, gender identity, sexual orientation), previous psychedelic use, and socioeconomic statuses. |
| 16 | Sociodemographic information should be recorded at application, screening, and enrollment phases. |
| 17 | Protocols should include practical steps to encourage participation from marginalised groups, for example by allowing for diagnoses to be made at screening. |
| 18 | Conventional modes of advertising for recruitment may not be applicable to psychedelic studies due to overwhelming public enthusiasm – indeed, researchers may face the converse problem of participant deceit to enrol. |

### 5.2. Approaches to control conditions

It is a fundamental assumption in pharmacological studies that the effects of therapeutic compounds consist of pharmacological target effects and the placebo effect (Price et al., 2008). In almost all human drug studies, the use of placebo controls is therefore essential to establish the effects of the drug under investigation above placebo, with the 'gold standard' for placebo control being the double-blind. Psychedelic research challenges the assumptions of this stance, because the distinct and highly recognizable subjective effects of psychedelics result in functional unblinding (the participants' and experimenters' ability to discern whether they have received the psychedelic or the placebo).

Given that treatment models with psychedelics do not neatly map onto existing research paradigms (Muthukumaraswarmy et al., 2025; Roseman, 2024; Pronovost-Morgan et al., 2023), standard assessment frameworks – which prioritise biological drug effects while treating extra-pharmacological factors as nuisance effects – require some adaptation when evaluating psychedelic studies. This challenge was evident in the recent controversy surrounding the Food and Drug Administration's rejection of findings of the Phase III trial for MDMA-assisted therapy for post-traumatic stress disorder (Mitchell et al., 2023), in part because of concerns surrounding 'functional unblinding' contaminating the efficacy and safety evaluation (see Section 5.2 – 'Approaches to control conditions'). To fully capture the effects of psychedelic drugs, both the effects of drug and its contextual factors should be evaluated as contributors to clinical outcome (see Roseman, 2024; Muthukumaraswarmy et al., 2025). However whilst such ideas are easy to describe, their execution is anything but.

Awareness of condition assignment introduces an expectancy effect within the psychedelic group, potentially heightened societal hype, as outlined in Section 2 (Butler et al., 2022; Noorani, 2020; Noorani et al., 2023). Determining the appropriate control conditions for a particular study requires careful consideration of the consequences of functional unblinding and mechanisms by which it might bias outcomes.

Conversely, in the placebo group of drug trials, participants' realisation that they may have 'missed out' on a potentially efficacious treatment can induce a disappointment-driven version of the nocebo effect (i.e. disappointment that may diminish any potential placebo benefits or perhaps even lead to worsened outcomes; Butler et al., 2022). A recent systematic review on trials of psychedelics or escitalopram for depression found that the placebo response in psychedelic trials was lower than that in escitalopram trials, a finding possibly attributable to a disappointment-driven nocebo effect (Hsu et al., 2024).

Importantly, few studies have been designed to estimate the magnitude of the placebo effect itself. Doing so requires a 'nil intervention' group – where outcomes data are collected in the absence of any intervention (and, arguably, even any knowledge that such intervention is available). This would allow benchmarking of improvements or decrements that may be induced by placebo or nocebo and allows the scaling of magnitude of therapeutic effects of active drugs above this. Such studies may be highly informative for understanding the relative benefits of active effects in psychedelic medicine and beyond. However they are also ethically and logistically problematic to achieve, not least because participants in a nil intervention group have little reason to remain in follow-up.

Methods for managing and mitigating expectancy and disappointment effects in psychedelic studies and trials, several of which can be implemented alongside each other, are discussed below. As argued by Aday et al. (2022), blinding is "not an all-or-nothing" phenomenon. Employing multiple methods together can reduce the impact of confounding variables. In addition, regardless of the control conditions selected, or whether studies are clinically- or mechanistically-focused, we suggest that data on the phenomenological experience of the drug, expectancy of effects both before and after the drug administration (e.g. by using the Stanford Expectations of Treatment Scale, Younger et al., 2012; for other possible scales, see review by Szigeti & Heifets, 2024), participants' guesses about condition assignment, and their confidence in those guesses, as well as the qualitative content of integration interviews should be collected and published alongside primary outcome endpoints. These measures will allow researchers and those interpreting research to objectively appraise the differences in expectancy and disappointment effects between drug groups and assess how these inter-individual differences contribute to primary outcomes.

Psychedelic studies to date have generally placed emphasis on the preparation of ‘set’ and ‘setting’ (Leary et al., 1963) prior to the drug experience (see Section 6 – Experiential considerations). It has been argued that these practices maximise the placebo component of response to psychedelic drugs via increased expectancy, contextual and interpersonal cueing (Szigeti & Heifets, 2024); i.e., psychedelics are perhaps simply great facilitators of the placebo effect and placebos in these studies are great nocebos (Dupuis & Veissière, 2022). While this may seem a cynical and deflationary perspective, these challenges to determine a clear placebo condition in psychedelic studies perhaps reveal a deeper question: is it actually meaningful to treat placebo as an epiphenomenon, to be subtracted in order to understand a drug’s ‘pure’ effects? Or would it make more sense to see *all* drugs as catalysts and modifiers to the specific conditions they are administered within?

Developments in the field of placebo research are in fact moving to reconceptulise the placebo effect as the ‘meaning response’ (Hutchinson & Moerman, 2018), to acknowledge that so-called placebo effects are not driven by the placebo pill, injection, or therapy administered, but rather the meaning applied to those interventions by the recipient. In other words, the ‘top-down’ contextualisation of the action of, e.g., ingesting a pill is the key factor in the observable positive outcomes. Psychedelics also are known to enhance the sense of meaning (Carhart-Harris et al., 2015; Griffiths et al., 2006; Preller et al., 2017), and have been used in trials with patients with terminal cancer to support them in finding meaning in their circumstances (see Yaden et al., (2022) for a review). Indeed, the influential REBUS (‘Relaxed Beliefs Under Psychedelics’) theory of computational mechanisms by which psychedelics act in the brain (Carhart-Harris & Friston, 2019) suggests that psychedelics increase the variability of top-down contextualisation of sensory information, targeting the same high-order contextualisation that may be at play in the placebo effect.

| **Recommendations** | |
|---|---|
| 19 | Enthusiasm about positive results from psychedelic studies should be tempered, but not disregarded, due to functional unblinding effects, and the interplay of expectancy and hype. |
| 20 | Psychedelic researchers should include measures of expectancy in study designs. |
| 21 | Regulators should consider the fact that randomised controlled trials may not be the ‘best fit’ for psychedelic-assisted therapy, and that this should contextualise rather than undermine any regulatory decisions. |
| 22 | Researchers and regulators should consider the ways in which psychedelic trials may highlight weaknesses in current understanding of pharmacological evidence-gathering, in which treatment effects are characterised as simply pharmacology plus treatment-related placebo effects. |
| 23 | Blinding efficacy should be examined in controlled studies of psychedelics (i.e. by asking participants after dosing which treatment arm they think they were assigned to). |

#### *5.2.1. Active comparators*

Inert placebos may be seen as a design weakness in psychedelic trials and hence there have been calls for active comparators. An analysis by Nayak et al. (2023) revealed that 35% of randomised psychedelic studies from 1940 to 2020 with drug control conditions used an active comparator. These are typically other (non-psychedelic) drugs with discernible subjective (psychoactive) effects, such as nicotinic acid (Pahnke, 1963), niacin (Raison et al., 2023), methylphenidate (Griffiths et al., 2006), and dextromethorphan (Carbonaro et al., 2018), the purpose of which is to mimic some of the subjective effects of psychedelics without directly influencing the target process.

While active comparators may be appropriate in some cases, their recommendation may represent a reductive response to the issue of unblinding, which, upon deeper examination, does not serve the purpose for which it is suggested. For example, in Pahnke (1963), although niacin helped achieve successful blinding in the early stages of administration due to its overlapping physiological effects with psilocybin, by the end of the session, all participants correctly identified whether they had received psilocybin or niacin, as the two drugs produced entirely distinct subjective experiences (a discrepancy likely amplified by the group setting in which drug administration took place).

Nevertheless, researchers successfully masked drug allocation in some participants in an ayahuasca trial involving experienced users by developing a placebo that closely mimicked the taste and somatic effects of ayahuasca (Uthaug et al., 2021). This demonstrates that the active comparator approach can be valuable when carefully and thoughtfully implemented. Researchers must be clear *a priori* on the mechanisms they believe to be responsible for therapeutic efficacy in a given study (while bearing in mind that there is still much to be discovered regarding the underlying mechanisms of action of psychedelic drugs).

Potential therapeutic mechanisms are many in psychedelic studies, although careful design can allow researchers to tease these apart. If the proposed therapeutic mechanism under study is the subjective phenomenological experience of the drug itself, then it is – by definition – futile to attempt blinding. Equally, the suggestion that compounds like niacin can help maintain the blind in psychedelics studies is simply at odds with practical experience and blind guess data. Conversely, introducing a comparator with an alternative pharmacological action may introduce further confounds when examining the therapeutic effects of psychedelics.

| **Recommendations** | |
|---|---|
| 24 | Active comparators, whilst occasionally useful, should not be viewed as an effective means to mitigate the challenges posed by functional unblinding. |

#### *5.2.2. Sub-efficacious doses*

A more common practice in recent clinical trials has been to use a very low dose of the study drug (e.g., 1 mg psilocybin, Goodwin et al., 2022). This allows researchers to inform all participants that they will receive a dose of an active drug, which may temper expectancy effects. This is a viable option for some clinical studies, but does not at all surmount the issue of subjective differences between micro and full doses during administration, nor the issue of unintended active, therapeutic effects of the low dose.

Goodwin et al. (2022) included ‘low’ (1mg), ‘middle’ (10mg), and ‘high’ (25mg) dose psilocybin groups, with the ‘high’ dose groups equating to the target therapeutic dose. Participants are less able to disambiguate between the high and middle doses, and between the low and middle doses, but blinding efforts are still imperfect (Holze et al., 2021). Despite promising anecdotal and observational reports (e.g. Anderson et al., 2019; Fadiman & Korb, 2019; Lea et al., 2020), empirical microdosing studies have typically found no consistent effects of such low doses of psychedelics relative to placebo on chosen outcome measures post-acutely (Cavanna et al., 2022; Murphy et al., 2023; Szigeti et al., 2021), bar a placebo-controlled LSD microdosing study showing increases in sleep duration, which, importantly, did not report participant condition guesses (Allen et al., 2024). However, no clinical studies using a low dose control have compared this to a high dose *and* inert placebo. As such it is unknown whether the low-dose approach confers any clinical benefit over an inert placebo, which is crucial for determining their therapeutic relevance and justifying their use in treatment protocols.

| **Recommendations** | |
|---|---|
| 25 | Sub-psychoactive doses of psychedelics may be a useful comparator in some trials, however they do not solve the challenges posed by functional unblinding, and introduce problems of their own. |

#### *5.2.3. Dose-dependency*

If multiple doses of a psychedelic drug which produce a discernible 'psychedelic' experience are used in a study, then a greater therapeutic or mechanistic effect of the highest dose compared to a medium dose can be taken as evidence of a pharmacological effect beyond expectancy and placebo (e.g. see the use of 1mg, 10mg, and 25mg of psilocybin by Goodwin et al., 2022). However, depending on the doses selected, there may still be subjective differences between doses regarding the phenomenology of the experience, and these will again be important to quantify in order to determine their impacts on primary outcomes. Dose adjustment with psychedelics is no small task, given the low central concentrations at which they exert their effects (see Section 3 – Pharmacological considerations) and of the lack of sophisticated human pharmacokinetic models of these effects. Both fixed and weight-adjusted doses show considerable variability in the metabolism and subjective intensity of psychedelic drugs, meaning a dose that is 'underwhelming' for one individual may prove extremely intense for another.

#### *5.2.4. Deception*

Another option, rarely utilised in the field, is to capitalise upon the placebo effect using deception. For example, researchers can inform participants that they will receive one of multiple therapeutic compounds (or one of several doses of drug) or placebo, but not be told which. Superiority studies or trials may actually give these compounds, but earlier phase studies may simply give placebo or active drugs.

If participants are led to believe (i.e. deceived) that they are part of a multi-arm trial, in which some of the compounds under test have similar therapeutic effects to psychedelics, but without their subjective effects (e.g. lisuride, or a theoretical 'non-hallucinogenic psychedelic', with similar plasticity-inducing properties with psychedelics but without the hallucinogenic effects), those who in fact received a placebo may still believe they have received a therapeutically active substance. This may serve to retain more placebo arm participants in follow-up, reducing bias in the long-term efficacy estimates. This approach is different to the low-dose comparison in that it maintains the true pharmacological comparison between drug and placebo, but reduces participants' ability to deduce their true drug condition and may mitigate some of the expectancy and disappointment effects attendant upon functional unblinding. It also allows an examination of the role of treatment guesses on outcomes, in the absence of any true pharmacological difference. However, this strategy raises serious ethical issues, which, in the context of trials with treatment-seekers, cannot be overcome. This approach is more acceptable in experimental studies, in which participants are neither expecting nor being deprived of an efficacious therapy if assigned to placebo under deceptive conditions.

| Recommendations | |
|---|---|
| 26 | Ethical forms of placebo deception could be considered in the design of psychedelic studies with non-treatment seeking participants. |

#### *5.2.5. Open-label extension*

In an attempt to improve participant retention and mitigate the disappointment effects participants experience in the placebo conditions of psychedelic trials, some trials feature an open-label extension phase (e.g., Rucker et al., 2021). This entails giving eligible participants, assigned to both the experimental and placebo condition, the option of an open label extension after the main trial has been concluded. This method helps to address the ethical concern of participants feeling deceived or disappointed by being in the placebo group, as they are eventually given access to the active treatment.

The potential to receive an active dose as part of an open-label extension can also improve retention within the main arm of a study. Additionally, it provides the added benefit of having a group of participants undergo two active rounds of the intervention, offering additional dose-response data relating to efficacy and safety. However, this method is resource-intensive and thus heavily dependent on available funding and capacity. Additionally, it limits the study's ability to collect long-term placebo data, as participants originally assigned to the placebo group can no longer serve as long-term controls once they receive the active treatment in the extension phase.

#### *5.2.6. Sequential parallel comparison design*

Along a similar vein to the open-label extension method, the sequential parallel comparison design (SPCD) approach may enable more robust detection of treatment effects in clinical trials (Colloca & Fava, 2024). This involves initial participant randomisation to active treatment and placebo groups at the first stage, then at the second stage, placebo non-responders are re-randomised to either continue in the placebo condition or active treatment condition. This increases the chances of detecting treatment effects by adjusting the study population based on early responses.

| Recommendations | |
|---|---|
| 27 | Open-label extensions or sequential parallel comparison design may help to mitigate some disappointment effects in placebo-controlled trials of psychedelics, and may help to improve participant retention. |

#### *5.2.7. The Zelen Design (Randomisation Prior to Consent)*

Another idea, recently proposed by Balázs Szigeti (Hu, 2025), to help mitigate the disappointment effects observed in the control conditions of psychedelic trials is to implement the Zelen design (Zelen, 1979). The key feature of this approach is to randomise participants to conditions prior to providing information about the study conditions and seeking consent. For example, prospective participants would enrol for a clinical trial for depression, without knowing the conditions. Upon enrollment, they will be told the trial is an open-label trial with psilocybin or the control drug, without awareness of the other condition. This design has the potential to reduce disappointment and expectancy bias in control participants, given that they are unaware of the existence of the experimental group, thereby reducing the possibility of them feeling they have 'missed out' on a potentially efficacious treatment.

There is the added possibility of further randomisation within this trial design, in which some participants in each arm of the trial are notified of their allocation status and the existence of the other arm, and some are not. This allows for further measurement of potential disappointment and expectancy effects.

#### *5.2.8. Blinding Psychological and Pharmacological Effects in Isolation*

Given that controlling for both the psychological and pharmacological effects of psychedelics is an ambitious aim, one approach can be to control for each in isolation. To control for pharmacological effects, one option may be to utilise virtual reality (VR) to simulate the psychological effects of psychedelic experiences without pharmacological manipulation (Colloca & Fava, 2024). VR approaches that emulate or facilitate psychedelic experiences are in development (see Aday et al., 2020, for review, and Kettner et al., 2025, for recent example), although there is some debate within our authorship about whether this is a viable option, and whether visual alterations from psychedelics are relevant to or inherent to any treatment effects.

Another approach to understanding the impact of the psychological effects of psychedelics involves intentional manipulation of these effects during the psychedelic experience. For example, researchers have examined the impact of task demands or interruptions on the subjective experience and its therapeutic efficacy (Roseman et al., 2024). Similarly, the effects of memory of the psychedelic experience have been investigated in a study in which the benzodiazepine midazolam was administered alongside psilocybin, resulting in the conclusion that memory of the psychedelic experience does factor into its therapeutic efficacy (Nicholas et al., 2024). Such approaches allow for a more fine-grained assessment of how specific elements of the psychedelic experience, such as memory and cognitive engagement, may mediate psychological outcomes.

Alternatively, to control for psychological effects, a potential approach is to administer psychedelics whilst participants are under anaesthetic (Colloca & Fava, 2024), as implemented in a randomised, placebo-controlled trial of ketamine for major depressive disorder (Lii et al., 2023). This method effectively isolates the pharmacological action of the drug, ensuring successful blinding and eliminating expectancy-related confounds, but introduces additional confounds related to the anaesthesia itself, such as its effects on neurobiology and subjective experience.

### 5.3. Measuring subjective experience

Acute subjective experiences induced by psychedelics encompass a diverse range of phenomena, including profound changes in perception, cognition, and emotion. They tend to include deeply personal insights and transformative affective content, which are highly salient and meaningful to the individual (Breeksema et al., 2020). These effects seem to be particularly elusive to linguistic expression (Preller & Vollenwieder, 2016; Shulgin et al., 1986), making them challenging to study using standardised scientific measures. However, unlike drugs such as amphetamines or sedatives, there are no characteristic behavioural patterns that map to subjective psychedelic experiences (see Tagliazucchi, 2022), meaning that behavioural measures are less applicable as proxies for the subjective experience than in other areas of psychopharmacology. In neuroscience more broadly, studying the relationships between subjective effects and neural correlates remains a challenge, but there are recent developments that psychedelic research can benefit from. In this section, we overview the main questionnaire-based options commonly used in the field, as well as two examples of such developments in the study of subjective experience that may be useful to researchers interested in exploring this avenue.

#### *5.3.1. Questionnaires*

Questionnaires are a primary tool for quantifying psychedelic effects. In contemporary research, three psychometric instruments commonly used to assess the broad subjective effects of psychedelics are the *5-Dimensional Altered States of Consciousness* (5D-ASC) Questionnaire; the *Mystical Experience*

*Questionnaire* (MEQ); the *Hallucinogen Rating Scale* (HRS). The *5D-ASC* (Dittrich et al., 2006) assesses three core dimensions: Oceanic Boundlessness, Dread of Ego Dissolution, and Visionary Restructuralisation. It also includes two supplementary dimensions: Auditory Alterations and Vigilance Reduction. A more recent version (the 11D-ASC) includes eleven dimensions (Studerus et al., 2010).

The MEQ-30 is now widely used to measure the positive subjective effects of psychedelic experiences. It assesses seven domains which were theoretically considered central to the 'mystical experience': Internal Unity; External Unity; Transcendence of Time and Space; Ineffability and Paradoxicality; Sense of Sacredness; Noetic Quality; Positive Mood (Griffiths et al., 2006; Pahnke, 1966). The HRS was originally developed to quantify the effects of DMT and ayahuasca. Its most recent edition includes 100 items distributed across six subscales: Somaesthesia, Affect, Volition, Cognition, Perception, and Intensity (Riba et al., 2001).

In addition to these, several psychometric tools have been developed to assess more specific dimensions of the psychedelic experience. For instance, the Challenging Experience Questionnaire evaluates negatively valenced experiences, such as paranoia or grief (Barrett et al., 2016). The Ego Dissolution Inventory (EDI) was designed to measure alterations in the sense of self (Nour et al., 2016). The newer Ego Dissolution Scale (EDS), which measures trait-like aspects of ego dissolution in everyday life, could be adapted in future research to also assess past experiences of ego dissolution during psychedelic states (Lynn et al., 2023). The Emotional Breakthrough Inventory was designed to assess the experience of confronting and overcoming difficult emotions (Roseman et al., 2019). The Acceptance and Avoidance-Promoting Experiences Questionnaire focuses on the degree of acceptance or tolerance toward aversive emotions (Wolff et al., 2022).

*Limitations and recommendations*

A major limitation of existing psychometric instruments, both within and outside of psychedelic research, is that they have been largely developed and validated in Western, Educated, Industrialised, Rich, and Democratic (WEIRD) populations (Yaden et al., 2024). Differences in the description and interpretation of subjective psychedelic experiences between WEIRD and non-WEIRD populations are not adequately captured by existing scales. For instance, ineffability is less emphasised in non-WEIRD contexts, and the meaning ascribed to similar perceptual experiences can vary, which measures may fail to reflect (Graziosi et al., 2023). Additionally, certain experiences commonly reported in non-WEIRD accounts – such as feelings of purification, the soul traveling or transfiguring into another being – are not represented in widely used scales (Graziosi et al., 2023). Such discrepancies hinder the scales' ability to accurately capture subjective experiences across different cultural contexts.

Even within WEIRD contexts, the methods used in the validation of these instruments pose challenges to their generalisability. Many validation studies rely on online surveys to obtain their samples. It is likely that these surveys often attract volunteers who are frequent psychedelic users or enthusiasts (Yaden et al., 2024). This raises concerns about the representativeness of the samples used for validation compared to those in which the scales are ultimately applied. In addition, many instruments show mixed success in maintaining their factor structures when adapted to other languages. For instance, the MEQ-30 retained its original (English version) factor structure when translated in French (Fauvel et al., 2023), Finnish (Kangaslampi et al., 2020), and Brazilian Portuguese (Schenberg et al., 2017), but not in Spanish (Bouso et al., 2016). The 5D-ASC, while widely used in English-language clinical research (Herrmann et al., 2023; Hovmand et al., 2024), has not been psychometrically validated in its full version in English (de Deus Pontual et al., 2023; Herrmann et al., 2023).

Furthermore, some constructs measured by these instruments are informed by cultural ideas that are not universally applicable. The construct of 'ego dissolution' assessed in commonly used scales (5D/11D-ASC,

EDI, EDS), reflects a specific and constraining cultural idea which limits its ability to fully capture the experiences it aims to describe, as touched on in the beginning of this paper. Commonly used scales, such as the MEQ-30 and 5D-ASC are derived from instruments developed using a top-down approach, rooted in theoretical assumptions about the nature of mystical experiences or ASCs. These tools originated from scales created before the scheduling of psychedelics in the 1970s. At the time, theoretical frameworks were the primary basis for developing psychometric instruments. This reliance on theory alone may have resulted in the omission or mischaracterisation of key aspects of the psychedelic experience – such as the narrow description of alterations to selfhood. Currently, novel measures are often developed exclusively for specific trials (Hovmand et al., 2024), resulting in isolated tools that lack the necessary validations and comparative analyses to connect them with existing literature. Bottom-up development of instruments, grounded in qualitative data and informed by descriptions of participants' experiences, may help create measures that more accurately capture the experience (see McAlpine et al., 2024, for an example of a validated, participant-informed scale used in psychedelic research developed using bottom-up methodology).

In addition, most current instruments assessing aspects of psychedelic drug effects were designed to characterise experience after it has finished. Participants are typically asked to evaluate their experience as a whole after the fact, resulting in single summary scores for each dimension. While this approach provides a broad overview, it can oversimplify the experience. This approach is especially problematic for substances like LSD, which can induce effects lasting 10–12 hours. Such approaches disregard the timeline and evolution of acute effects, potentially obscuring critical temporal nuances. To address these issues, it would be useful to consider developing shorter, acute scales that can be administered at different points during the psychedelic experience (de Deus Pontual et al., 2023). Supplementing these measures with emerging qualitative methodologies can help combat some limitations. Cognitive interviewing, for instance, can be used to examine how participants understand items in questionnaires (Yaden et al., 2024). This allows participants to explain their interpretation of concepts in measures, uncovering discrepancies between the intended and perceived meanings of items. Recent developments in neurophenomenological approaches such as microphenomenology and experience sampling (overviewed below) also hold significant promise for advancing the understanding of subjective experiences.

| **Recommendations** | |
|---|---|
| 28 | Future scales which attempt to capture psychedelic experience should be developed using inductive approaches, which use participants' experiences as the primary data (as opposed to 'top-down' approaches). |
| 29 | Future scales could be developed which are shorter and comprehensible during the acute psychedelic experience. This would allow for a less reductive and more time-sensitive quantification of psychedelic experiences. |

#### *5.3.2. Neurophenomenology*

Neurophenomenology bridges the subjective, first-person experience with the objective, third-person analysis of neuronal dynamics (see Varela, 1996). Unlike conventional approaches that often rely on group averages, neurophenomenology prioritises person-specific data, providing deeper insights into individual neural and experiential processes. To further our understanding of the subjective experience in psychedelics, research must go beyond "objective" data and incorporate a comprehensive method for exploring the structure of ongoing human experience. A recent approach of neurophenomenological approaches to psychedelics and other non-ordinary states of consciousness outlines the benefits of such an approach to significantly advance the science of consciousness and mental health applications of these states (Houot et al., 2021; Timmermann, Bauer, et al., 2023). Three emerging approaches that have been applied in studies exploring the neural correlates of subjective experience – which some of the authors of this paper are implementing in the context of psychedelic research (e.g., Timmermann et al., 2019; Sanders et al, 2024) – are *microphenomenology* (Petitmengin, 2006; Petitmengin et al., 2019), *experience sampling* (Smallwood et al., 2021; Kuc et al., in prep), and *temporal experience tracing* (Jachs et al, 2022).

*Microphenomenology*

Microphenomenology incorporates in-depth interviews that uncover fine-grained details of participants' experiences, offering insights that may be missed by coarser qualitative and psychometric methods. A central problem researchers face in studying subjective experience is that considerable information is lost between raw, unexamined first-person experience and the data after being recorded, interpreted, or reflected upon. Our introspective views of our own thoughts and sensations are shaped by cultural beliefs, theories, context and judgements (Petitmengin, 2006). This can lead us to unknowingly describe our interpretation of an experience rather than the experience itself. Accessing the raw 'felt sense' of experience (Gendlin, 1992) is the goal of the microphenomenological approach (Petitmengin, 2006).

The method focuses on the precise exploration of the subtle, fleeting aspects of consciousness, often dedicating over an hour of interview to the description of a moment of experience on the order of seconds or minutes. Delving deeply into such a brief experience offers the unique benefit of encouraging participants to explore aspects of their experience they had not consciously noticed in the first instance. Microphenomenological analysis uncovers overarching patterns or structures in the experiences that can be generalised across subjects (Petitmengin, 2006; Petitmengin et al., 2019; Valenzuela-Moguillansky & Vásquez-Rosati, 2019). Using this technique we can uncover two dimensions of lived experience, the temporal or dynamic dimension (*diachronic*) which captures the unfolding of experience in time; or the non-temporal dimension (*synchronic*) which corresponds to specific characteristics of the experience at a given moment in time (e.g., sensory modalities, emotional tone, bodily feelings, perceptual position.) In a recent example, Sanders et al. (Sanders et al, 2024) employed micro-phenomenological interviews to study the DMT experience focusing on the emergence of experiences of social cognition induced by DMT. This approach enabled them to determine that these complex experiences were almost always preceded by multisensory sensory effects, and then immersive embodied effects.

Microphenomenology can also be integrated with neuroimaging or electrophysiology to locate potential neurophysiological correlates of the experience. In the context of psychedelic neuroimaging studies, post-hoc interviews can be conducted on the participant's experience in the MRI/MEG scanner/while wearing an EEG cap, identifying time-locked neural signatures that correspond to the described experience (with the additional possibility of within-subjects comparisons of neurophenomenology on and off the study drug). For example, Timmermann et al., (2019) investigated the neural correlates of the DMT experience using a combination of microphenomenology and EEG, identifying three primary experiential dimensions: visual, bodily, and emotional/metacognitive aspects. When integrated with EEG data, they found that changes in bodily awareness were negatively associated with central and parietal beta-band power. Additionally, the intensity of visual imagery was inversely related to alpha power and positively correlated with delta and theta power.

*Experience sampling*

Experience sampling (Smallwood et al., 2021) involves collecting quasi-real-time data at multiple points during the psychedelic experience, reducing the reliance on retrospective summaries and providing a dynamic account of subjective states. This approach has been used by the Liechti Lab to assess the acute subjective effects of multiple psychedelics, including LSD (Holze et al., 2019; 2021), psilocybin (Holze et al., 2022), mescaline (Klaiber et al., 2024), and DMT (Vogt et al., 2023). Their specific approach was simple:

participants were presented with visual analogue scales at multiple timepoints (e.g. every 30-60 minutes) on which they rated valence, liking, fear, ego dissolution, and temporal effects of their experience. Similar measures have been applied in other studies with higher frequency, with participants in a DMT study giving ratings of intensity every minute, for example (Timmermann et al., 2019). However, it is important to strike a balance between the need for intensive assessment as required for experience sampling analyses and participant burden, as frequent measurements can have the unwanted effect of interfering with the acute psychedelic experience that is under study: 11D-ASC scores in participants administered with DMT have been found to be higher (accompanied by a greater reduction in depressive symptoms) in those who were not required to provide peri-dosing ratings compared to those who were (Roseman et al., 2024).

As well as experience sampling during the acute psychedelic experience, experience sampling techniques can be employed in the days and weeks following a psychedelic experience, as well as at baseline, to map enduring changes in experience, as used by Kuc et al. (in prep) in participants who partook in a 5-MeO-DMT retreat. This work uses smartphone applications that allow participants to journal their experiences over a more extended period of time, which, when combined with other information that can be obtained from smartphones such as location, movement, nutrition, and sleep quality, can yield high-dimensional data with significant predictive power. With this approach, experience sampling again offers richer longitudinal data with greater ecological validity than one-off post hoc questionnaire measures, recording fluctuations in cognition, mood, behaviour, and, crucially, use of language. Furthermore, linguistic modelling of verbal descriptions of subjective experience have been seen to be more strongly predictive of mental wellbeing than many traditional questionnaire-based measures (Zhang et al., 2022). This approach requires some additional considerations under data protection and compliance/attrition, but could provide deeper insights into the long-term effects of psychedelics and their relationship to long-term therapeutic outcomes.

*Temporal Experience Tracing*

Temporal Experience Tracing (TET) is a recently developed method that could be considered a ‘middle-ground’ between experience sampling and retrospective questionnaire as it expands the former by introducing a time dimension to the reporting of subjective effects. Such an approach has benefits for attempting to establish some temporal progression to the effects of psychedelics while also relying heavily on the memory of participants to report on these dynamics. See (Lewis-Healey et al., 2024) for a recent example incorporating TET and EEG recordings in naturalistic use of DMT.

### 5.4. Considerations for neuroimaging

Psychedelic neuroimaging has been a key driver of the renewed scientific interest in psychedelics. Functional magnetic resonance imaging (fMRI) and electrophysiological tools such as magnetoencephalography (MEG) are notable developments that were unavailable to researchers during the first surge in psychedelic research in the early twentieth century. Images of the “brain on LSD” from the seminal study by Carhart-Harris et al. (2016), as well as the now-famous image of brain functional networks on psilocybin versus placebo from Petri et al. (2014), sparked a renewed interest in psychedelics, catalysing efforts to understand how psychedelics modulate consciousness and promote enduring therapeutic effects.

However, conducting neuroimaging to understand the broad spectrum of psychedelic effects presents unique methodological hurdles and practical considerations. The complexity of neuroimaging in all psychopharmacological research is captured by the humorously elaborate portmanteau "clinical/cognitive neuropsychopharmacology," underscoring the diverse, interdisciplinary expertise needed – spanning phenomenology, participant care, clinical oversight, drug licensing, cognitive neuroscience, physics, and mathematics – which often surpasses the resources of individual teams. Consequently, analyses are frequently outsourced, sometimes creating a disconnect between data collection and interpretation.

Nonetheless, this data sharing has spurred creative analyses, from whole-brain modelling (Deco et al., 2018; Herzog et al., 2023) to proposed information theoretic approaches (Luppi, Mediano, et al., 2021; Varley et al., 2024; Liardi et al., 2024)

Psychedelic effects unfold across multiple interacting temporal phases – acute intoxication, subacute after-effects, long-term outcomes – each potentially exhibiting distinct neural signatures. Designing reproducible studies is further complicated by contextual ('set' and 'setting'; Roseman et al., 2024) and individual factors (Haijen et al., 2018). Consequently, studies often focus on either acute-state effects or pre/post comparisons examining enduring changes, frequently relying on resting-state data (Wall et al., 2023). While valuable, this approach may overlook specific cognitive processes and limit reliability (Finn et al., 2022). Addressing these complexities requires frameworks integrating findings across timescales. Recent neuroplasticity-centred hypotheses (Sumner & Lukasiewicz, 2023; de Vos et al., 2021; Nutt et al., 2023) suggest transient serotonergic modulation triggers both immediate shifts in consciousness and lasting adaptations. A comprehensive understanding thus necessitates integrating neural signatures spanning pre-dose predictors, acute correlates, and long-term outcomes.

To these ends, multimodal approaches are increasingly employed; the publicly available LSD dataset (Carhart-Harris, Muthukumaraswamy, et al., 2016), featuring acute fMRI, MEG, and behavioural measures, exemplifies this, supporting diverse inquiries linking phenomenology to neural activity (e.g. Atasoy et al., 2017; Girn et al., 2022; Luppi, Carhart-Harris, et al., 2021; Mediano et al., 2024; Roseman et al., 2016; Shinozuka et al., 2025; Tagliazucchi et al., 2016). Despite the richness of such datasets, data are often parsed across publications, with fewer studies jointly modelling multiple timepoints, modalities, or conditions (though see Siegel et al., 2024; Lyons et al., 2024). Below are some of the considerations – given the current evidence and our own experiences – specific to conducting neuroimaging studies with psychedelics, alongside recommendations for investigating the neuroplasticity hypothesis.

#### *5.4.1. fMRI*

Functional MRI has revealed acute changes in brain activity under psychedelics, including regionally specific alterations in cerebral blood flow and increased global functional connectivity under psilocybin, LSD, and DMT (e.g., Bedford et al., 2023; Carhart-Harris et al., 2011; Carhart-Harris, Erritzoe, et al., 2012; Carhart-Harris, Muthukumaraswamy, et al., 2016; Gaddis et al., 2022; Preller et al., 2018; Siegel et al., 2024; Stoliker et al., 2024; Timmermann, Roseman, et al., 2023). These alterations correlate with a range of phenomenological and electrophysiological measures (e.g., Mortaheb et al.; 2024; Siegel et al., 2024; Timmermann, Roseman, et al., 2023). fMRI also probes longer-term changes potentially underlying clinical efficacy (e.g., Daws et al., 2022; Mertens et al., 2020).

Nevertheless, there are a number of considerations and potential pitfalls when using fMRI to study any pharmacological challenge, particularly serotonergic psychedelics. Firstly, fMRI measures neural activity indirectly via the Blood-Oxygenation-Level Dependent (BOLD) signal, which reflects slow changes in blood oxygenation. This relative measure has long-term drifts needing substantial filtering (limiting temporal precision) and is susceptible to non-neural physiological confounds. Since many psychedelics alter vasculature, heart rate, and respiration (see Wsół 2023 for review), observed BOLD changes might reflect vascular effects rather than purely neural activity. Therefore, monitoring of physiological measures such as heart and respiratory rates are important, as are more involved approaches such as quantifying vascular reactivity (e.g., with breath hold paradigms; Carhart-Harris et al., 2011) and combining BOLD with more direct measures of blood flow such as arterial spin labelling or other imaging modalities.

Secondly, head motion during acute dosing can create complex spatiotemporal artefacts which are challenging to correct. Author experience varies; some report participants finding it harder to stay still, while others occasionally observe the opposite, suggesting significant inter-individual variability and drug differences (e.g., some – but not all – of us have noticed *less* head motion with DMT than at baseline). Aggressive motion correction, temporal censoring, and participant removal based on motion thresholds may be necessary. It is also worth considering the impact of possible selection bias in recruiting participants who are more likely to remain still throughout the scan. Imaging only pre- and post-dosing circumvents acute motion issues but limits the research questions that can be explored by the experiment.

Thirdly, the choice of task paradigm is critical. Active cognitive tasks can be demanding during profound attentional and phenomenological shifts induced by psychedelics, and may even alter drug effects on brain networks (Siegel et al., 2024). This practical challenge has led to a dominance of "resting-state" fMRI, where participants lie still with no explicit task. While reducing burden, this approach generates substantial interpretive limitations; it becomes difficult to know the internal focus of participants, which varies greatly within and between individuals, especially under psychedelics. This uncontrolled mentation makes isolating drug-specific neural signatures difficult. Furthermore, without structured stimuli, separating signal from physiological noise (heartbeat, respiration) becomes harder. This ambiguity fuels problematic reverse inference – attributing observed network changes (e.g., decreased DMN connectivity) solely to specific mental phenomena (e.g., ego dissolution), ignoring the DMN's role in other processes (memory, social cognition) and the fact that psychedelics impact multiple networks globally (Carhart-Harris, Muthukumaraswamy, et al., 2016; Mediano et al., 2024). In effect, resting under psychedelics becomes an unmonitored, highly variable cognitive "task," confounding interpretation (Finn et al., 2022).

An emerging solution lies in naturalistic paradigms, using stimuli such as movies (Mediano et al., 2024; Stoliker et al., 2025) or music (Kaelen et al., 2015). Such paradigms increase ecological validity whilst offering clearer interpretive anchors. They can improve data quality (enhancing engagement, reducing motion), are often better tolerated, and provide time-locked events for modelling neural responses. This allows distinguishing general drug effects from context-dependent responses triggered by specific stimuli, helping constrain the vast range of possible mental states while preserving aspects of spontaneous experience. As argued by Finn et al. (2022), such engaging paradigms that perturb the brain in controlled yet rich ways improve the interpretability of functional connectivity. Task-based fMRI also remains valuable; for instance, emotional face-processing tasks reveal context-dependent modulation of amygdala activity correlating with therapeutic outcomes in depression (Roseman, Demetriou, et al., 2018) or attenuated responses in healthy volunteers during acute intoxication under psilocybin (Armand et al., 2024), highlighting the importance of state, population, and timescale.

Analytical choices further impact interpretation. Common reliance on low-resolution parcellations (e.g., Schaefer or Yeo atlases; Yeo et al., 2011) averages activity across potentially distinct regions, risking loss of crucial local variations and hindering interpretability, especially when mapping subtle shifts in consciousness. Finer-grained (e.g., DiFuMo; Dadi et al., 2020) or gradient-based atlases allowing overlapping network membership may better capture psychedelic-induced reconfigurations. While static, time-averaged functional connectivity is widespread, dynamic approaches (e.g., sliding windows, hidden Markov models) aim to capture transient brain states potentially linked to fluctuating subjective experiences (Tagliazucchi et al., 2016; Luppi, Carhart-Harris, et al., 2021; Mortaheb et al., 2024). However, sliding windows can inadvertently reintroduce motion/physiological artefacts (Lindquist et al., 2024). Furthermore, conventional connectivity relies on linear correlations, potentially missing nonlinear dependencies, synergies, and redundancies better captured by multivariate methods like Partial Information Decomposition (Mediano et al., 2021; Luppi et al., 2024; Varley, 2024) or dynamical systems perspectives (Singleton et al., 2022, Girn et al 2023; Varley et al., 2024; Stoliker et al., 2025). These suggest psychedelics may induce a more flexible, metastable state by flattening the brain's control energy landscape (Singleton et al., 2022) and reshaping local interactions (Varley et al., 2024), facilitating transitions between global activity patterns.

#### *5.4.2. M/EEG*

Complementing fMRI's spatial resolution, electroencephalography (EEG) and magnetoencephalography (MEG) provide superior temporal resolution (milliseconds), essential for capturing the rapid, fluctuating neural dynamics characteristic of psychedelic states. EEG’s relative portability and lower cost compared to MRI also facilitate more complex and naturalistic experimental designs. Key resting-state findings with classical serotonergic psychedelics include a highly robust suppression of alpha band (8–12 Hz) power (Muthukumaraswamy et al., 2013; Nikolic et al., 2024; Ort et al., 2023; Pallavicini et al., 2021; Timmermann et al., 2019; Timmermann, Roseman, et al., 2023; Riba et al., 2004), although this may not be universal across all psychedelic compounds (e.g., 5-MeO-DMT; Blackburne et al., 2024). Another consistent observation is an increase in spontaneous signal complexity, frequently measured using Lempel-Ziv complexity or similar metrics (Schartner et al., 2017; Timmermann, Roseman, et al., 2023; Murray et al., 2024). This is often framed within the "entropic brain" hypothesis (Carhart-Harris et al., 2014), suggesting a transition towards a more flexible, less constrained, and potentially more information-rich mode of cortical function, correlating with subjective reports of richness and novelty.

However, M/EEG studies face their own challenges. Like fMRI, head movement can be problematic. More specific to M/EEG are artefacts arising from increased facial muscle contractions, particularly jaw-clenching, swallowing, or yawning (see Muthukumaraswamy, 2014, for review of MEG in psychopharmacology), which appear more frequently under psychedelics (Marini, 1981). These electromyographic (EMG) signals contaminate the scalp recording, particularly in the gamma frequency band (30-100 Hz), a finding likely explaining the inconsistent findings regarding gamma power, with studies reporting mixed results (for increases, see Timmermann, Roseman, et al., 2023; Schenberg et al., 2015; for decreases, see Muthukumaraswamy et al., 2013). Disentangling genuine neural gamma oscillations from muscle artefacts remains a significant methodological hurdle requiring sophisticated cleaning techniques (Pallavicini et al., 2019) and careful interpretation.

Leveraging M/EEG's advantages for study design, researchers can move beyond resting-state analyses to probe specific cognitive and perceptual functions altered by psychedelics. The high temporal resolution of M/EEG is well-suited for event-related paradigms that isolate neural responses to discrete stimuli or cognitive events. Furthermore, its relative deployability allows for measurements in environments potentially less intimidating or restrictive than an fMRI scanner, which may be beneficial for participant comfort and for studying effects sensitive to environmental context. This flexibility facilitates the implementation of various task-based paradigms. For example, M/EEG studies using mismatch negativity (MMN) and event-related potentials (ERPs) have shown that serotonergic psychedelics modulate predictive and emotional processing in nuanced ways, with effects depending on context, task design, and chronicity of use (Umbricht et al., 2002; Timmermann et al., 2017; Duerler et al., 2022; Orłowski et al., 2023, 2024; De Filippo et al., 2024).

Furthermore, M/EEG enables innovative approaches assessing brain dynamics more directly. Perturbational methods like TMS-EEG can probe cortical excitability and connectivity changes under psychedelics (Ort et al., 2023), revealing how baseline signal complexity relates (or doesn't) to the brain's response to external input. The deployability of EEG is a major advantage, enabling studies outside the traditional lab, such as investigating psychedelic effects on sleep architecture using overnight recordings (cf. Thomas et al., 2022, in rodents showing psilocin alters sleep) or exploring real-time social interactions using hyperscanning setups (e.g., forthcoming work by Timmermann et al.). These approaches are vital for understanding how psychedelic effects manifest and consolidate in more ecologically valid settings.

| Recommendations | |
|---|---|
| 30 | Neuroimaging paradigms in psychedelic research should be broadened beyond acute, resting-state studies; there is a need for more psychedelic studies including task fMRI, music/movie fMRI, and further EEG studies in naturalistic contexts. |
| 31 | Coarse parcellations (e.g., seven-network atlases) risk obscuring local variations and hinder interpretability. Fine-grained atlases (e.g., DiFuMo) and gradient-based segmentations can capture overlapping or gradient-like transitions in brain organisation, which are essential for characterising intersubject variability that typifies responses to psychedelics. |
| 32 | Psychedelics produce acute, subacute, and long-term effects. Incorporating M/EEG, MEG, or fMRI recordings across multiple timepoints can link short-term network changes to enduring behavioural or clinical outcomes, shedding light on neuroplastic and therapeutic processes. |
| 33 | Linking phenomenological data (e.g., moment-to-moment ratings, structured experience sampling) with brain metrics helps contextualise neural findings and provides clearer maps between subjective effects, contextual factors, and dynamic neurobiology. |
| 34 | Neuroscientists using psychedelics should develop guidelines to collect and process data to minimise the influence of artifacts commonly encountered in these studies and standardise analytic procedures that could enable consensus pertaining to the effects of psychedelics in the brain. |

#### *5.4.3. Measures to investigate proposed plasticity hypothesis*

A driving narrative in psychedelic neuroscience at present is that post-acute changes are, in part, driven by enhanced neuroplasticity post-dose (Aleksandrova and Phillips, 2021; Nutt et al., 2023), as evidenced by preclinical *in vitro* and *in vivo* animal work (Calder and Hasler, 2023; Vargas et al., 2023; Nardou et al., 2023). However, the application of this neuroplasticity hypothesis in explaining the therapeutic mechanisms of psychedelics in humans is limited by the broad and multifaceted nature of neuroplasticity. There are also limitations in techniques to measure different neuroplastic processes – such as structural or functional plasticity, and hyper/hypoplasticity and metaplasticity (Agnorelli et al., 2024) – resulting in the current lack of direct human evidence to support this hypothesis and bridge the pre-clinical findings with post-acute findings thought to be driven by neuroplasticity i.e., modularity (Daws et al., 2022; Lyons et al., 2024). For an in depth breakdown of psychoplastogens and neuroplasticity see (Agnorelli et al., 2024).

Peripheral expression of brain-derived neurotrophic factor (BDNF), a neuroplasticity-associated neurotrophin, has been used as a molecular marker of neuroplasticity and is hypothesised to be a potential driver of therapeutic effects (de Almeida et al., 2019; Hutten et al., 2020; Holze et al., 2021; Becker et al., 2022; Rocha et al., 2021; Holze et al., 2022). Yet, across studies, peripheral BDNF increases (de Almeida et al., 2019; Hutten et al., 2020; Holze et al., 2021; Becker et al., 2022) or remains unchanged (Rocha et al., 2021; Holze et al., 2022) post-treatment. Additionally, ketamine-induced elevations in rodent peripheral plasma BDNF levels have been shown to not correspond to central BDNF levels, putting into question the utility of peripheral BDNF as a proxy for central BDNF concentrations following psychopharmacological treatment (Le Nedelec et al., 2018). A meta-analysis of all human psychedelic studies measuring peripheral BDNF illustrated there is no evidence that peripheral BDNF is elevated following administration of psychedelics in humans, suggesting this is not a reliable marker of post-dose changes in neuroplasticity (Calder et al., 2024). This reflects broader challenges in linking molecular and psychological levels of inquiry. We recommend a cautious approach to generalising between these levels; even if BDNF reliably indexed neuroplasticity, increased neuroplasticity does not necessarily imply therapeutic benefit.

There is a need for more robust human *in vivo* neuroplasticity biomarkers; neuroimaging and electrophysiological techniques may help probe these processes. Sensory plasticity electrophysiological paradigms present participants with trains of rapidly repeating auditory or visual stimuli, which are thought to induce long-term potentiation (Clapp et al., 2012; Spriggs et al., 2018), the synaptic process underpinning learning (Bliss & Lomo, 1970). Recent research has utilised these sensory functional plasticity indices in psychedelic (Hutten et al., 2024; Skosnik et al., 2023; Murphy et al., 2024) and ketamine studies (Sumner et al., 2020). To measure structural plasticity in humans, PET imaging, using tracers [11C]UCB-J, a synaptic vesicle glycoprotein 2A (SV2A) ligand, can quantify pre-synaptic density as an index of synaptic plasticity (Finnema et al., 2016). [11C]UCB-J is sensitive to ketamine and SSRI treatment in major depressive disorder (MDD) and healthy populations respectively (Holmes et al., 2022; Johansen et al., 2023), and psilocybin in pigs (Raval et al., 2021). Additionally, diffusion tensor imaging and structural magnetic resonance imaging techniques may be used to quantify structural plasticity (Agnorelli et al., 2024).

Regardless of the current dearth of robust human neuroplasticity evidence, there is a recent significant body of research on designing so-called “non-hallucinogenic psychoplastogens”. These analogues are designed to mimic the neuroplasticity-inducing effects of psychedelics whilst omitting the acute subjective experience (Cameron et al., 2021; Lewis et al., 2023), relying on the assumption that therapeutic effects arise from rapid neuroplasticity (Olson, 2020), rather than from the entropic brain states and altered beliefs induced by psychedelics (Carhart-Harris & Friston, 2019). Future studies could use these candidate markers to clarify how psychedelics and non-hallucinogenic psychoplastogens modulate neuroplasticity in humans, and how such changes relate to long-term functional and therapeutic outcomes.

## 6. Experiential considerations

Previous studies (Griffiths et al., 2006; Grof, 1980) report that participants regularly rate their psychedelic experiences as among the most meaningful of their lives. There is also increasing evidence that the quality of the psychedelic experience is an important predictor of therapeutic outcomes (Ko et al., 2022; Yaden & Griffiths, 2021; Herrmann et al., 2023; Nicholas et al., 2024; Roseman et al., 2024; Roseman, Nutt, and Carhart-Harris, 2018). It has also been seen that if memory of the psychedelic experience is inhibited, the therapeutic effect is reduced (Nicholas et al., 2024). In some cases, deeply-entrenched patterns of psychopathology (e.g., depression, anxiety, substance use) – which typically take several years of psychotherapy to address – have been alleviated after just one or two sessions of psychedelic-assisted therapy (Goodwin et al., 2022; Gasser et al., 2014; Carhart-Harris, Bolstridge, et al., 2016; Johnson et al., 2014; Bogenschutz et al., 2018). Given the potentially life-altering nature of the psychedelic experience, a significant consideration – as yet unstandardised – is how to manage contextual factors that influence the participant’s experience before, during and after the dosing session(s).

Since the terms were first described by Leary et al. (1963), it has been well established that the main factors influencing participant experience are ‘set’ (the inner milieu – e.g., personality, mood, beliefs, recent experiences) and ‘setting’ (e.g., the outer environment – lights, temperature, music, people). Factors related to *set* such as negative mindset, lack of preparation, psychological vulnerability (e.g., young age or psychiatric history), lack of psychological support, and a major prior life event predict a negative or difficult psychedelic experience (Simonsson et al., 2023; Bremler et al. 2023). Similarly, attributes related to *setting* such as disagreeable physical or social environment, adverse contextual conditions, were associated with negative experiences (Simonsson et al., 2023; Bremler et al. 2023; Carbonaro et al., 2016). This means that, while each psychedelic experience is highly individualised, there is much that can be done by researchers and clinicians to optimise participant experience.

Pronovost-Morgan et al. (2025) also recommend that studies on psychedelic substances should provide a comprehensive account of the set and setting in which the research was conducted. They conceptualise a set of 'ReSPCT' guidelines (Reporting of Setting in Psychedelic Clinical Trials) to provide a standardised framework for documenting extra-pharmacological factors that shape psychedelic experiences. The guidelines identify 30 key setting variables categorised into physical environment (e.g, location, ambience, music), dosing session procedures (e.g., presence of facilitators, participant autonomy), therapeutic framework (preparation, integration, cultural competence), and subjective experiences (trust, physical and psychological safety). Below are some of the key experiential considerations we recommend given our previous experiences with psychedelic studies.

### 6.1. Physical environment

Carefully preparing the physical environment in which the participant experiences the psychedelic state is therefore of paramount importance. A typical dosing room may consist of comfortable furniture (a bed or sofa), soft lighting, and windows / some connection to nature. Music is commonly played during dosing sessions, and may influence therapeutic outcomes (O'Callaghan et al., 2020). Music that is incongruent with the participants' unfolding experience may hinder outcomes (Kaelen et al., 2018; Watts et al., 2017). O'Callaghan et al. (2020) suggest that unfamiliar music may be preferable over familiar music that may already have prior associations. There are available playlists made by researchers who have conducted previous psychedelic studies (e.g., Hess, 2017; Kaelen et al., 2018). However, music preferences are deeply personal and culturally specific, and it may be beneficial for participants to be able to make choices about their auditory environment – e.g., being in silence or listening to environmental sounds such as birdsong or rainfall (Hoffer, 1965; Noorani et al., 2018; O'Callaghan et al., 2020). A recent study suggests that periods of silence can in fact facilitate therapist-patient interactions in psychedelic-assisted therapy (Gloecker et al., 2024) – more research is necessary to explore these effects.

Another consideration here is the need to move between environments, particularly with longer-acting psychedelics such as psilocybin, LSD, or mescaline; or neuroimaging studies of acute effects wherein the duration of effects may be spent partly in a scanner and partly in a testing room. It must be considered whether participants will have any accessibility needs, or might encounter any strangers, which could potentially be distressing. For longer experiences, planning for basic necessities such as bathroom visits and food and drink (with advance knowledge of dietary requirements and allergens) is also essential.

### 6.2. Psychotherapy

Current psychiatric medications are generally prescribed such that the patient takes their drug(s) in the quantities and schedule indicated, with no particular emphasis on the context in which the drug is taken: the pharmacological action of the drug alone is the crux of the treatment effect. However, as mentioned above, psychedelic drugs appear to act in large part by increasing suggestibility and sensitivity to environmental context (Carhart-Harris et al., 2015; Stoliker et al, 2025). Therefore, psychedelics may be seen (in a clinical context) as catalysts for psychotherapeutic outcomes – although more research is needed to determine the optimal therapeutic approaches to use with each psychedelic and for specific patient needs.

Subjects may also report benefits to their wellbeing as a result of their participation in psychedelic studies that are not clinical trials and therefore do not contain a recognisable psychotherapeutic component. Even in these circumstances, however, there will generally be a trained and experienced facilitator present with the participant; and having an empathetic, protective presence in the room during the experience likely acts, in these instances, as a proxy for the sort of guidance a therapist would provide in a more formal therapeutic context, or a guide would provide in a ceremonial context.

The approach to psychedelic-assisted therapy is far from standardised, and even the principle of attempting to standardise it is contentious. Commercially orientated trials aimed at regulatory approval for psilocybin (for example) specifically do not provide ‘psychotherapy’ at all, but instead specify a limited form of ‘psychological support’. This is in anticipation of regulatory confusion (as was seen with MDMA therapy for PTSD) if too much ‘psychotherapy’ is provided alongside the drug intervention. Semantics aside, the approach used, in our experience, is often non-specific. Implicitly, this is in recognition of the fact that no two participants are alike, and no two psychedelic experiences are alike either. The attempts to ‘standardise’ therapy in this seems manifestly rather absurd. Even so, some researchers have attempted to identify common themes across modalities that have proven particularly effective in these settings (Levin et al., 2024). One aspect in particular that has been shown to significantly improve outcomes for psychedelic-assisted therapy is the ‘working alliance’: the relationship between therapist(s) and patient, or facilitator(s) and participant (Clarkson, 2003; Levin et al., 2024; Murphy et al., 2022). It is also important to acknowledge that almost all psychiatric interventions also include some form of (structured on unstructured) psychological support and that the issues encountered in the elucidation of the (neuroplasticity vs psychotherapeutic) mechanisms are the result that the evidence indicates that psychedelics may ‘oversensitise’ patients to the influence of psychological support (Levin et al., 2024; Murphy et al., 2022). This calls for the development of models that account for the ‘synergy’ of neurobiological and psychotherapeutic mechanisms rather than their independent influence.

A psychedelic-assisted therapy approach worthy of mention is the psycholytic approach, which involves the use of low-to-medium doses of psychedelics in multiple therapy sessions to facilitate psychodynamic processing (Leuner, 1967). Although the majority of psychedelic-assisted psychotherapy research related to psycholytic therapy rather than high-dose psychedelic therapy (Passie, 1997), this method has largely fallen out of favour, likely due to its intensive time requirements and lower cost-effectiveness compared to high-dose models (Passie et al., 2022). Nevertheless, it remains a relatively underexplored avenue that may warrant renewed research attention, particularly in chronic or treatment-resistant populations.

### 6.3. Working alliance

The ‘working alliance’ (Clarkson, 2003) generally encompasses the sense of trust and safety the participant/patient has in the facilitator/therapist, as well as their mutual understanding (Levin et al., 2024). It is a critical component of the psychedelic setting. This is particularly relevant in studies involving psychedelic-assisted therapy, but insights regarding the establishment of a therapeutic relationship are useful in basic science studies as well, as all researchers must create a safe environment for their participants. In two psilocybin trials for depression stronger therapeutic alliance before and after psilocybin sessions predicted greater emotional breakthroughs, mystical experiences, and sustained reductions in depression (Levin et al., 2024; Murphy et al., 2022). The therapeutic relationship not only influenced the quality of the psychedelic experience but also had a direct effect on long-term symptom improvement, independent of the acute drug experience (Levin et al., 2024; Murphy et al., 2022). In our experiences, this mutual understanding can comprise a variety of factors, including alignment of treatment goals, illness conceptions, and expectancies. As a result, initial conversations between therapist and patient about goals, conceptions, and expectations are essential in laying the groundwork for the therapeutic ‘working alliance’ (Clarkson, 2003).

A robust psychotherapeutic alliance can account for a larger proportion of effect than the antidepressant agent prescribed (Krupnick et al., 1996). An analysis of a multicentre placebo-controlled trial in antidepressants found the ‘most effective’ third of prescribers achieved better outcomes with placebo than ineffective prescribers using an antidepressant (McKay et al., 2006). The act of prescribing a medication itself has psychological and relational significance: the participant is given the medication laden with the expectations of both participant and prescriber (Konstantinidou et al., 2023). Researchers need to be aware of these factors when designing protocols and make efforts to measure them.

The therapeutic relationship is central to positive experiences and treatment outcomes, both in psychedelic-assisted therapy (Levin et al., 2024) and psychotherapy more broadly (Norcross & Lambert, 2018; Clarkson, 1996, 2003; Rogers, 1957; Flückiger et al., 2018). The building of this alliance is supported by the therapist setting clear expectations for the work to be done, practising active listening, as well as developing their own self-awareness and emotional regulation (Clarkson, 1996, 2003; McWilliams, 2004; Rogers, 1957). Key considerations include adequately preparing the participant; being empathetically present during the dosing; and managing the integration of insights.

| **Recommendations** | |
|---|---|
| 35 | The dosing environment is a priority in psychedelic research – careful planning of the set-up of the room (e.g., lighting, music), and practical considerations (e.g., food, accessibility, transport to and from the facility) may mitigate the risk of adverse reactions. |
| 36 | The working alliance between facilitator and participant is also a key priority for the safety and comfort of the participant, and may predict positive outcomes in clinical trials. |

### 6.4. Preparation and integration

Preparing the participant for the psychedelic experience – whether they are naive or experienced users of psychedelics – is integral to establishing the facilitator-participant alliance and ensuring positive outcomes. This process begins from the moment of contact with the participant; i.e., from the point of recruitment. It is helpful to meet participants before dosing sessions with the purpose of building trust and providing psychoeducation (Goodwin et al., 2022). The way in which information about the study is shared, and the way in which consent is taken, is therefore of particular and unique importance in psychedelic studies.

In psychology or neuroscience experiments with human subjects, there will typically be a standard set of instructions for each participant. However, in the interest of building the relationship, we propose that the instructions given to participants in psychedelic studies should be tailored specifically to each participant. For example, there will be significant differences in the discussions that will connect with participants who are experienced users of psychedelic drugs versus those who are naive to the experience. Explaining the psychedelic experience to an experienced user may feel patronising and risk alienating them, while failing to prepare a naive participant for the nature of psychedelic states may leave them unprepared for the intensity of the experience.

In the preparation phase, it is important to manage expectations and gently challenge beliefs that frame psychedelics as a euphoric fast-track to psychological well-being, which could lead to disappointment and reduce the likelihood of positive outcomes. If the participant is not prepared to face challenging psychological content that may be uncovered during a trip, they may exhibit heightened reactions to challenging material arising. Similarly, participants should be advised to be open to the experience, especially during the onset of psychoactive effects and the occurrence of challenging experiences. Resistance or a desire to control such experiences can render them more distressing: Wood et al. (2024) found that, “I tried to let go, accept or surrender to the experience" and, "I tried to observe my mind and not fight it", were rated the most useful strategies in resolving a challenging psychedelic experience.

With this in mind, psychedelic preparation should include a range of strategies for managing challenging experiences. This might include, for example, teaching participants techniques to manage anxiety – such as mindfulness and breathing exercises. Some studies also emphasise the importance of the physical presence, and the option of physical touch by the facilitator as a means of reassurance during challenging experiences (Penn et al., 2021; McLane et al., 2021). However, there are others who have cautioned that this poses the potential for boundary violations, with some high-profile cases of such violations having actually been committed (McGuire et al., 2024; McNamee et al., 2023). This is something that inevitably must be managed with care; informed consent is essential.

We would advise tailoring the nature and consent around physical touch to the individual, based on clear discussions during preparation, and actively planning a protocol of what to do if they begin to feel anxious (e.g., they may reach out with their hand if they wish for it to be held). It is of course more sensitive to manage situations in which participants refuse consent during initial consent procedures, but change their mind and need support during the psychedelic dosing session. As McGuire et al. (2024) suggest, more research is required around the influence of supportive touch in psychedelic experiences; for now it is important that there be an experienced facilitator, therapist, or clinician overseeing any episodes wherein the participant experiences anxiety or discomfort.

Equally, the self-reflections that follow the psychedelic dosing session(s) – during the 'integration' phase – can help the participant actively process their psychedelic experience and integrate any insights gained into their lives going forward (Bathje et al., 2022). Bathje et al (2022) suggest that therapeutic support may help with this integration process – they review several models of integration for psychedelic therapy, proposing a synthesised model describing the areas of existence that may be influenced by the psychedelic experience, which the therapist must explore with the participant. This model identifies the following consistent dimensions: somatic, emotional, existential, lifestyle, communal, and nature. However, integration practices are also yet to be standardised and are in need of further validation and research.

Finally, the magnitude of psychological insights and revelations that occur during psychedelic sessions can have a significant impact in patients' lives. Case studies have emphasised that such revelations could be both helpful and misleading, emphasising the 'double-edged sword' nature of these experiences. A good example of these risks concerns the induction of false memories and dramatic changes in spiritual or metaphysical beliefs following psychedelic sessions (Timmermann et al, 2022). It has been proposed that such issues raise ethical considerations related to the informed consent of such transformations and the potential they have to induce forms of social coercion. It has been proposed that an approach based on the expert facilitation of therapists and guides - during the session as well as in preparation and integration stages - is key to develop a 'gentle touch' concerning such insights (Timmermann et al, 2022).

| **Recommendations** | |
|---|---|
| 37 | It is essential to prepare the participant for the psychedelic experience by providing detailed information about possible subjective effects, managing overly positive expectations for therapeutic outcomes, and teaching methods to alleviate anxiety. |

| Recommendations | |
|---|---|
| 38 | The informed consent process should be tailored to each participant, for example to consider their prior experiences with psychedelics, their expectations for the experience and its outcomes, or their preferences surrounding reassuring physical touch. |
| 39 | The training of psychedelic therapists and facilitators should incorporate a nuanced understanding of the opportunities and pitfalls of psychedelic insights and how to aid individuals navigate these to ensure the safety and effectiveness of psychedelic practices. |

## 7. Conclusions

Studies examining the therapeutic potential of psychedelics, often with concurrent psychological support, have burgeoned globally. As yet, it is unclear if psychedelics will enter conventional medical practice, and there is a wide agreement that more – and better – research is required. There are a range of specific methodological complexities faced by researchers looking to set up psychedelic studies. These complexities are wide-ranging, from specific considerations about individual participant safety, to fundamental questions about how evidence for treatments in neuropsychiatric disorders is collected and appraised.

Together, the authors of this paper have amassed a breadth of experience in the design, implementation, delivery, and interpretation of psychedelic studies. In this paper, we have summarised the current state of the field of psychedelic research. Below (and throughout), we have summarised a set of informal recommendations aimed at those looking to undertake studies of psychedelics. These recommendations span topics including study approval processes, legal regulation, placebo and blinding, inclusive recruitment, and psychotherapeutic support. Such recommendations represent guidance and not intended as proscriptive criteria, and we hope that they are improved upon iteratively as the field of psychedelic research continues to develop.

**Table 2. Summary of Recommendations**

| # | Relevant section | Recommendation |
|---|---|---|
| 1 | **2.1.1** | Psychedelics should be reclassified as Schedule II / 2 drugs, with ongoing consideration of how regulatory frameworks can stymie research. |
| 2 | **2.1.2** | Researchers should be aware of the impact of their findings on patients, public opinion, corporate agendas, and Indigenous populations to whom psychedelics are culturally significant. Findings should be disseminated in a careful and balanced manner to avoid stoking unwarranted hype or stigma. |
| 3 | **2.1.3** | Valuable insights may be gained from existing traditional practices using psychedelics (e.g., those used by Indigenous communities), such as the possible limitations of dualist assumptions that underlie some elements of Western medical practice and psychometric measurements of selfhood. |

| Table 2. Summary of Recommendations | | |
|---|---|---|
| 4 | **2.2** | Researchers may benefit from considering unconventional sources of funding (e.g., from crowd-funding or private donations) as well as in kind donations (e.g., study drug donations for research purposes) in order to finance their psychedelic studies. |
| 5 | **2.2** | Public investments into psychedelic research should be encouraged, given the promise of medical benefits for patients who currently lack treatment options, as well as the possible economic benefits. |
| 6 | **2.3** | Regulators should look to streamline approval processes across bodies in order to help facilitate research for patient benefit. |
| 7 | **2.3** | Researchers should be aware of and budget the time required to undergo the process of obtaining all necessary approvals. |
| 8 | **4.1.1** | Given that psychedelic administration in controlled settings increasing the risk of psychosis remains unsubstantiated, and participants with a family history of psychosis are not contraindicated from benefiting from psychedelic therapies, their exclusion from psychedelic trials must be carefully considered. |
| 9 | **4.1.1** | Exclusion criteria considering suicidality should be more specific to suicide planning or previous suicide attempts, as opposed to the much broader criterion of suicidal ideation. Even in cases of past planning or attempts, automatic exclusion may not be justified. |
| 10 | **4.1.4** | Inclusion or exclusion of participants on grounds of previous psychedelic use are not universally delineable: they should be tailored to the goals of the given study. |
| 11 | **4.2** | Wider 'suitability for inclusion' factors, such as social support network, therapy and therapist suitability, and ability to commit to often lengthy trials, may be necessary to consider at screening, at least for clinical trials. |
| 12 | **4.2** | Researchers should record and report all adverse events and contribute this information to public registries that can be used by other researchers to prepare their risk assessments and ethical reviews. |
| 13 | **4.2.1** | Approval frameworks should focus on holistic and prophylactic safety factors rather than simply pharmacological interventions for distress. |
| 14 | **4.2.2** | There should be ongoing clinical and research consideration of the delicate distinction between allowing negative emotions that arise to be addressed in a safe and contained context, and permitting a distressing experience with potential long lasting harms to continue. |
| 15 | **5.1.1** | Psychedelic trialists should proactively aim to be inclusive, for example via recruiting participants with representative proportions of protected characteristics (e.g. ethnicity, gender identity, sexual orientation), previous psychedelic use, and socioeconomic statuses. |

| Table 2. Summary of Recommendations | | |
|---|---|---|
| 16 | **5.1.1** | Sociodemographic information should be recorded at application, screening, and enrollment phases. |
| 17 | **5.1.1** | Protocols should include practical steps to encourage participation from marginalised groups, for example by allowing for diagnoses to be made at screening. |
| 18 | **5.1.2** | Conventional modes of advertising for recruitment may not be applicable to psychedelic studies due to overwhelming public enthusiasm – indeed, researchers may face the converse problem of participant deceit to enrol. |
| 19 | **5.2** | Enthusiasm about positive results from psychedelic studies should be tempered, but not disregarded, due to functional unblinding effects, and the interplay of expectancy and hype. |
| 20 | **5.2** | Psychedelic researchers should include measures of expectancy in study designs. |
| 21 | **5.2** | Regulators should consider the fact that randomised controlled trials may not be the 'best fit' for psychedelic-assisted therapy, and that this should contextualise rather than undermine any regulatory decisions. |
| 22 | **5.2** | Researchers and regulators should consider the ways in which psychedelic trials may be highlighting weaknesses in current understanding of pharmacological evidence-gathering, in which treatment effects are characterised as simply pharmacology plus treatment-related placebo effects. |
| 23 | **5.2** | Blinding efficacy should be examined in controlled studies of psychedelics (i.e. by asking participants after dosing which treatment arm they think they were assigned to). |
| 24 | **5.2.1** | Active comparators can be useful in psychedelic trials, but should not be viewed as an effective means to remove the challenges posed by functional unblinding. |
| 25 | **5.2.2** | Sub-psychoactive doses of psychedelics may be a useful comparator in some trials, however they do not solve all the challenges posed by functional unblinding, and introduce problems of their own |
| 26 | **5.2.4** | Ethical forms of placebo deception could be considered in the design of psychedelic studies with non-treatment seeking participants. |
| 27 | **5.2.5, 5.2.6, 5.2.7** | Open-label extensions, sequential parallel comparison design, or the Zelen Design may help to mitigate some disappointment effects in placebo-controlled trials of psychedelics, and may help to improve participant retention. |

| **Table 2. Summary of Recommendations** | | |
|---|---|---|
| 28 | **5.3.1** | Future scales which attempt to capture psychedelic experience should be developed using inductive approaches, which use participants' experiences as the primary data (as opposed to 'top-down' approaches). |
| 29 | **5.3.1** | Future scales could be developed which are shorter and comprehensible during the acute psychedelic experience. This would allow for a less reductive and more time-sensitive quantification of psychedelic experiences. |
| 30 | **5.4** | Neuroimaging paradigms in psychedelic research should be broadened beyond acute, resting-state studies; there is a need for more psychedelic studies including task fMRI, music/movie fMRI, and further EEG studies in naturalistic contexts. |
| 31 | **5.4** | Coarse parcellations (e.g., seven-network atlases) risk obscuring local variations and hinder interpretability. Fine-grained atlases (e.g., DiFuMo) and gradient-based segmentations can capture overlapping or gradient-like transitions in brain organisation, which are essential for characterising intersubject variability that typifies responses to psychedelics. |
| 32 | **5.4** | Psychedelics produce acute, subacute, and long-term effects. Incorporating M/EEG, MEG, or fMRI recordings across multiple timepoints can link short-term network changes to enduring behavioural or clinical outcomes, shedding light on neuroplastic and therapeutic processes. |
| 33 | **5.4** | Linking phenomenological data (e.g., moment-to-moment ratings, structured experience sampling) with brain metrics helps contextualise neural findings and provides clearer maps between subjective effects, contextual factors, and dynamic neurobiology. |
| 34 | **5.4** | Neuroscientists using psychedelics should develop guidelines to collect and process data to minimise the influence of artifacts commonly encountered in these studies and standardise analytic procedures that could enable consensus pertaining to the effects of psychedelics in the brain. |
| 35 | **6.3** | The dosing environment is a priority in psychedelic research – careful planning of the set-up of the room (e.g., lighting, music), and practical considerations (e.g., food, accessibility, transport to and from the facility) may mitigate the risk of adverse reactions. |
| 36 | **6.3** | The working alliance between facilitator and participant is also a key priority for the safety and comfort of the participant, and may predict positive outcomes in clinical trials. |
| 37 | **6.4** | It is essential to prepare the participant for the psychedelic experience by providing detailed information about possible subjective effects, managing |

| Table 2. Summary of Recommendations | | |
|---|---|---|
| | | overly positive expectations for therapeutic outcomes, teaching methods to alleviate anxiety, and discussing consent around physical touch. |
| 38 | **6.4** | The informed consent process should be tailored to each participant, for example to consider their prior experiences with psychedelics, their expectations for the experience and its outcomes, or their preferences surrounding reassuring physical touch. |
| 39 | **6.4** | The training of psychedelic therapists and facilitators should incorporate a nuanced understanding of the opportunities and pitfalls of psychedelic insights and how to aid individuals navigate these to ensure the safety and effectiveness of psychedelic practices. |

## Acknowledgements

MJG is a PhD student funded by the ESRC. AB and RL would like to acknowledge support from the Beckley Foundation. PK is a PhD student partially funded by atai Impact, the philanthropic arm of atai Life Sciences. MAM was supported by the National Institute for Health and Care Research (NIHR) Maudsley Biomedical Research Centre (BRC), and has received grant funding from the National Institutes of Health Research, Alzheimer's Research UK, the Wellcome Trust, UK Research and Innovation, Lundbeck, and SoseiHeptares. JR has been awarded grant funding (received and managed by King's College London) from Compass Pathways, Beckley PsyTech, Multidisciplinary Association for Psychedelic Studies, Transcend Therapeutics, National Institute for Health Research, Wellcome Trust, Biomedical Research Centre at the South London, and Maudsley NHS Foundation Trust, and acknowledges support from a fellowship (CS-2017-17-007) from the National Institute for Health Research (NIHR) from 2018-2023. CB is a PhD student funded by the NIHR Maudsley BRC. CB works under the Psychoactive Trials Group, which has received research funding from Compass Pathways Ltd, Beckley Psytech Ltd, Transcend Therapeutics and NHIR.

## Conflicts of Interest

PK was previously employed as a Clinical Research scientist at atai Life Sciences. MAM is a consultant for Boehringer-Ingelheim, SoseiHeptares, and Neurocrine; sits on the advisory board for Quolet; and has received speaker fees from Takeda. JR has undertaken paid advisory boards for Clerkenwell Health (Past), Beckley PsyTech (Past), Delica Therapeutics (Past), and paid articles for Janssen, and has received assistance for attendance at conferences from Compass Pathways (past) and Janssen.

# Appendix 1 – Sample Ethics Application

**Welcome to the Integrated Research Application System**

**IRAS Project Filter**

The integrated dataset required for your project will be created from the answers you give to the following questions. The system will generate only those questions and sections which (a) apply to your study type and (b) are required by the bodies reviewing your study. Please ensure you answer all the questions before proceeding with your applications.

Please complete the questions in order. If you change the response to a question, please select 'Save' and review all the questions as your change may have affected subsequent questions.

**Please enter a short title for this project** (maximum 70 characters)
Precision imaging of neural responses to LSD

**1. Is your project research?**

(•) Yes ( ) No

**2. Select one category from the list below:**

( ) Ionising Radiation for combined review of clinical trial of an investigational medicinal product

( ) Ionising Radiation and Devices form for combined review of combined trial of an investigational medicinal product and an investigational medical device

( ) Clinical investigation or other study of a medical device

( ) Other clinical trial to study a novel intervention or randomised clinical trial to compare interventions in clinical practice

(•) Basic science study involving procedures with human participants

( ) Study administering questionnaires/interviews for quantitative analysis, or using mixed quantitative/qualitative methodology

( ) Study involving qualitative methods only

( ) Study limited to working with human tissue samples (or other human biological samples) and data (specific project only)

( ) Study limited to working with data (specific project only)

( ) Research tissue bank

( ) Research database

**If your work does not fit any of these categories, select the option below:**

( ) Other study

**2a. Will the study involve the use of any medical device without a UKCA/CE UKNI/CE Mark, or a UKCA/CE UKNI/CE marked device which has been modified or will be used outside its intended purposes?**

( ) Yes (•) No

**2b. Please answer the following question(s):**

a) Does the study involve the use of any ionising radiation? ( ) Yes (•) No

b) Will you be taking new human tissue samples (or other human biological samples)? (•) Yes ( ) No

c) Will you be using existing human tissue samples (or other human biological samples)? ◯ Yes ◉ No

d) Will the study involve any other clinical procedures with participants (e.g. MRI, ultrasound, physical examination)? ◯ Yes ◉ No

**3. In which countries of the UK will the research sites be located?***(Tick all that apply)*

- [x] England
- [ ] Scotland
- [ ] Wales
- [ ] Northern Ireland

**3a. In which country of the UK will the lead NHS R&D office be located:**

◉ England
◯ Scotland
◯ Wales
◯ Northern Ireland
◯ This study does not involve the NHS

**4. Which applications do you require?**

- [x] IRAS Form
- [ ] Confidentiality Advisory Group (CAG)
- [ ] HM Prison and Probation Service (HMPPS)

**Most research projects require review by a REC within the UK Health Departments' Research Ethics Service. Is your study exempt from REC review?**

◯ Yes ◉ No

**5. Will any research sites in this study be NHS organisations?**

◉ Yes ◯ No

**5a. Are all the research costs and infrastructure costs (funding for the support and facilities needed to carry out the research, e.g. NHS support costs) for this study provided by an NIHR Biomedical Research Centre, NIHR Applied Research Collaboration, NIHR Patient Safety Research Collaboration, or an NIHR HealthTech Research Centre in all study sites?**

**Please see information button for further details.**

◯ Yes ◉ No

***Please see information button for further details.***

**5b. Do you wish to make an application for the study to be considered for NIHR Research Delivery Network (RDN) Support and inclusion in the NIHR RDN Portfolio?**

**Please see information button for further details.**

○ Yes ◉ No

*The NIHR Research Delivery Network (RDN) enables the health and care system to attract, optimise and deliver research across England e.g. by supporting the successful delivery of high-quality research, as an active partner in the research system.*

*If you select yes to this question, information from your IRAS submission will automatically be shared with the NIHR RDN.*

**6. Do you plan to include any participants who are children?**

○ Yes ◉ No

**7. Do you plan at any stage of the project to undertake intrusive research involving adults lacking capacity to consent for themselves?**

○ Yes ◉ No

*Answer Yes if you plan to recruit living participants aged 16 or over who lack capacity, or to retain them in the study following loss of capacity. Intrusive research means any research with the living requiring consent in law. This includes use of identifiable tissue samples or personal information, except where application is being made to the Confidentiality Advisory Group to set aside the common law duty of confidentiality in England and Wales. Please consult the guidance notes for further information on the legal frameworks for research involving adults lacking capacity in the UK.*

**8. Do you plan to include any participants who are prisoners or young offenders in the custody of HM Prison Service or who are offenders supervised by the probation service in England or Wales?**

○ Yes ◉ No

**9. Is the study or any part of it being undertaken as an educational project?**

○ Yes ◉ No

**10. Will this research be financially supported by the United States Department of Health and Human Services or any of its divisions, agencies or programs?**

○ Yes ◉ No

**11. Will identifiable patient data be accessed outside the care team without prior consent at any stage of the project (including identification of potential participants)?**

○ Yes ◉ No

**Integrated Research Application System**
**Application Form for Basic science study involving procedures with human participants**

## IRAS Form (project information)

*Please refer to the E-Submission and Checklist tabs for instructions on submitting this application.*

The Chief Investigator should complete this form. Guidance on the questions is available wherever you see this symbol displayed. We recommend reading the guidance first. The complete guidance and a glossary are available by selecting Help.

Please define any terms or acronyms that might not be familar to lay reviewers of the application.

**Short title and version number:** (maximum 70 characters - this will be inserted as header on all forms)
Precision imaging of neural responses to LSD

*Please complete these details after you have booked the REC application for review.*

**REC Name:**
PR Committee

**REC Reference Number:**
24/PR/0909

**Submission date:**
16/07/2024

## PART A: Core study information

### 1. ADMINISTRATIVE DETAILS

**A1. Full title of the research:**

Using high-resolution precision neuroimaging to investigate the effects of LSD on brain activity in healthy, hallucinogen-experienced volunteers

**A3-1. Chief Investigator:**

| | Title | Forename/Initials | Surname |
|---|---|---|---|
| | [redacted] | | [redacted] |
| Post | [redacted] | | |
| Qualifications | [redacted] | | |
| ORCID ID | [redacted] | | |
| Employer | [redacted] | | |
| Work Address | [redacted] | | |
| Post Code | [redacted] | | |
| Work E-mail | [redacted] | | |
| * Personal E-mail | [redacted] | | |
| Work Telephone | [redacted] | | |

* Personal Telephone/Mobile 07792580531

Fax

** This information is optional. It will not be placed in the public domain or disclosed to any other third party without prior consent.*
*A copy of a current CV (maximum 2 pages of A4) for the Chief Investigator must be submitted with the application.*

**A4. Who is the contact on behalf of the sponsor for all correspondence relating to applications for this project?**
*This contact will receive copies of all correspondence from REC and HRA/R&D reviewers that is sent to the CI.*

| | Title | Forename/Initials | Surname |
|---|---|---|---|
| | [redacted] | | [redacted] |
| Address | [redacted] | | |
| [redacted] | [redacted] | | |
| [redacted] | [redacted] | | |
| [redacted] | [redacted] | | |
| [redacted] | [redacted] | | |

**A5-1. Research reference numbers.** *Please give any relevant references for your study:*

| | |
|---|---|
| Applicant's/organisation's own reference number, e.g. R & D (if available): | N/A |
| Sponsor's/protocol number: | 7TLSDV1 |
| Protocol Version: | 1.0 |
| Protocol Date: | 09/04/2024 |
| Funder's reference number (enter the reference number or state not applicable): | Not applicable |
| Project website: | NA |

**Registry reference number(s):**
*The UK Policy Framework for Health and Social Care Research sets out the principle of making information about research publicly available. Furthermore: Article 19 of the World Medical Association Declaration of Helsinki adopted in 2008 states that "every clinical trial must be registered on a publicly accessible database before recruitment of the first subject"; and the International Committee of Medical Journal Editors (ICMJE) will consider a clinical trial for publication only if it has been registered in an appropriate registry. Please see guidance for more information.*

International Standard Randomised Controlled Trial Number (ISRCTN):

ClinicalTrials.gov Identifier (NCT number):

**Additional reference number(s):**

| Ref.Number Description | Reference Number |
|---|---|
| | |

**A5-2. Is this application linked to a previous study or another current application?**

○ Yes ◉ No

*Please give brief details and reference numbers.*

## 2. OVERVIEW OF THE RESEARCH

*To provide all the information required by review bodies and research information systems, we ask a number of specific questions. This section invites you to give an overview using language comprehensible to lay reviewers and members of the public. Please read the guidance notes for advice on this section.*

**A6-1. Summary of the study.** *Please provide a brief summary of the research (maximum 300 words) using language easily understood by lay reviewers and members of the public. Where the research is reviewed by a REC within the UK Health Departments' Research Ethics Service, this summary will be published on the Health Research Authority (HRA) website following the ethical review. Please refer to the question specific guidance for this question.*

This study will use functional Magnetic Resonance Imaging (fMRI) to investigate – in more detail than ever before – the effects of the psychedelic compound lysergic acid diethylamide (LSD) on brain activity and connectivity. LSD has been shown to significantly improve outcomes for patients with depression, post-traumatic stress disorder and substance addictions – especially those for whom traditional psychiatric drugs are ineffective. LSD also evokes profound changes in consciousness and is often associated with deeply meaningful experiences that prompt lasting changes in perspective.

These experiences tend to be individualised and personal, varying significantly between users, so it is of particular interest here to adopt a 'precision medicine' approach, in which the intertwining influences within subjects are thoroughly characterised (in contrast to between group analyses, which usually do not afford the same level of nuance). We therefore intend to leverage recent developments in precision neuroimaging to map the individualised neural response under LSD, to an unprecedented degree of granularity. Discovering how LSD changes brain activity to produce such states will significantly inform the neurobiology of consciousness and will improve our understanding of how LSD helps patients with the mental disorders listed above – as well as giving us insights into how it may be applied to treat patients with other major mental illnesses.

The study will be performed at neuroimaging research centres at University College London and King's College London. Participants will undergo fMRI and cognitive testing as well as collection of blood samples, with and without the psychedelic compound LSD. Data collection for the study will last approximately 10 months, and is being funded by the charity the Beckley Foundation.

**A6-2. Summary of main issues.** *Please summarise the main ethical, legal, or management issues arising from your study and say how you have addressed them.*

*Not all studies raise significant issues. Some studies may have straightforward ethical or other issues that can be identified and managed routinely. Others may present significant issues requiring further consideration by a REC, HRA, or other review body (as appropriate to the issue). Studies that present a minimal risk to participants may raise complex organisational or legal issues. You should try to consider all the types of issues that the different reviewers may need to consider.*

1. Purpose and design

Psychedelic compounds such as LSD has been shown to have significant potential to improve outcomes for patients with depression, post-traumatic stress disorder and substance addictions – especially those for whom traditional psychiatric drugs are ineffective. The neural mechanisms of action at the macroscopic scale of LSD are not well. This study will provide address a large gap in the existing literature, by using high-field 7T MRI to investigate changes in macroscopic neural function under LSD at far greater resolution than previously allowing us to study alterations at the level of cortical layers. It has also been designed and powered to allow a precision neuroimaging approach; focusing on highly individualised changes and how they relate to quantification of alterations in individual experience, rather than group-level changes as has been the focus in previous research. This work will substantially improve our understanding of how LSD alters the brain's functional neuroanatomy, which will help us understand how it may be applied to treat patients with other major mental illnesses in future.

The design of the study has been the product of a collaboration between basic neuroscientists, MR physicists, psychologists and clinicians with expertise in neuroimaging, psychiatric disorders and pharmacological experiments, as well as with previous users of LSD and other psychedelic compounds.

2. Risks, burdens and benefits

(i). LSD is a Schedule 1 controlled substance under the 2001 Misuse of Drugs Regulations. This mandates that all research carried out using LSD must be authorised by the UK Home Office. This trial will be in possession of the relevant licenses before the trial commences recruitment.

(ii). The most common adverse reaction to LSD is acute anxiety and dysphoria (a 'bad trip'). However, in healthy volunteers, evidence suggests that these experiences are short-lived. There is reliable evidence that these 'bad trip' experiences can be significantly and rapidly mitigated by administration of Diazepam. Contrary to sensational media reports, there is no evidence that there is any risk of prolonged psychotic reactions to LSD in individuals without a personal or family history of psychiatric illness6.

No incidents of prolonged emotional or perceptual disturbances have been reported in modern clinical studies involving the administration of LSD-like hallucinogens to humans. Greater than one hundred healthy individuals have received LSD-like hallucinogens in modern clinical studies and several thousand received LSD in research in the 1950s and 60s. A high level of attention will be paid to the screening process to ensure that we do not include individuals with psychiatric vulnerabilities or histories of reacting adversely to hallucinogens. A medical doctor will be present on study days.

LSD possesses an extremely low level of physiological toxicity and no known risk of mortality. There is no animal or human data to suggest that LSD, even in very high doses, is neurotoxic, and there is no evidence that it contributes to compulsive drug seeking behaviors in animals, or dependency in humans. The lethal dose of LSD in humans is estimated to be 14mg, but no deaths resulting directly from the ingestion have been reported, and doses higher than 14mg have been survived. The typical dose of LSD ranges from 60-250µg, and previous research has shown that doses of 200µg are safely tolerated by healthy volunteers. We intend to give 100µg-200µg (0.1-0.2mg) for this neuroimaging study, administered orally. Due to interindividual variability in responses to LSD and the previous experience of the volunteers with this drug, we will individualise the dosage (within the specified range) based on the participant's prior experiences of self-administration.

(iii). LSD has a characteristic subjective psychoactive effect that is noticeable to both participants and observers, so blinding is not feasible with this dosage of LSD, especially with non-drug naïve participants. Therefore, no formal placebo condition will be conducted - however, there will be a matched baseline condition on separate days before and after the dosing session. The focus here will be on individual changes associated with LSD, and variability within an individual over time relating to their psychedelic experience. Detailed experience sampling will be performed throughout both the dosing and the baseline scanning sessions. Analysis will be performed by a researcher blind to treatment allocation.

3. Inclusion / exclusion

For the reasons mentioned above, we will only include healthy volunteers who have no personal history of psychiatric illness and no family history of psychosis, and all of them will have used hallucinogens before, without reacting adversely. Previous research has shown that hallucinogens are well tolerated in fMRI and MEG settings

4. Consent
Given the criteria for inclusion in the study (i.e., no history of psychiatric illness, previous familiarity with LSD or equivalent hallucinogens), all participants will be capable of providing informed consent.

5. Confidentiality
Person identifiable information will be restricted to the minimum necessary for us enroll and consent participants and meet inclusion/exclusion criteria. This information will be secured and restricted to the principal and co-investigators and clinicians and will not be used in subsequent analyses, all of which will not require this information.

6. Conflict of interest
There are no conflicts of interest (e.g., no commercial interests) for the researchers involved in designing or conducting the research, data collected from the study will only be used for non-commercial research purposes (e.g., academic papers).

7. Use of tissue samples in future research
Any blood samples will be stored at King's College London and will not be shared with other groups or analysed for purposely other than those highlighted in the protocol.

## 3. PURPOSE AND DESIGN OF THE RESEARCH

**A7. Select the appropriate methodology description for this research.** *Please tick all that apply*:

- [x] Case series/ case note review
- [ ] Case control

- [ ] Cohort observation
- [ ] Controlled trial without randomisation
- [ ] Cross-sectional study
- [ ] Database analysis
- [ ] Epidemiology
- [x] Feasibility/ pilot study
- [ ] Laboratory study
- [ ] Metanalysis
- [ ] Qualitative research
- [ ] Questionnaire, interview or observation study
- [ ] Randomised controlled trial
- [ ] Other (please specify)

**A10. What is the principal research question/objective?** *Please put this in language comprehensible to a lay person.*

Examining the relationship between of LSD-induced changes in brain activity and subjective experience sampling of the psychedelic state.

Discovering how LSD changes brain activity to produce such states will significantly inform the neurobiology of consciousness and will improve our understanding of how LSD helps patients with the mental disorders listed above – as well as giving us insights into how it may be applied to treat patients with other major mental illnesses.

**A11. What are the secondary research questions/objectives if applicable?** *Please put this in language comprehensible to a lay person.*

1. Studying cortical laminar changes in response to LSD
2. Assessing effective connectivity underlying psychedelic states associated with LSD

**A12. What is the scientific justification for the research?** *Please put this in language comprehensible to a lay person.*

LSD is a hallucinogen (i.e., it can elicit hallucinations), and is structurally similar to the endogenous neurotransmitter serotonin (5-HT)22, meaning that it influences brain activity by interacting with the same neuronal receptors that 5-HT binds with. LSD was used extensively in the 1950s and early 1960s in combination with psychotherapy to treat a range of mental illnesses, including depression and alcohol dependence (a recent meta-analysis concluded that there is a significant evidence-base for the efficacy of LSD for alcohol dependence). Between the 1950s and mid-1960s there were more than 1000 clinical papers discussing 40,000 patients, several dozen books and 6 international conferences on LSD-assisted psychotherapy, before it later - rather abruptly - became a highly controlled substance.

There has been a recent renaissance of psychedelic neuroscience research: a growing number of studies have been conducted in recent years involving the administration of LSD-like hallucinogens to healthy human volunteers. Physiologically, these drugs are well tolerated at a range of doses. The prevalence of LSD-related adverse events is very low, especially in healthy samples. In fact, independent assessments of relative drug harms have consistently ranked LSD as one of the least harmful controlled drugs, well below some legal drugs and licensed medicines, e.g. alcohol, tobacco and amphetamine.

Prior functional neuroimaging studies investigating LSD and other hallucinogens have typically taken a group approach: collecting data and analysing results to produce an average group response. However, it is becoming increasingly clear that functional neuroimaging studies need to operate at the individual level, because of the considerable variability in individual neuroanatomy, functional activity, and experience. Therefore, recent techniques have focused on increasing the amount of data collected from individual participants so that individualised analyses can be performed. Advances in MR field strength (from 3Tesla to 7Tesla) allow for highly precise functional localisation of neural phenomena. While previous functional MRI studies have highlighted the general changes in neural function associated with psychedelics, they have been limited both by lack of precision in terms of scanner resolution and individual variability. Imaging at a higher resolution also allows for investigating mechanisms of top-down vs bottom-up information flow across regions and layers of the cortex under psychedelics that have been conjectured previously but so far have not been possible to test.

**A13. Please summarise your design and methodology.** *It should be clear exactly what will happen to the research participant, how many times and in what order. Please complete this section in language comprehensible to the lay person. Do not simply reproduce or refer to the protocol. Further guidance is available in the guidance notes.*

This will be a study involving fMRI scanning of 30 healthy volunteers, both prior, during and after dosing with LSD. All volunteers will attend a screening visit before being enrolled in the study. Volunteers will have two other scanning days without any drug.

Volunteers will attend seven times including four separate scanning days at the Wellcome Centre for Human Neuroimaging.

Day 1 will involve a screening visit.

The first scanning session will be on Day 8 and will involve detailed acquisition of resting state fMRI at 7Tesla and MEG as well as functional localiser tasks without drug dosing. This will enable us to acquire robust baseline data that can be used to define individualised functional neuroanatomy without LSD.

The fist dosing session will be on Day 9, participants will be dosed with LSD. Participants will receive a fixed dose of 100ug of LSD (up to 200ug has been safely administered in previous studies). Participants will undergo MRI and MEG scans.

On Day 10, participants will be invited in for a detailed interview about their experience on Day 9.

The second dosing session will be on Day 25 (2 weeks after the first dosing session). Participants will be able to decide on a dose - in liaison with a medical doctor - choosing between 100ug, 150ug or 200ug, based on their previous experience. Participants will undergo MRI and MEG scans.

On Day 26, participants will, again, undertake a detailed interview about their experience on day 25.

The final scanning session will be on Day 33 without drug as in Day 8.

Blood samples will be taken on Days 8, 10, 26 and 33.

Each scanning day will involve approximately 2 hours of MRI scanning, in one-hour sessions with a break in between. In addition to MRI, participants will also be scanned using MEG (approximately 1 hour) during each of the 4 scanning sessions to read out electrophysiological activity with high temporal precision. Given that blinding is not possible with this dosage of LSD and non-drug naïve participants, no formal placebo condition will be conducted. The focus will be on individual changes associated with LSD, and variability within an individual over time relating to their psychedelic experience. Therefore, detailed experience sampling will be performed throughout both the dosing and the baseline scanning sessions. The first dosing session will involve cognitive tasks in the scanner, whereas the second dosing session will involve qualitative observation that will allow a more eco-valid sampling of the patient's experience, without interference. Prior to the baseline and follow-up scans (on days with no drug dosing), blood samples (12ml) will be collected from participants who opt in, and stored for future epigenomic research.

Volunteers will arrive at the scanning centre at a pre-arranged time. All participants will be assessed for MR suitability before entering the MR room, and will be monitored throughout the day. Subjective ratings will be taken before entering the scanner, as well as during and after the scan. At the end of each scan, volunteers will be safely assisted out of the apparatus. They will remain under supervision in a comfortable environment for the remainder of the day. Volunteers will be allowed to go home once the behavioural effects of the drug have subsided (8-12 hours after ingesting the drug). A medical doctor will assess the volunteer’s suitability to go home. Volunteers will be booked a taxi or picked up by a friend/relative; they will not be allowed to drive or cycle home. Volunteers will be asked to contact us when they return home and we will telephone them the following day to enquire about their general wellbeing.

**A14-1. In which aspects of the research process have you actively involved, or will you involve, patients, service users, and/or their carers, or members of the public?**

- [x] Design of the research
- [ ] Management of the research
- [ ] Undertaking the research
- [x] Analysis of results
- [x] Dissemination of findings

☐ None of the above

*Give details of involvement, or if none please justify the absence of involvement.*
In the design of the study protocol, we conducted a survey of users of hallucinogens (i.e., potential participants, given the inclusion/exclusion criteria) to ensure that the study is feasible and tolerable and relevant aspects of the design (such as multi-dimensional experience sampling) are relevant and suitable to assessing altered experiences with in the psychedelic state.

Members of the public will be actively involved in dissemination of the findings of this study. Psychedelic research is increasingly capturing public interest, particularly for its potential for advancing psychiatric medicine and psychotherapy. Our data will also be made open source and easily accessible so that anyone who would like to use or access the raw data will be able to do so.

## 4. RISKS AND ETHICAL ISSUES

## RESEARCH PARTICIPANTS

**A15. What is the sample group or cohort to be studied in this research?**

Select all that apply:

☐ Blood
☐ Cancer
☐ Cardiovascular
☐ Congenital Disorders
☐ Dementias and Neurodegenerative Diseases
☐ Diabetes
☐ Ear
☐ Eye
☑ Generic Health Relevance
☐ Infection
☐ Inflammatory and Immune System
☐ Injuries and Accidents
☐ Mental Health
☐ Metabolic and Endocrine
☐ Musculoskeletal
☐ Neurological
☐ Oral and Gastrointestinal
☐ Paediatrics
☐ Renal and Urogenital
☐ Reproductive Health and Childbirth
☐ Respiratory
☐ Skin
☐ Stroke

Gender: Male and female participants

Lower age limit: 18 Years

Upper age limit: 80 Years

**A17-1. Please list the principal inclusion criteria (list the most important, max 5000 characters).**

Sample size: 30 completers
Physically and mentally healthy (confirmed by requesting GP records of diagnostic history)
Males & females
18+ years old
Able to communicate effectively with researchers to understand the requirements/ activities for the study
Prior experience with hallucinogenic doses of LSD without adverse event

**A17-2. Please list the principal exclusion criteria (list the most important, max 5000 characters).**

Volunteers are ineligible for inclusion if they:
1. Have a current or previously diagnosed psychiatric or neurological disorder
2. Have one or more immediate family members with a current or previously diagnosed psychotic disorder
3. Have a medically significant condition which renders them unsuitable for the study (e.g., diabetes, severe cardiovascular disease, hepatic or renal failure etc)
4. Show MR contraindications (e.g., metal implants, pacemakers, claustrophobia etc)
5. Have no prior experience with a hallucinogenic drug (e.g. LSD, magic mushrooms, DMT, ayahuasca).
6. Have previously experienced an adverse response to a hallucinogenic drug
7. Excessive use of alcohol or other drugs
8. Are currently pregnant
9. Are also participating in or have recently (in the last two months) participated in other pharmacological research studies.

## RESEARCH PROCEDURES, RISKS AND BENEFITS

**A18. Give details of all non-clinical intervention(s) or procedure(s) that will be received by participants as part of the research protocol.** *These include seeking consent, interviews, non-clinical observations and use of questionnaires.*

Please complete the columns for each intervention/procedure as follows:

1. Total number of interventions/procedures to be received by each participant as part of the research protocol.
2. If this intervention/procedure would be routinely given to participants as part of their care outside the research, how many of the total would be routine?
3. Average time taken per intervention/procedure (minutes, hours or days)
4. Details of who will conduct the intervention/procedure, and where it will take place.

| Intervention or procedure | 1 | 2 | 3 | 4 |
|---|---|---|---|---|
| Providing Participant Information sheet and receiving informed consent | 7 | 0 | 15 | Named researchers<br>UCL, Wellcome Centre for Human Neuroimaging |
| Assessment of inclusion/exclusion criteria | 3 | 0 | 15 | Clinical lead (psychiatrist),<br>UCL, Wellcome Centre for Human Neuroimaging |
| Questionnaires (e.g., Mystical Experience Scale), outside the scanner | 1 | 0 | 60 | Named researchers,<br>UCL, Wellcome Centre for Human Neuroimaging |
| Passive cognitive tasks inside the fMRI scanner.<br>Tasks are: a standard oddball/mismatch negativity paradigm, movie-watching (short excerpt of neutral valence film) | 1 | 0 | 120 | Named researchers<br>UCL, Wellcome Centre for Human Neuroimaging |
| Passive cognitive tasks inside the MEG scanner.<br>Tasks are: a standard oddball/mismatch negativity paradigm, movie-watching (short excerpt of neutral valence film) | 1 | 0 | 60 | |
| Microphenomenological interview | 2 | 0 | 120 | Named researchers<br>UCL, Wellcome Centre for Human |

Neuroimaging

**A19. Give details of any clinical intervention(s) or procedure(s) to be received by participants as part of the research protocol.** *These include uses of medicinal products or devices, other medical treatments or assessments, mental health interventions, imaging investigations and taking samples of human biological material. Include procedures which might be received as routine clinical care outside of the research.*

Please complete the columns for each intervention/procedure as follows:

1. Total number of interventions/procedures to be received by each participant as part of the research protocol.
2. If this intervention/procedure would be routinely given to participants as part of their care outside the research, how many of the total would be routine?
3. Average time taken per intervention/procedure (minutes, hours or days).
4. Details of who will conduct the intervention/procedure, and where it will take place.

| Intervention or procedure | 1 | 2 | 3 | 4 |
|---|---|---|---|---|
| Blood sample | 4 | 0 | 15 | Clinician supervised by clinical lead.<br>UCL, Wellcome Centre for Human Neuroimaging |
| LSD administration | 2 | 0 | 15 | Clinician supervised by clinical lead.<br>UCL, Wellcome Centre for Human Neuroimaging |

**A21. How long do you expect each participant to be in the study in total?**

Up to 45 hours over 35 days.

**A22. What are the potential risks and burdens for research participants and how will you minimise them?**

*For all studies, describe any potential adverse effects, pain, discomfort, distress, intrusion, inconvenience or changes to lifestyle. Only describe risks or burdens that could occur as a result of participation in the research. Say what steps would be taken to minimise risks and burdens as far as possible.*

Risks and burdens of MRI scanning:
1. All metallic objects and be screened for prior to participation to avoid any injuries when in the MRI magnetic field.
2. High-resolution scanning can cause some dizziness on entry into the scanner - the participant will enter the scanner very slowly by a trained radiographer in order to avoid discomfort. The participant will also have a baseline scan without the study drug in the first instance to ensure they are comfortable in the high resolution scanner.

Risks and burdens of study drug:

Pharmacovigilance, adverse events & reactions
1. Drug quality
The drug will be purchased from Alembic Pharmaceuticals, and supplied with a certificate of analysis, certifying its purity.

2. Definitions:
1. Adverse Event (AE): Any untoward medical occurrence in a patient or clinical trial subject administered a medicinal product and which does not necessarily have a causal relationship with this treatment
2. Adverse Reaction (AR): All untoward and unintended responses to a medicinal product related to any dose administered
3. Unexpected Adverse Reaction (UAR): An AR, the nature or severity of which is not consistent with the applicable product information.
4. Serious Adverse Event (SAE) or Serious Adverse Reaction: Any untoward medical occurrence or effect that at any dose i) results in death ii) is life-threatening iii) requires hospitalization iv) results in persistent or significant disability or incapacity
5. Suspected Unexpected Serious Adverse Reactions (SUSAR): any suspected adverse reaction related to a medicinal product that is both unexpected and serious

Most AEs and ARs that occur in this study, whether they are serious or not, will be expected treatment-related toxicities due to the drugs in this study. The assignment of the causality will be made by the team including the principle investigator responsible for the care of the participant using the following definitions:

1. Unrelated: There is no evidence of any causal relationship
2. Unlikely: There is little evidence to suggest there is a causal relationship and there is another reasonable explanation for the event.
3. Possible: There is some evidence to suggest a causal however the influence of other factors may have contributed to the event
4. Probable: There is evidence to suggest a causal relationship and the influence of other factors is unlikely
5. Definitely: There is clear evidence to suggest a causal relationship and other possible contributing factors can be ruled out
6. Not Assessable: There is insufficient or incomplete evidence to make a clinical judgment of the causal relationship.

Should any uncertainty exist regarding causality we will discuss within the team and further afield with local clinicians and representatives with the pharmaceutical company if necessary.

As a precaution, oral Midazolam will be available in the event of an adverse drug reaction; this will be administered by the clinician on call during the dosing session.

**A24. What is the potential for benefit to research participants?**

1. A psychiatric and psychological assessment of a participant's mental health.
2. Ongoing psychological and psychiatric support throughout the trial process
3. Optional neuroimaging investigations that may pick up incidental, treatable conditions early

**RECRUITMENT AND INFORMED CONSENT**

***In this section we ask you to describe the recruitment procedures for the study. Please give separate details for different study groups where appropriate.***

**A27-1. How will potential participants, records or samples be identified? Who will carry this out and what resources will be used?***For example, identification may involve a disease register, computerised search of GP records, or review of medical records. Indicate whether this will be done by the direct healthcare team or by researchers acting under arrangements with the responsible care organisation(s).*

The participant pool will be specialised, requiring prior use of LSD and ideally experience with being in an MRI scanner, as well as the fact that the testing will be highly individualised, with a small sample size (n=30). We (the named researchers) will therefore recruit a diverse convenience sample of healthy adult participants from personal correspondence regarding their experience in the use of psychedelics and prior experience being in an MRI scanner, by word of mouth.

**A27-2. Will the identification of potential participants involve reviewing or screening the identifiable personal information of patients, service users or any other person?**

○ Yes ◉ No

*Please give details below:*

**A28. Will any participants be recruited by publicity through posters, leaflets, adverts or websites?**

○ Yes ◉ No

**A29. How and by whom will potential participants first be approached?**

Through personal communication by the named researchers.

**A30-1. Will you obtain informed consent from or on behalf of research participants?**

(●) Yes ( ) No

*If you will be obtaining consent from adult participants, please give details of who will take consent and how it will be done, with details of any steps to provide information (a written information sheet, videos, or interactive material). Arrangements for adults unable to consent for themselves should be described separately in Part B Section 6, and for children in Part B Section 7.*

*If you plan to seek informed consent from vulnerable groups, say how you will ensure that consent is voluntary and fully informed.*

One of the named researchers will take consent from participants prior to participation. This will involve signing a consent form after having read a written information sheet and an explanation of the study from the researcher present.

*If you are not obtaining consent, please explain why not.*

*Please enclose a copy of the information sheet(s) and consent form(s).*

**A30-2. Will you record informed consent (or advice from consultees) in writing?**

(●) Yes ( ) No

**A31. How long will you allow potential participants to decide whether or not to take part?**

The participants will have 24 hours after screening to decide whether or not to take part in the LSD dosing session. They will also be free to withdraw at any point.

**A32. Will you recruit any participants who are involved in current research or have recently been involved in any research prior to recruitment?**

( ) Yes
(●) No
( ) Not Known

**A33-1. What arrangements have been made for persons who might not adequately understand verbal explanations or written information given in English, or who have special communication needs?***(e.g. translation, use of interpreters)*

Sufficient written and verbal English will be an inclusion criterion, as this will be a requisite for the cognitive assessments and tasks to be carried out as a part of the study.

**A35. What steps would you take if a participant, who has given informed consent, loses capacity to consent during the study?** *Tick one option only.*

(●) The participant and all identifiable data or tissue collected would be withdrawn from the study. Data or tissue which is not identifiable to the research team may be retained.

( ) The participant would be withdrawn from the study. Identifiable data or tissue already collected with consent would be retained and used in the study. No further data or tissue would be collected or any other research procedures carried out on or in relation to the participant.

( ) The participant would continue to be included in the study.

( ) Not applicable – informed consent will not be sought from any participants in this research.

( ) Not applicable – it is not practicable for the research team to monitor capacity and continued capacity will be assumed.

*Further details:*

## CONFIDENTIALITY

**In this section, personal data means any data relating to a participant who could potentially be identified. It includes pseudonymised data capable of being linked to a participant through a unique code number.**

### Storage and use of personal data during the study

**A36. Will you be undertaking any of the following activities at any stage (including in the identification of potential participants)?***(Tick as appropriate)*

- [x] Access to medical records by those outside the direct healthcare team
- [ ] Access to social care records by those outside the direct social care team
- [x] Electronic transfer by magnetic or optical media, email or computer networks
- [ ] Sharing of personal data with other organisations
- [ ] Export of personal data outside the EEA
- [x] Use of personal addresses, postcodes, faxes, emails or telephone numbers
- [x] Publication of direct quotations from respondents
- [ ] Publication of data that might allow identification of individuals
- [x] Use of audio/visual recording devices
- [x] Storage of personal data on any of the following:
  - [x] Manual files (includes paper or film)
  - [ ] NHS computers
  - [ ] Social Care Service computers
  - [ ] Home or other personal computers
  - [x] University computers
  - [ ] Private company computers
  - [x] Laptop computers

*Further details:*
- Electronic transfer by magnetic or optical media, email or computer networks: Using encrypted methods where participant identifiable data is used.
- Neuroimaging data and information from scales and assessments will be shared with the research team and collaborators.
- Anonymised data will be made open source after the first publication.
- Audio/video recordings will only be accessible by the research team and if we wish to share them (e.g., with a collaborator), consent will be obtained first.
- Information will be shared with research collaborators at University College London in a pseudoanonymised format.
- Use of personal addresses, postcodes, faxes, emails or telephone numbers. Used to communicate with participants: These data will be stored separately to participant research data.
- Publication of direct quotations from respondents. This may be done in research reports or presentations, but will be brief and anonymised. Further direct consent would be sought for more extensive publications of participant narratives, as these may be self-identifying.
- Storage of personal data: Fully encrypted devices with mandatory password protection.
- We will request GP records of diagnostic history at baseline screening to ensure
- Data will be stored securely in a digital format on password-protected computers at King's College London and University College London
- Data will be transferred electronically between members of the research team in an encrypted format
- The study is subject to monitoring/audit by individuals from the sponsoring organisation

**A37. Please describe the physical security arrangements for storage of personal data during the study?**

- All paper records will be kept in locked offices. Access to the university building is governed by electronic access cards and 24 hour security.
- All electronic records will be kept on secure university computers at KCL and UCL that comply with data security standards pertinent to this type of data.
- All desktop computers will be centrally managed by the university with appropriate security protocols and encryption to prevent unauthorised data access.
- All laptop computers will be encrypted and password protected.
- Paper documentation will be stored in locked filing cabinets - only the research team will have access to this and identifiable electronic data.

**A38. How will you ensure the confidentiality of personal data?***Please provide a general statement of the policy and procedures for ensuring confidentiality, e.g. anonymisation or pseudonymisation of data.*

Data will be pseudonymised: personal data from participants collected during this study will be linked to a participant number. The link back document listing each participant against their assigned code will be stored in an encrypted database in a secure cluster at King's College London (KCL), separately to the main dataset. All data will be backed up to a secure cluster at KCL. Confidentiality of personal data will be maintained and will not be made public. Each participant will be informed that personal data can be looked upon by representatives of the ethical and regulatory authorities and of the sponsor to check that the study is being conducted correctly. All will have a duty of confidentiality.

**A40. Who will have access to participants' personal data during the study?** *Where access is by individuals outside the direct care team, please justify and say whether consent will be sought.*

Members of the study team at King's College London and University College London, as well as representatives from the Sponsor. These data will be stored on password-protected university computers.

## Storage and use of data after the end of the study

**A41. Where will the data generated by the study be analysed and by whom?**

The data will be generated at the Wellcome Centre for Human Neuroimaging at University College London (UCL) and analysed at UCL and King's College London.

**A42. Who will have control of and act as the custodian for the data generated by the study?**

| | Title | Forename/Initials | Surname |
|---|---|---|---|
| | [redacted] | | [redacted] |
| Post | [redacted] | | |
| Qualifications | [redacted] | | |
| Work Address | [redacted] | | |
| Post Code | [redacted] | | |
| Work Email | [redacted] | | |
| [redacted] | | | |
| Fax | | | |

**A43. How long will personal data be stored or accessed after the study has ended?**

- ◯ Less than 3 months
- ◯ 3 – 6 months
- ◯ 6 – 12 months

○ 12 months – 3 years

◉ Over 3 years

*If longer than 12 months, please justify:*
Archiving regulations require this.

Contact details will be kept for 10 years (consent will be obtained from participants for this).

**A44. For how long will you store research data generated by the study?**

Years: 15
Months:

**A45. Please give details of the long term arrangements for storage of research data after the study has ended.***Say where data will be stored, who will have access and the arrangements to ensure security.*

Following completion, the data will be archived within a secure, fire-proof location within the research department and accessible by authorised recall. Each site will be responsible for their own archiving arrangements (as per their Trust/local policies).

## INCENTIVES AND PAYMENTS

**A46. Will research participants receive any payments, reimbursement of expenses or any other benefits or incentives for taking part in this research?**

◉ Yes ○ No

*If Yes, please give details. For monetary payments, indicate how much and on what basis this has been determined.*
The participants will receive the standard payment offered to participants in MRI research at the Wellcome Centre for Human Neuroimaging (£10-12 per hour).

**A47. Will individual researchers receive any personal payment over and above normal salary, or any other benefits or incentives, for taking part in this research?**

○ Yes ◉ No

**A48. Does the Chief Investigator or any other investigator/collaborator have any direct personal involvement (e.g. financial, share holding, personal relationship etc.) in the organisations sponsoring or funding the research that may give rise to a possible conflict of interest?**

○ Yes ◉ No

## NOTIFICATION OF OTHER PROFESSIONALS

**A49-1. Will you inform the participants' General Practitioners (and/or any other health or care professional responsible for their care) that they are taking part in the study?**

○ Yes ◉ No

*If Yes, please enclose a copy of the information sheet/letter for the GP/health professional with a version number and date.*

## References

Abramson, H. A., Jarvik, M. E., Kaufman, M. R., Kornetsky, C., Levine, A., & Wagner, M. (1955). Lysergic Acid Diethylamide (LSD-25): I. Physiological and Perceptual Responses. *The Journal of Psychology*, *39*(1), 3–60. https://doi.org/10.1080/00223980.1955.9916156

Aday, J. S., Davoli, C. C., & Bloesch, E. K. (2020). Psychedelics and virtual reality: Parallels and applications. *Therapeutic Advances in Psychopharmacology*, *10*, 2045125320948356. https://doi.org/10.1177/2045125320948356

Aday, J. S., Heifets, B. D., Pratscher, S. D., Bradley, E., Rosen, R., & Woolley, J. D. (2022). Great Expectations: Recommendations for improving the methodological rigor of psychedelic clinical trials. *Psychopharmacology*, *239*(6), 1989–2010. https://doi.org/10.1007/s00213-022-06123-7

Agnorelli, C., Spriggs, M., Godfrey, K., Sawicka, G., Bohl, B., Douglass, H., Fagiolini, A., Parastoo, H., Carhart-Harris, R., Nutt, D., & Erritzoe, D. (2024). *Neuroplasticity and Psychedelics: A comprehensive examination of classic and non-classic compounds in pre and clinical models* (arXiv:2411.19840). arXiv. https://doi.org/10.48550/arXiv.2411.19840

Aleksandrova, L. R., & Phillips, A. G. (2021). Neuroplasticity as a convergent mechanism of ketamine and classical psychedelics. *Trends in pharmacological sciences*, *42*(11), 929–942. https://doi.org/10.1016/j.tips.2021.08.003

Allen, C. J. (2019). RIGHTING IMBALANCE: Striving for Well-Being in the Andes. *Science, Religion and Culture*, *6*(1). https://doi.org/10.17582/journal.src/2019.6.1.6.14

Allen, N., Jeremiah, A., Murphy, R., Sumner, R., Forsyth, A., Hoeh, N., Menkes, D. B., Evans, W., Muthukumaraswamy, S., Sundram, F., & Roop, P. (2024). LSD increases sleep duration the night after microdosing. *Translational psychiatry*, *14*(1), 191. https://doi.org/10.1038/s41398-024-02900-4

Amarnani, A. D., Free, M. F., & Baweja, R. (2024). Psilocybin and the Development of Serotonin Toxicity. *Primary Care Companion for CNS Disorders*, *26*(1). https://doi.org/10.4088/PCC.23CR03648

Armand, S., Larsen, K., Madsen, M. K., Ozenne, B., Preller, K. H., Knudsen, G. M., Stenbæk, D. S., & Fisher, P. M. (2024). Amygdala response to emotional faces following acute administration of psilocybin in healthy individuals. *Neuroscience Applied*, *3*, 103934. https://doi.org/10.1016/j.nsa.2023.103934

Aron, A., & Aron, E. N. (1986). *Love and the expansion of self: Understanding attraction and satisfaction* (pp. x, 172). Hemisphere Publishing Corp/Harper & Row Publishers.

Aronson, H., Silverstein, A. B., & Klee, G. D. (1959). Influence of lysergic acid diethylamide (LSD-25) on subjective time. *A.M.A. Archives of General Psychiatry*, *1*, 469–472. https://doi.org/10.1001/archpsyc.1959.03590050037003

Atasoy, S., Roseman, L., Kaelen, M., Kringelbach, M. L., Deco, G., & Carhart-Harris, R. L. (2017). Connectome-harmonic decomposition of human brain activity reveals dynamical repertoire re-organization under LSD. *Scientific Reports*, *7*(1), 17661. https://doi.org/10.1038/s41598-017-17546-0

Barber, M., Gardner, J., & Carter, A. (2024). History, Hype, and Responsible Psychedelic Medicine: A Qualitative Study of Psychedelic Researchers. *Journal of Bioethical Inquiry,* 1–17. https://doi.org/10.1007/s11673-024-10386-4

Barker, S. A. (2018). N, N-Dimethyltryptamine (DMT), an Endogenous Hallucinogen: Past, Present, and Future Research to Determine Its Role and Function. *Frontiers in Neuroscience*, *12*. https://doi.org/10.3389/fnins.2018.00536

Barker, S. A. (2022). Administration of N,N-dimethyltryptamine (DMT) in psychedelic therapeutics and research and the study of endogenous DMT. *Psychopharmacology*, *239*(6), 1749–1763. https://doi.org/10.1007/s00213-022-06065-0

Barrett, F. S., Bradstreet, M. P., Leoutsakos, J.-M. S., Johnson, M. W., & Griffiths, R. R. (2016). The Challenging Experience Questionnaire: Characterization of challenging experiences with psilocybin mushrooms. *Journal of Psychopharmacology (Oxford, England)*, *30*(12), 1279–1295. https://doi.org/10.1177/0269881116678781

Barsuglia, J., Davis, A. K., Palmer, R., Lancelotta, R., Windham-Herman, A. M., Peterson, K., Polanco, M., Grant, R., & Griffiths, R. R. (2018). Intensity of mystical experiences occasioned by 5-MeO-DMT and comparison with a prior psilocybin study. *Frontiers in Psychology*, *9*(DEC), 430539. https://doi.org/10.3389/FPSYG.2018.02459/BIBTEX

Bathje, G. J., Majeski, E., & Kudowor, M. (2022). Psychedelic integration: An analysis of the concept and its practice. *Frontiers in Psychology*, *13*. https://doi.org/10.3389/fpsyg.2022.824077

Baumann, S., Carhart-Harris, R., Nutt, D., Erritzoe, D., & Szigeti, B. (2022). *Evidence for tolerance in psychedelic*

*microdosing from the self-blinding microdose trial.* PsyArXiv Preprints. https://doi.org/10.31234/osf.io/s4qhf
Becker, A. M., Holze, F., Grandinetti, T., Klaiber, A., Toedtli, V. E., Kolaczynska, K. E., Duthaler, U., Varghese, N., Eckert, A., Grünblatt, E., & Liechti, M. E. (2022). Acute Effects of Psilocybin After Escitalopram or Placebo Pretreatment in a Randomized, Double-Blind, Placebo-Controlled, Crossover Study in Healthy Subjects. *Clinical pharmacology and therapeutics*, *111*(4), 886–895. https://doi.org/10.1002/cpt.2487
Bedford, P., Hauke, D. J., Wang, Z., Roth, V., Nagy-Huber, M., Holze, F., Ley, L., Vizeli, P., Liechti, M. E., Borgwardt, S., Müller, F., & Diaconescu, A. O. (2023). The effect of lysergic acid diethylamide (LSD) on whole-brain functional and effective connectivity. *Neuropsychopharmacology*, *48*(8), 1175–1183. https://doi.org/10.1038/s41386-023-01574-8
Bender, D., & Hellerstein, D. J. (2022). Assessing the risk-benefit profile of classical psychedelics: A clinical review of second-wave psychedelic research. *Psychopharmacology*, *239*(6), 1907–1932. https://doi.org/10.1007/s00213-021-06049-6
Blackburne, G., McAlpine, R.G., Fabus, M., Liardi, A., Kamboj, S.K., Mediano, P.A.M. & Skipper, J.I. (2024) *Complex slow waves radically reorganise human brain dynamics under 5-MeO-DMT.*p.2024.10.04.616717. https://doi.org/10.1101/2024.10.04.616717.
Bliss, T. V., & Lomo, T. (1970). Plasticity in a monosynaptic cortical pathway. *The Journal of physiology*, *207*(2), 61P.
Blond, B. N., & Schindler, E. A. D. (2023). Case report: Psychedelic-induced seizures captured by intracranial electrocorticography. *Frontiers in neurology*, *14*, 1214969. https://doi.org/10.3389/fneur.2023.1214969
Bogenschutz, M. P., Forcehimes, A. A., Pommy, J. A., Wilcox, C. E., Barbosa, P. C. R., & Strassman, R. J. (2015). Psilocybin-assisted treatment for alcohol dependence: A proof-of-concept study. *Journal of Psychopharmacology (Oxford, England)*, *29*(3), 289–299. https://doi.org/10.1177/0269881114565144
Bogenschutz, M. P., Podrebarac, S. K., Duane, J. H., Amegadzie, S. S., Malone, T. C., Owens, L. T., Ross, S., & Mennenga, S. E. (2018). Clinical Interpretations of Patient Experience in a Trial of Psilocybin-Assisted Psychotherapy for Alcohol Use Disorder. *Frontiers in Pharmacology*, *9*, 100. https://doi.org/10.3389/fphar.2018.00100
Bousman, C. A., Stevenson, J. M., Ramsey, L. B., Sangkuhl, K., Hicks, J. K., Strawn, J. R., Singh, A. B., Ruaño, G., Mueller, D. J., Tsermpini, E. E., Brown, J. T., Bell, G. C., Leeder, J. S., Gaedigk, A., Scott, S. A., Klein, T. E., Caudle, K. E., & Bishop, J. R. (2023). Clinical Pharmacogenetics Implementation Consortium (CPIC) Guideline for CYP2D6, CYP2C19, CYP2B6, SLC6A4, and HTR2A Genotypes and Serotonin Reuptake Inhibitor Antidepressants. *Clinical Pharmacology & Therapeutics*, *114*(1), 51–68. https://doi.org/10.1002/CPT.2903
Bouso, J. C., Pedrero-Pérez, E. J., Gandy, S., & Alcázar-Córcoles, M. Á. (2016). Measuring the subjective: Revisiting the psychometric properties of three rating scales that assess the acute effects of hallucinogens. *Human Psychopharmacology*, *31*(5), 356–372. https://doi.org/10.1002/hup.2545
Breeksema, J. J., Kuin, B. W., Kamphuis, J., van den Brink, W., Vermetten, E., & Schoevers, R. A. (2022). Adverse events in clinical treatments with serotonergic psychedelics and MDMA: A mixed-methods systematic review. *Journal of Psychopharmacology*, *36*(10), 1100–1117.
Breeksema, J. J., Niemeijer, A. R., Krediet, E., Vermetten, E., & Schoevers, R. A. (2020). Psychedelic Treatments for Psychiatric Disorders: A Systematic Review and Thematic Synthesis of Patient Experiences in Qualitative Studies. *CNS Drugs*, *34*(9), 925–946. https://doi.org/10.1007/s40263-020-00748-y
Bremler, R., Katati, N., Shergill, P., Erritzoe, D., & Carhart-Harris, R. L. (2023). Case analysis of long-term negative psychological responses to psychedelics. *Scientific Reports*, *13*(1), 15998. https://doi.org/10.1038/s41598-023-41145-x
Brown, R. T., Nicholas, C. R., Cozzi, N. V., Gassman, M. C., Cooper, K. M., Muller, D., Thomas, C. D., Hetzel, S. J., Henriquez, K. M., Ribaudo, A. S., & Hutson, P. R. (2017). Pharmacokinetics of Escalating Doses of Oral Psilocybin in Healthy Adults. *Clinical Pharmacokinetics*, *56*(12), 1543–1554. https://doi.org/10.1007/s40262-017-0540-6
Bück, R. W. (1961). Mushroom poisoning since 1924 in the United States. *Mycologia*, *53*(5), 537-538.
Butler, M., Bird, C., Maggio, C., Durden, A., Modlin, N., Campbell-Coker, K., Edwards, M., Pick, S., Millman, L. S. M., Lowery, E., Bhagavan, C., Kanaan, R., Golder, D., Mildon, B., Mehta, M., Rucker, J., & Nicholson, T. R. (2024). Probing the functional magnetic resonance imaging response to psilocybin in functional neurological disorder (PsiFUND): Study protocol. *Wellcome Open Research*, 9, 401. https://doi.org/10.12688/wellcomeopenres.22543.1
Butler, M., Jelen, L., & Rucker, J. (2022). Expectancy in placebo-controlled trials of psychedelics: If so, so what? *Psychopharmacology*, *239*(10), 3047–3055. https://doi.org/10.1007/s00213-022-06221-6

Butler, M., Seynaeve, M., Nicholson, T. R., Pick, S., Kanaan, R. A., Lees, A., Young, A. H., & Rucker, J. (2020). Psychedelic treatment of functional neurological disorder: A systematic review. *Therapeutic Advances in Psychopharmacology*, *10*, 2045125320912125. https://doi.org/10.1177/2045125320912125

Cai, H., Xie, X.-M., Zhang, Q., Cui, X., Lin, J.-X., Sim, K., Ungvari, G. S., Zhang, L., & Xiang, Y.-T. (2021). Prevalence of Suicidality in Major Depressive Disorder: A Systematic Review and Meta-Analysis of Comparative Studies. *Frontiers in Psychiatry*, *12*, 690130. https://doi.org/10.3389/fpsyt.2021.690130

Calder, A. E., & Hasler, G. (2023). Towards an understanding of psychedelic-induced neuroplasticity. *Neuropsychopharmacology*, *48*(1), 104–112. https://doi.org/10.1038/s41386-022-01389-z

Calder, A.E., Hase, A. & Hasler, G. (2024) Effects of psychoplastogens on blood levels of brain-derived neurotrophic factor (BDNF) in humans: a systematic review and meta-analysis. *Molecular Psychiatry.* 1–14. https://doi.org/10.1038/s41380-024-02830-z.

Cameron, L. P., Tombari, R. J., Lu, J., Pell, A. J., Hurley, Z. Q., Ehinger, Y., Vargas, M. V., McCarroll, M. N., Taylor, J. C., Myers-Turnbull, D., Liu, T., Yaghoobi, B., Laskowski, L. J., Anderson, E. I., Zhang, G., Viswanathan, J., Brown, B. M., Tjia, M., Dunlap, L. E., Rabow, Z. T., … Olson, D. E. (2021). A non-hallucinogenic psychedelic analogue with therapeutic potential. *Nature*, *589*(7842), 474–479. https://doi.org/10.1038/s41586-020-3008-z

Carbonaro, T. M., Bradstreet, M. P., Barrett, F. S., MacLean, K. A., Jesse, R., Johnson, M. W., & Griffiths, R. R. (2016). Survey study of challenging experiences after ingesting psilocybin mushrooms: Acute and enduring positive and negative consequences. *Journal of Psychopharmacology*, *30*(12), 1268–1278. https://doi.org/10.1177/0269881116662634

Carbonaro, T. M., Johnson, M. W., Hurwitz, E., & Griffiths, R. R. (2018). Double-blind comparison of the two hallucinogens psilocybin and dextromethorphan: Similarities and differences in subjective experiences. *Psychopharmacology*, *235*(2), 521–534. https://doi.org/10.1007/s00213-017-4769-4

Carhart-Harris, R. L., Bolstridge, M., Rucker, J., Day, C. M. J., Erritzoe, D., Kaelen, M., Bloomfield, M., Rickard, J. A., Forbes, B., Feilding, A., Taylor, D., Pilling, S., Curran, V. H., & Nutt, D. J. (2016). Psilocybin with psychological support for treatment-resistant depression: An open-label feasibility study. *The Lancet. Psychiatry*, *3*(7), 619–627. https://doi.org/10.1016/S2215-0366(16)30065-7

Carhart-Harris, R. L., Erritzoe, D., Haijen, E., Kaelen, M., & Watts, R. (2018). Psychedelics and connectedness. *Psychopharmacology*, *235*(2), 547–550. https://doi.org/10.1007/s00213-017-4701-y

Carhart-Harris, R. L., Erritzoe, D., Williams, T., Stone, J. M., Reed, L. J., Colasanti, A., Tyacke, R. J., Leech, R., Malizia, A. L., Murphy, K., Hobden, P., Evans, J., Feilding, A., Wise, R. G., & Nutt, D. J. (2012). Neural correlates of the psychedelic state as determined by fMRI studies with psilocybin. *Proceedings of the National Academy of Sciences of the United States of America*, *109*(6), 2138–2143. https://doi.org/10.1073/pnas.1119598109

Carhart-Harris, R. L., & Friston, K. J. (2019). REBUS and the Anarchic Brain: Toward a Unified Model of the Brain Action of Psychedelics. *Pharmacological Reviews*, *71*(3), 316–344. https://doi.org/10.1124/pr.118.017160

Carhart-Harris, R. L., Giribaldi, B., Watts, R., Baker-Jones, M., Murphy-Beiner, A., Murphy, R., Martell, J., Blemings, A., Erritzoe, D., & Nutt, D. J. (2021). Trial of Psilocybin versus Escitalopram for Depression. *New England Journal of Medicine*, *384*(15), 1402–1411. https://doi.org/10.1056/NEJMoa2032994

Carhart-Harris, R. L., & Goodwin, G. M. (2017). The Therapeutic Potential of Psychedelic Drugs: Past, Present, and Future. *Neuropsychopharmacology*, *42*(11), 2105–2113. https://doi.org/10.1038/npp.2017.84

Carhart-Harris, R. L., Kaelen, M., Bolstridge, M., Williams, T. M., Williams, L. T., Underwood, R., Feilding, A., & Nutt, D. J. (2016). The paradoxical psychological effects of lysergic acid diethylamide (LSD). *Psychological Medicine*, *46*(7), 1379–1390. https://doi.org/10.1017/S0033291715002901

Carhart-Harris, R. L., Kaelen, M., Whalley, M. G., Bolstridge, M., Feilding, A., & Nutt, D. J. (2015). LSD enhances suggestibility in healthy volunteers. *Psychopharmacology*, *232*(4), 785–794. https://doi.org/10.1007/s00213-014-3714-z

Carhart-Harris, R. L., Leech, R., Erritzoe, D., Williams, T. M., Stone, J. M., Evans, J., Sharp, D. J., Feilding, A., Wise, R. G., & Nutt, D. J. (2013). Functional connectivity measures after psilocybin inform a novel hypothesis of early psychosis. *Schizophrenia Bulletin*, *39*(6), 1343–1351. https://doi.org/10.1093/schbul/sbs117

Carhart-Harris, R. L., Leech, R., Hellyer, P. J., Shanahan, M., Feilding, A., Tagliazucchi, E., Chialvo, D. R., & Nutt, D. (2014). The entropic brain: A theory of conscious states informed by neuroimaging research with psychedelic drugs. *Frontiers in Human Neuroscience*, *8*, 20. https://doi.org/10.3389/fnhum.2014.00020

Carhart-Harris, R. L., Leech, R., Williams, T. M., Erritzoe, D., Abbasi, N., Bargiotas, T., Hobden, P., Sharp, D. J., Evans, J., Feilding, A., Wise, R. G., & Nutt, D. J. (2012). Implications for psychedelic-assisted psychotherapy:

Functional magnetic resonance imaging study with psilocybin. *The British Journal of Psychiatry: The Journal of Mental Science*, *200*(3), 238–244. https://doi.org/10.1192/bjp.bp.111.103309
Carhart-Harris, R. L., Muthukumaraswamy, S., Roseman, L., Kaelen, M., Droog, W., Murphy, K., Tagliazucchi, E., Schenberg, E. E., Nest, T., Orban, C., Leech, R., Williams, L. T., Williams, T. M., Bolstridge, M., Sessa, B., McGonigle, J., Sereno, M. I., Nichols, D., Hellyer, P. J., … Nutt, D. J. (2016). Neural correlates of the LSD experience revealed by multimodal neuroimaging. *Proceedings of the National Academy of Sciences*, *113*(17), 4853–4858. https://doi.org/10.1073/pnas.1518377113
Carhart-Harris, R. L., Williams, T. M., Sessa, B., Tyacke, R. J., Rich, A. S., Feilding, A., & Nutt, D. J. (2011). The administration of psilocybin to healthy, hallucinogen-experienced volunteers in a mock-functional magnetic resonance imaging environment: A preliminary investigation of tolerability. *Journal of Psychopharmacology*, *25*(11), 1562–1567. https://doi.org/10.1177/0269881110367445
Cavanna, F., Muller, S., de la Fuente, L. A., Zamberlan, F., Palmucci, M., Janeckova, L., Kuchar, M., Pallavicini, C., & Tagliazucchi, E. (2022). Microdosing with psilocybin mushrooms: A double-blind placebo-controlled study. *Translational Psychiatry*, *12*(1), 1–11. https://doi.org/10.1038/s41398-022-02039-0
Chaves, C., Dos Santos, R. G., Dursun, S. M., Tusconi, M., Carta, M. G., Brietzke, E., & Hallak, J. E. C. (2024). Why N,N-dimethyltryptamine matters: unique features and therapeutic potential beyond classical psychedelics. *Frontiers in psychiatry*, *15*, 1485337. https://doi.org/10.3389/fpsyt.2024.1485337
Clapp, W. C., Hamm, J. P., Kirk, I. J., & Teyler, T. J. (2012). Translating long-term potentiation from animals to humans: a novel method for noninvasive assessment of cortical plasticity. *Biological psychiatry*, *71*(6), 496–502. https://doi.org/10.1016/j.biopsych.2011.08.021
Clarkson, P. (1996). Ending Psychotherapy: Recurring Themes. *Self & Society*, *24*(5), 26–30. https://doi.org/10.1080/03060497.1996.11085689
Clarkson, P. (2003). *The therapeutic relationship, 2nd ed* (pp. xxxix, 429). Whurr Publishers.
Colloca, L., & Fava, M. (2024). What should constitute a control condition in psychedelic drug trials? *Nature Mental Health*, *2*(10), 1152–1160. https://doi.org/10.1038/s44220-024-00321-2
Dadi, K., Varoquaux, G., Machlouzarides-Shalit, A., Gorgolewski, K. J., Wassermann, D., Thirion, B., & Mensch, A. (2020). Fine-grain atlases of functional modes for fMRI analysis. *NeuroImage*, *221*, 117126. https://doi.org/10.1016/j.neuroimage.2020.117126
Dahmane, E., Hutson, P. R., & Gobburu, J. V. S. (2021). Exposure-Response Analysis to Assess the Concentration-QTc Relationship of Psilocybin/Psilocin. *Clinical Pharmacology in Drug Development*, *10*(1), 78–85. https://doi.org/10.1002/CPDD.796
Davis, A. K., Barsuglia, J. P., Lancelotta, R., Grant, R. M., & Renn, E. (2018). The epidemiology of 5-methoxy- N, N-dimethyltryptamine (5-MeO-DMT) use: Benefits, consequences, patterns of use, subjective effects, and reasons for consumption. *Journal of psychopharmacology (Oxford, England)*, *32*(7), 779–792. https://doi.org/10.1177/0269881118769063
Daws, R.E., Timmermann, C., Giribaldi, B., Sexton, J.D., Wall, M.B., Erritzoe, D., Roseman, L., Nutt, D. & Carhart-Harris, R. (2022) Increased global integration in the brain after psilocybin therapy for depression. *Nature Medicine*. 28 (4), 844–851. http://doi.org/10.1038/s41591-022-01744-z.
de Almeida, R.N., Galvão, A.C. de M., da Silva, F.S., Silva, E.A. dos S., Palhano-Fontes, F., Maia-de-Oliveira, J.P., de Araújo, L.-S.B., Lobão-Soares, B. & Galvão-Coelho, N.L. (2019) Modulation of Serum Brain-Derived Neurotrophic Factor by a Single Dose of Ayahuasca: Observation From a Randomized Controlled Trial. *Frontiers in Psychology.* 10, 1234. https://doi.org/10.3389/fpsyg.2019.01234.
de Deus Pontual, A. A., Senhorini, H. G., Corradi-Webster, C. M., Tófoli, L. F., & Daldegan-Bueno, D. (2023). Systematic Review of Psychometric Instruments Used in Research with Psychedelics. *Journal of Psychoactive Drugs*, *55*(3), 359–368. https://doi.org/10.1080/02791072.2022.2079108
De Filippo, R., & Schmitz, D. (2024). Synthetic surprise as the foundation of the psychedelic experience. *Neuroscience and biobehavioral reviews*, *157*, 105538. https://doi.org/10.1016/j.neubiorev.2024.105538
de Laportalière, T. T., Jullien, A., Yrondi, A., Cestac, P., & Montastruc, F. (2023). Reporting of harms in clinical trials of esketamine in depression: A systematic review. *Psychological Medicine*, *53*(10), 4305–4315. https://doi.org/10.1017/S0033291723001058
de Mori, B. B. (2017). Paths of healing, voices of sorcerers. *Terrain. Anthropologie & Sciences Humaines*, *68*. https://doi.org/10.4000/terrain.16425
de Vos, C. M. H., Mason, N. L., & Kuypers, K. P. C. (2021). Psychedelics and Neuroplasticity: A Systematic Review Unraveling the Biological Underpinnings of Psychedelics. *Frontiers in psychiatry*, *12*, 724606. https://doi.org/10.3389/fpsyt.2021.724606

de Wit, H., Molla, H. M., Bershad, A., Bremmer, M., & Lee, R. (2022). Repeated low doses of LSD in healthy adults: A placebo-controlled, dose-response study. *Addiction Biology*, *27*(2), e13143. https://doi.org/10.1111/adb.13143

Deco, G., Cruzat, J., Cabral, J., Knudsen, G. M., Carhart-Harris, R. L., Whybrow, P. C., Logothetis, N. K., & Kringelbach, M. L. (2018). Whole-Brain Multimodal Neuroimaging Model Using Serotonin Receptor Maps Explains Non-linear Functional Effects of LSD. *Current Biology*, *28*(19), 3065-3074.e6. https://doi.org/10.1016/j.cub.2018.07.083

Dittrich, A., Lamparter, D., & Maurer, M. (2006). 5D-ABZ: Fragebogen zur Erfassung Aussergewöhnlicher Bewusstseinszustände. *Eine Kurze Einführung [5D-ASC: Questionnaire for the Assessment of Altered States of Consciousness. A Short Introduction]. Zurich, Switzerland: PSIN PLUS*.

Dolder, P. C., Schmid, Y., Steuer, A. E., Kraemer, T., Rentsch, K. M., Hammann, F., & Liechti, M. E. (2017). Pharmacokinetics and Pharmacodynamics of Lysergic Acid Diethylamide in Healthy Subjects. *Clinical Pharmacokinetics*, *56*(10), 1219–1230. https://doi.org/10.1007/s40262-017-0513-9

Domínguez-Clavé, E., Soler, J., Pascual, J. C., Elices, M., Franquesa, A., Valle, M., Alvarez, E., & Riba, J. (2019). Ayahuasca improves emotion dysregulation in a community sample and in individuals with borderline-like traits. *Psychopharmacology*, *236*(2), 573–580. https://doi.org/10.1007/s00213-018-5085-3

Dornbierer, D. A., Marten, L., Mueller, J., Aicher, H. D., Mueller, M. J., Boxler, M., Kometer, M., Kosanic, D., von Rotz, R., Puchkov, M., Kraemer, T., Landolt, H.-P., Seifritz, E., & Scheidegger, M. (2023). Overcoming the clinical challenges of traditional ayahuasca: A first-in-human trial exploring novel routes of administration of N,N-Dimethyltryptamine and harmine. *Frontiers in Pharmacology*, *14*, 1246892. https://doi.org/10.3389/fphar.2023.1246892

Dos Santos, R. G. (2013). A Critical Evaluation of Reports Associating Ayahuasca with Life-Threatening Adverse Reactions. *Journal of Psychoactive Drugs*, *45*(2), 179–188. https://doi.org/10.1080/02791072.2013.785846

Dos Santos, R. G., Balthazar, F. M., Bouso, J. C., & Hallak, J. E. (2016). The current state of research on ayahuasca: A systematic review of human studies assessing psychiatric symptoms, neuropsychological functioning, and neuroimaging. *Journal of psychopharmacology (Oxford, England)*, *30*(12), 1230–1247. https://doi.org/10.1177/0269881116652578

Dos Santos, R. G., Bouso, J. C., & Hallak, J. E. C. (2017). Ayahuasca, dimethyltryptamine, and psychosis: a systematic review of human studies. *Therapeutic Advances in Psychopharmacology*, *7*(4), 141. https://doi.org/10.1177/2045125316689030

Doss, M. K., Kloft, L., Mason, N. L., Mallaroni, P., Reckweg, J. T., van Oorsouw, K., Tupper, N., Otgaar, H., & Ramaekers, J. G. (2024). Ayahuasca enhances the formation of hippocampal-dependent episodic memory without impacting false memory susceptibility in experienced ayahuasca users: An observational study. *Journal of psychopharmacology*, 2698811241301216. Advance online publication. https://doi.org/10.1177/02698811241301216

Doss, M. K., Považan, M., Rosenberg, M. D., Sepeda, N. D., Davis, A. K., Finan, P. H., Smith, G. S., Pekar, J. J., Barker, P. B., Griffiths, R. R., & Barrett, F. S. (2021). Psilocybin therapy increases cognitive and neural flexibility in patients with major depressive disorder. *Translational Psychiatry*, *11*(1), 574. https://doi.org/10.1038/s41398-021-01706-y

Dourron, H. M., Bradley, M., Simonsson, O., Copes, H., Grossman, D., Shallcross, R. A., & Hendricks, P. S. (2025). Naturalistic Psychedelic Use and Psychotic Symptoms: A Cross-Sectional Study of Individuals with a Personal or Family History of Psychotic or Bipolar Disorders. *Psychedelic Medicine*. https://doi.org/10.1089/psymed.2024.0036

Dourron, H. M., Nichols, C. D., Simonsson, O., Bradley, M., Carhart-Harris, R., & Hendricks, P. S. (2023). 5-MeO-DMT: An atypical psychedelic with unique pharmacology, phenomenology & risk? *Psychopharmacology.* https://doi.org/10.1007/s00213-023-06517-1

Doyle, M. A., Ling, S., Lui, L. M. W., Fragnelli, P., Teopiz, K. M., Ho, R., Di Vincenzo, J. D., Rosenblat, J. D., Gillissie, E. S., Nogo, D., Ceban, F., Jawad, M. Y., & McIntyre, R. S. (2022). Hallucinogen persisting perceptual disorder: A scoping review covering frequency, risk factors, prevention, and treatment. *Expert Opinion on Drug Safety*, *21*(6), 733–743. https://doi.org/10.1080/14740338.2022.2063273

D'Souza, D. C., Syed, S. A., Flynn, L. T., Safi-Aghdam, H., Cozzi, N. V., & Ranganathan, M. (2022). Exploratory study of the dose-related safety, tolerability, and efficacy of dimethyltryptamine (DMT) in healthy volunteers and major depressive disorder. *Neuropsychopharmacology: Official Publication of the American College of Neuropsychopharmacology*, *47*(10), 1854–1862. https://doi.org/10.1038/s41386-022-01344-y

Duerler, P., Brem, S., Fraga-González, G., Neef, T., Allen, M., Zeidman, P., Stämpfli, P., Vollenweider, F. X., & Preller, K. H. (2021). Psilocybin Induces Aberrant Prediction Error Processing of Tactile Mismatch Responses-A Simultaneous EEG-FMRI Study. *Cerebral cortex (New York, N.Y. : 1991)*, 32(1), 186–196. https://doi.org/10.1093/cercor/bhab202
Dupuis, D., & Veissière, S. (2022). Culture, context, and ethics in the therapeutic use of hallucinogens: Psychedelics as active super-placebos? *Transcultural Psychiatry*, *59*(5), 571–578. https://doi.org/10.1177/13634615221131465
Dyck, E. (2005). Flashback: Psychiatric Experimentation with LSD in Historical Perspective. *The Canadian Journal of Psychiatry*, *50*(7), 381–388. https://doi.org/10.1177/070674370505000703
Dyck, E. (2008). *Psychedelic psychiatry: LSD from clinic to campus* (pp. xiii, 199). Johns Hopkins University Press.
Ermakova, A. O., Dunbar, F., Rucker, J., & Johnson, M. W. (2021). A narrative synthesis of research with 5-MeO-DMT. *Journal of Psychopharmacology (Oxford, England), 36*(3), 273. https://doi.org/10.1177/02698811211050543
Erne, L., Vogt, S. B., Müller, L., Nuraj, A., Becker, A., Klaiber, A., ... & Liechti, M. E. (2024). Acute dose-dependent effects and self-guided titration of continuous N, N-dimethyltryptamine infusions in a double-blind placebo-controlled study in healthy participants. Neuropsychopharmacology, 1-9.
*Erowid 5-MeO-DMT Vault: Dosage.* (2025). Retrieved February 26, 2025, from https://erowid.org/chemicals/5meo_dmt/5meo_dmt_dose.shtml
*Erowid DMT Vault: Dosage*. (2025). Retrieved February 26, 2025, from https://erowid.org/chemicals/dmt/dmt_dose.shtml
Evans, J., & Adams, J. H. (2024). *Guruism and Cultic Social Dynamics in Psychedelic Practices and Organisations* (pp. 1–17). Springer Berlin Heidelberg. https://doi.org/10.1007/7854_2024_535
Evans, J., Robinson, O. C., Argyri, E. K., Suseelan, S., Murphy-Beiner, A., McAlpine, R., Luke, D., Michelle, K., & Prideaux, E. (2023). Extended difficulties following the use of psychedelic drugs: A mixed methods study. *PLOS ONE*, *18*(10), e0293349. https://doi.org/10.1371/journal.pone.0293349
Falchi-Carvalho, M., Barros, H., Bolcont, R., Laborde, S., Wießner, I., Silva, S. R. B., Montanini, D., Barbosa, D. C., Teixeira, E., Florence-Vilela, R., Almeida, R., Macedo, R. K. A. de, Arichelle, F., Pantrigo, É. J., Costa-Macedo, J. V., Arcoverde, E., Galvão-Coelho, N., Araujo, D. B., & Palhano-Fontes, F. (2024). The Antidepressant Effects of Vaporized N,N-Dimethyltryptamine: An Open-Label Pilot Trial in Treatment-Resistant Depression. *Psychedelic Medicine*. https://doi.org/10.1089/psymed.2024.0002
Fauvel, B., Kangaslampi, S., Strika-Bruneau, L., Roméo, B., & Piolino, P. (2023). Validation of a French Version of the Mystical Experience Questionnaire with Retrospective Reports of the Most Significant Psychedelic Experience among French Users. *Journal of Psychoactive Drugs*, *55*(2), 170–179. https://doi.org/10.1080/02791072.2022.2059796
Feulner, L., Sermchaiwong, T., Rodland, N., & Galarneau, D. (2023). Efficacy and Safety of Psychedelics in Treating Anxiety Disorders. *Ochsner journal*, *23*(4), 315–328. https://doi.org/10.31486/toj.23.0076
Finn, E. S., Glerean, E., Hasson, U., & Vanderwal, T. (2022). Naturalistic imaging: The use of ecologically valid conditions to study brain function. *NeuroImage*, *247*, 118776. https://doi.org/10.1016/j.neuroimage.2021.118776
Finnema, S.J., Nabulsi, N.B., Eid, T., Detyniecki, K., Lin, S.-F., Chen, M.-K., Dhaher, R., Matuskey, D., Baum, E., Holden, D., Spencer, D.D., Mercier, J., Hannestad, J., Huang, Y. & Carson, R.E. (2016) Imaging synaptic density in the living human brain. *Science Translational Medicine*. 8 (348), 348ra96. https://doi.org/10.1126/scitranslmed.aaf6667.
*F.I.V.E, 5-MeO-DMT Information & Vital Education*.(2025). Retrieved 10 March, 2025, from https://five-meo.education/basic-information/
Flückiger, C., Del Re, A. C., Wampold, B. E., & Horvath, A. O. (2018). The alliance in adult psychotherapy: A meta-analytic synthesis. *Psychotherapy (Chicago, Ill.)*, *55*(4), 316–340. https://doi.org/10.1037/pst0000172
Fotiou, E., & Gearin, A. K. (2019). Purging and the body in the therapeutic use of ayahuasca. *Social Science & Medicine (1982)*, *239*, 112532. https://doi.org/10.1016/j.socscimed.2019.112532
Freidel, N., Kreuder, L., Rabinovitch, B. S., Chen, F. Y., Huang, R. S. T., & Lewis, E. C. (2023). Psychedelics, epilepsy, and seizures: A review. *Frontiers in Pharmacology*, *14*, 1326815. https://doi.org/10.3389/fphar.2023.1326815
Gabay, A. S., Carhart-Harris, R. L., Mazibuko, N., Kempton, M. J., Morrison, P. D., Nutt, D. J., & Mehta, M. A. (2018). Psilocybin and MDMA reduce costly punishment in the Ultimatum Game. *Scientific Reports*, *8*(1), 8236. https://doi.org/10.1038/s41598-018-26656-2
Gable, R. S. (2007). Risk assessment of ritual use of oral dimethyltryptamine (DMT) and harmala alkaloids. *Addiction (Abingdon, England)*, *102*(1), 24–34. https://doi.org/10.1111/j.1360-0443.2006.01652.x
Gaddis, A., Lidstone, D. E., Nebel, M. B., Griffiths, R. R., Mostofsky, S. H., Mejia, A. F., & Barrett, F. S. (2022). Psilocybin induces spatially constrained alterations in thalamic functional organizaton and connectivity.

*NeuroImage*, *260*, 119434. https://doi.org/10.1016/j.neuroimage.2022.119434
Garcia-Romeu, A., Barrett, F. S., Carbonaro, T. M., Johnson, M. W., & Griffiths, R. R. (2021). Optimal dosing for psilocybin pharmacotherapy: Considering weight-adjusted and fixed dosing approaches. *Journal of Psychopharmacology*, *35*(4), 353–361. https://doi.org/10.1177/0269881121991822
De la Garza, M. (2012). *Sueño y éxtasis en el mundo náhuatl y maya*. Instituto de Investigaciones Filológicas-Fondo de Cultura Económica.
Gashi, L., Sandberg, S., & Pedersen, W. (2021). Making "bad trips" good: How users of psychedelics narratively transform challenging trips into valuable experiences. *International Journal of Drug Policy*, *87*, 102997. https://doi.org/10.1016/j.drugpo.2020.102997
Gasser, P., Holstein, D., Michel, Y., Doblin, R., Yazar-Klosinski, B., Passie, T., & Brenneisen, R. (2014). Safety and Efficacy of Lysergic Acid Diethylamide-Assisted Psychotherapy for Anxiety Associated With Life-threatening Diseases. *The Journal of Nervous and Mental Disease*, *202*(7), 513. https://doi.org/10.1097/NMD.0000000000000113
Gendlin, E. T. (1992). The wider role of bodily sense in thought and language. In M. Sheets-Johnstone (Ed.), *Giving the body its due* (pp. 192–207). State University of New York Press.
Gerault, A., & Picart, D. (1996). Intoxication mortelle à la suite de la consommation volontaire et en groupe de champignons hallucinogènes [Fatal Poisoning After a Group of People Voluntarily Consumed Hallucinogenic Mushrooms]. *Bulletin Trimestriel de La Société Mycologique de France*, *112*(1), 1–14.
Gillin, J. C., Kaplan, J., Stillman, R., & Wyatt, R. J. (1976). The psychedelic model of schizophrenia: the case of N,N-dimethyltryptamine. *The American journal of psychiatry*, *133*(2), 203–208. https://doi.org/10.1176/ajp.133.2.203
Girn, M., Rosas, F. E., Daws, R. E., Gallen, C. L., Gazzaley, A., & Carhart-Harris, R. L. (2023). A complex systems perspective on psychedelic brain action. *Trends in cognitive sciences*, *27*(5), 433–445. https://doi.org/10.1016/j.tics.2023.01.003
Girn, M., Roseman, L., Bernhardt, B., Smallwood, J., Carhart-Harris, R., & Nathan Spreng, R. (2022). Serotonergic psychedelic drugs LSD and psilocybin reduce the hierarchical differentiation of unimodal and transmodal cortex. *NeuroImage*, *256*, 119220. https://doi.org/10.1016/j.neuroimage.2022.119220
Gloeckler, S. G., Thibault Lévesque, J., Lehmann, A., Farzin, H., & Greenway, K. T. (2024). Music and non-music approaches in psilocybin-assisted psychotherapy: The sound of silence. *Journal of Psychedelic Studies* (published online ahead of print 2024). https://doi.org/10.1556/2054.2024.00421
Gonzalez, D., Cantillo, J., Perez, I., Carvalho, M., Aronovich, A., Farre, M., Feilding, A., Obiols, J. E., & Bouso, J. C. (2021). The Shipibo Ceremonial Use of Ayahuasca to Promote Well-Being: An Observational Study. *Frontiers in pharmacology*, *12*, 623923. https://doi.org/10.3389/fphar.2021.623923
Good, M., Joel, Z., Benway, T., Routledge, C., Timmermann, C., Erritzoe, D., Weaver, R., Allen, G., Hughes, C., Topping, H., Bowman, A., & James, E. (2023). Pharmacokinetics of N,N-dimethyltryptamine in Humans. *European Journal of Drug Metabolism and Pharmacokinetics*, *48*(3), 311–327. https://doi.org/10.1007/s13318-023-00822-y
Goodwin, G. M., Aaronson, S. T., Alvarez, O., Arden, P. C., Baker, A., Bennett, J. C., Bird, C., Blom, R. E., Brennan, C., Brusch, D., Burke, L., Campbell-Coker, K., Carhart-Harris, R., Cattell, J., Daniel, A., DeBattista, C., Dunlop, B. W., Eisen, K., Feifel, D., … Malievskaia, E. (2022). Single-Dose Psilocybin for a Treatment-Resistant Episode of Major Depression. *The New England Journal of Medicine*, *387*(18), 1637–1648. https://doi.org/10.1056/NEJMoa2206443
Gordon, A. R., Carrithers, B. M., Pagni, B. A., Kettner, H., Marrocu, A., Nayak, S., Weiss, B. J., Carhart-Harris, R. L., Roberts, D. E., Erritzoe, D., & Zeifman, R. J. (2024). *The Effect of Psychedelics on Individuals with a Personality Disorder: Results from two Prospective Cohort Studies*. Research Square. https://doi.org/10.21203/rs.3.rs-4203641/v2
Grabski, M., McAndrew, A., Lawn, W., Marsh, B., Raymen, L., Stevens, T., Hardy, L., Warren, F., Bloomfield, M., Borissova, A., Maschauer, E., Broomby, R., Price, R., Coathup, R., Gilhooly, D., Palmer, E., Gordon-Williams, R., Hill, R., Harris, J., … Morgan, C. J. A. (2022). Adjunctive Ketamine With Relapse Prevention-Based Psychological Therapy in the Treatment of Alcohol Use Disorder. *The American Journal of Psychiatry*, *179*(2), 152–162. https://doi.org/10.1176/appi.ajp.2021.21030277
Graziosi, M., Singh, M., Nayak, S. M., & Yaden, D. B. (2023). Acute Subjective Effects of Psychedelics within and Beyond WEIRD Contexts. *Journal of Psychoactive Drugs*, *55*(5), 558–569. https://doi.org/10.1080/02791072.2023.2255274
Gresch, P. J., Smith, R. L., Barrett, R. J., & Sanders-Bush, E. (2005). Behavioral Tolerance to Lysergic Acid

Diethylamide is Associated with Reduced Serotonin-2A Receptor Signaling in Rat Cortex. *Neuropsychopharmacology*, *30*(9), 1693–1702. https://doi.org/10.1038/sj.npp.1300711

Griffiths, R. R., Johnson, M. W., Carducci, M. A., Umbricht, A., Richards, W. A., Richards, B. D., Cosimano, M. P., & Klinedinst, M. A. (2016). Psilocybin produces substantial and sustained decreases in depression and anxiety in patients with life-threatening cancer: A randomized double-blind trial. *Journal of Psychopharmacology*, *30*(12), 1181–1197. https://doi.org/10.1177/0269881116675513

Griffiths, R. R., Richards, W. A., McCann, U., & Jesse, R. (2006). Psilocybin can occasion mystical-type experiences having substantial and sustained personal meaning and spiritual significance. *Psychopharmacology*, *187*(3), 268–283. https://doi.org/10.1007/s00213-006-0457-5

Grof, S. (1980). *LSD Psychotherapy*. Hunter House. https://books.google.co.uk/books?id=DYU3AQAAIAAJ

Haijen, E. C., Kaelen, M., Roseman, L., Timmermann, C., Kettner, H., Russ, S., ... & Carhart-Harris, R. L. (2018). Predicting responses to psychedelics: a prospective study. Frontiers in pharmacology, 9, 897.

Hall, W. (2022). Why was early therapeutic research on psychedelic drugs abandoned?. Psychological Medicine, 52(1), 26-31.

Halman, A., Kong, G., Sarris, J., & Perkins, D. (2024). Drug-drug interactions involving classic psychedelics: A systematic review. *Journal of psychopharmacology (Oxford, England)*, *38*(1), 3–18. https://doi.org/10.1177/02698811231211219

Härter, P. M. (2021). The Influence of Psychedelic Drugs on the 'Sense of Self'. *Maastricht Student Journal of Psychology and Neuroscience*, *9*(1), 10-36.

Hartmann, W. E., Kim, E. S., Kim, J. H. J., Nguyen, T. U., Wendt, D. C., Nagata, D. K., & Gone, J. P. (2013). In Search of Cultural Diversity, Revisited: Recent Publication Trends in Cross-Cultural and Ethnic Minority Psychology. *Review of General Psychology, 17*(3), 243-254. https://doi.org/10.1037/a0032260

Hartogsohn, I. (2017). Constructing drug effects: A history of set and setting. *Drug Science, Policy and Law*, *3*, 2050324516683325. https://doi.org/10.1177/2050324516683325

Hartogsohn, I. (2020). *American Trip: Set, Setting, and the Psychedelic Experience in the Twentieth Century*. The MIT Press. https://doi.org/10.7551/mitpress/11888.001.0001

Hasler, F., Bourquin, D., Brenneisen, R., Bär, T., & Vollenweider, F. X. (1997). Determination of psilocin and 4-hydroxyindole-3-acetic acid in plasma by HPLC-ECD and pharmacokinetic profiles of oral and intravenous psilocybin in man. *Pharmaceutica acta Helvetiae*, *72*(3), 175–184. https://doi.org/10.1016/s0031-6865(97)00014-9

Hasler, F., Grimberg, U., Benz, M. A., Huber, T., & Vollenweider, F. X. (2004). Acute psychological and physiological effects of psilocybin in healthy humans: A double-blind, placebo-controlled dose-effect study. *Psychopharmacology*, *172*(2), 145–156. https://doi.org/10.1007/s00213-003-1640-6

Hayes, S. C., Law, S., Malady, M., Zhu, Z., & Bai, X. (2020). The centrality of sense of self in psychological flexibility processes: What the neurobiological and psychological correlates of psychedelics suggest. *Journal of Contextual Behavioral Science*, *15*, 30–38. https://doi.org/10.1016/j.jcbs.2019.11.005

Heal, D. J., Smith, S. L., Belouin, S. J., & Henningfield, J. E. (2023). Psychedelics: Threshold of a Therapeutic Revolution. *Neuropharmacology*, *236*, 109610. https://doi.org/10.1016/j.neuropharm.2023.109610

Hendricks, P. S., Johnson, M. W., & Griffiths, R. R. (2015). Psilocybin, psychological distress, and suicidality. *Journal of Psychopharmacology (Oxford, England)*, *29*(9), 1041–1043. https://doi.org/10.1177/0269881115598338

Hendricks, P. S., Thorne, C. B., Clark, C. B., Coombs, D. W., & Johnson, M. W. (2015). Classic psychedelic use is associated with reduced psychological distress and suicidality in the United States adult population. *Journal of Psychopharmacology (Oxford, England)*, *29*(3), 280–288. https://doi.org/10.1177/0269881114565653

Herrera, T. (1992). De los que saben de hongos. *Ciencias*, *28*, 37–40.

Herrmann, Z., Earleywine, M., De Leo, J., Slabaugh, S., Kenny, T., & Rush, A. J. (2023). Scoping Review of Experiential Measures from Psychedelic Research and Clinical Trials. *Journal of Psychoactive Drugs*, *55*(4), 501–517. https://doi.org/10.1080/02791072.2022.2125467

Herzog, R., Mediano, P. A., Rosas, F. E., Lodder, P., Carhart-Harris, R., Perl, Y. S., ... & Cofre, R. (2023). A whole-brain model of the neural entropy increase elicited by psychedelic drugs. Scientific reports, 13(1), 6244.

Hess, P. (2017, December 4). *Here's Scientist Bill Richards's Playlist for Tripping on Mushrooms*. Inverse. https://www.inverse.com/article/38980-psilocybin-mushroom-playlist-research

Hinkle, J. T., Graziosi, M., Nayak, S. M., & Yaden, D. B. (2024). Adverse Events in Studies of Classic Psychedelics: A Systematic Review and Meta-Analysis. *JAMA Psychiatry*, *81*(12), 1225–1235. https://doi.org/10.1001/jamapsychiatry.2024.2546

Hintzen, A. (2006). *Die (Psycho-) Pharmakologie von Lysergsäurediäthylamid (LSD-25): Eine Literaturübersicht*

*(1939-2005) unter besonderer Berücksichtigung der psychiatrischen Forschung* [PhD Thesis].
Hirschfeld, T., Prugger, J., Majić, T., & Schmidt, T. T. (2023). Dose-response relationships of LSD-induced subjective experiences in humans. *Neuropsychopharmacology: Official Publication of the American College of Neuropsychopharmacology*, *48*(11), 1602–1611. https://doi.org/10.1038/s41386-023-01588-2
Hirschfeld, T., & Schmidt, T. T. (2021). Dose–response relationships of psilocybin-induced subjective experiences in humans. *Journal of Psychopharmacology, 35*(4), 384–397. https://doi.org/10.1177/0269881121992676
Hoch, P. (1956). Studies in Routes of Administration and Counteracting Drugs. In L. Cholden (Ed.), *Lysergic Acid Diethylamide and Mescaline in Experimental Psychiatry* (pp. 8–12). Grune & Stratton.
Hoffer, A. (1965). D-Lysergic acid diethylamide (LSD): A review of its present status. *Clinical Pharmacology and Therapeutics*, *6*, 183–255. https://doi.org/10.1002/cpt196562183
Hollister, L. E., & Moore, F. F. (1965). Increased plasma free fatty acids following psychotomimetic drugs. *Journal of Psychiatric Research*, *3*(3), 199–203. https://doi.org/10.1016/0022-3956(65)90029-4
Holmes, S. E., Finnema, S. J., Naganawa, M., DellaGioia, N., Holden, D., Fowles, K., Davis, M., Ropchan, J., Emory, P., Ye, Y., Nabulsi, N., Matuskey, D., Angarita, G. A., Pietrzak, R. H., Duman, R. S., Sanacora, G., Krystal, J. H., Carson, R. E., & Esterlis, I. (2022). Imaging the effect of ketamine on synaptic density (SV2A) in the living brain. *Molecular psychiatry*, *27*(4), 2273–2281. https://doi.org/10.1038/s41380-022-01465-2
Holze, F., Duthaler, U., Vizeli, P., Müller, F., Borgwardt, S., & Liechti, M. E. (2019). Pharmacokinetics and subjective effects of a novel oral LSD formulation in healthy subjects. *British Journal of Clinical Pharmacology*, *85*(7), 1474–1483. https://doi.org/10.1111/bcp.13918
Holze, F., Ley, L., Müller, F., Becker, A. M., Straumann, I., Vizeli, P., Kuehne, S. S., Roder, M. A., Duthaler, U., Kolaczynska, K. E., Varghese, N., Eckert, A., & Liechti, M. E. (2022). Direct comparison of the acute effects of lysergic acid diethylamide and psilocybin in a double-blind placebo-controlled study in healthy subjects. *Neuropsychopharmacology*, *47*(6), 1180–1187. https://doi.org/10.1038/s41386-022-01297-2
Holze, F., Vizeli, P., Ley, L., Müller, F., Dolder, P., Stocker, M., Duthaler, U., Varghese, N., Eckert, A., Borgwardt, S., & Liechti, M. E. (2021). Acute dose-dependent effects of lysergic acid diethylamide in a double-blind placebo-controlled study in healthy subjects. *Neuropsychopharmacology*, *46*(3), 537–544. https://doi.org/10.1038/s41386-020-00883-6
Home Office. (2022) *List of the most commonly encountered drugs currently controlled under the misuse of drugs legislation.* Last accessed 11th Dec 2024: https://www.gov.uk/government/publications/controlled-drugs-list--2/list-of-most-commonly-encountered-drugs-currently-controlled-under-the-misuse-of-drugs-legislation
Honk, L., Stenfors, C. U. D., Goldberg, S. B., Hendricks, P. S., Osika, W., Dourron, H. M., Lebedev, A., Petrovic, P., & Simonsson, O. (2024). Longitudinal associations between psychedelic use and psychotic symptoms in the United States and the United Kingdom. *Journal of Affective Disorders*, *351*, 194–201. https://doi.org/10.1016/j.jad.2024.01.197
Houot, A. (2021). Phenomenology for Psychedelic Researchers: A Review of Current Methods & Practices. *Journal of Consciousness Exploration & Research*, *12*(3), Article 3. https://jcer.com/index.php/jcj/article/view/977
Hovmand, O. R., Poulsen, E. D., & Arnfred, S. (2024). Assessment of the acute subjective psychedelic experience: A review of patient-reported outcome measures in clinical research on classical psychedelics. *Journal of Psychopharmacology (Oxford, England)*, *38*(1), 19–32. https://doi.org/10.1177/02698811231200019
Hsu, T.-W., Tsai, C.-K., Kao, Y.-C., Thompson, T., Carvalho, A. F., Yang, F.-C., Tseng, P.-T., Hsu, C.-W., Yu, C.-L., Tu, Y.-K., & Liang, C.-S. (2024). Comparative oral monotherapy of psilocybin, lysergic acid diethylamide, 3,4-methylenedioxymethamphetamine, ayahuasca, and escitalopram for depressive symptoms: Systematic review and Bayesian network meta-analysis. *BMJ (Clinical Research Ed.)*, *386*, e078607. https://doi.org/10.1136/bmj-2023-078607
Hu, J. C. (2025, March 24). The power of "know-cebo": 5 Questions for data scientist Balázs Szigeti [Substack newsletter]. *The Microdose*. https://themicrodose.substack.com/p/the-power-of-know-cebo-5-questions
Hughes, M. E., & Garcia-Romeu, A. (2024). Ethnoracial inclusion in clinical trials of psychedelics: a systematic review. *EClinicalMedicine*, *74*, 102711. https://doi.org/10.1016/j.eclinm.2024.102711
Husain, M. (2024). Why we need a revolution in clinical research. *Brain*, *147*(9), 2897–2898. https://doi.org/10.1093/brain/awae265
Husain, M. (2025). On the responsibilities of intellectuals and the rise of bullshit jobs in universities. *Brain*, *148*(3), 687–688. https://doi.org/10.1093/brain/awaf045
Hutchinson, P., & Moerman, D. E. (2018). The Meaning Response, "Placebo," and Methods. *Perspectives in Biology and Medicine*, *61*(3), 361–378. https://doi.org/10.1353/pbm.2018.0049

Hutten, N. R. P. W., Mason, N. L., Dolder, P. C., Theunissen, E. L., Holze, F., Liechti, M. E., Feilding, A., Ramaekers, J. G., & Kuypers, K. P. C. (2020). Mood and cognition after administration of low LSD doses in healthy volunteers: A placebo controlled dose-effect finding study. *European Neuropsychopharmacology*, *41*, 81–91. https://doi.org/10.1016/j.euroneuro.2020.10.002

Hutten, N. R. P. W., Quaedflieg, C. W. E. M., Mason, N. L., Theunissen, E. L., Liechti, M. E., Duthaler, U., Kuypers, K. P. C., Bonnelle, V., Feilding, A., & Ramaekers, J. G. (2024). Inter-individual variability in neural response to low doses of LSD. *Translational Psychiatry*, *14*(1), 1–10. https://doi.org/10.1038/s41398-024-03013-8

Isbell, H., Wolbach, A. B., Wikler, A., & Miner, E. J. (1961). Cross tolerance between LSD and psilocybin. *Psychopharmacologia*, *2*(3), 147–159. https://doi.org/10.1007/BF00407974

Jachs, B., Garcia, M. C., Canales-Johnson, A., & Bekinschtein, T. A. (2022). *Drawing the experience dynamics of meditation* (p. 2022.03.04.482237). bioRxiv. https://doi.org/10.1101/2022.03.04.482237

Jarvik, M. E., Abramson, H. A., & Hirsch, M. W. (1955). Lysergic Acid Diethylamide (LSD-25): VI. Effect upon Recall and Recognition of Various Stimuli. *The Journal of Psychology*, *39*(2), 443–454. https://doi.org/10.1080/00223980.1955.9916194

Johansen, A., Armand, S., Plavén-Sigray, P., Nasser, A., Ozenne, B., Petersen, I.N., Keller, S.H., Madsen, J., Beliveau, V., Møller, K., Vassilieva, A., Langley, C., Svarer, C., Stenbæk, D.S., Sahakian, B.J. & Knudsen, G.M. (2023) Effects of escitalopram on synaptic density in the healthy human brain: a randomized controlled trial. *Molecular Psychiatry.* 1–8. https://doi.org/10.1038/s41380-023-02285-8.

Johansen, P.-Ø., & Krebs, T. S. (2015). Psychedelics not linked to mental health problems or suicidal behavior: A population study. *Journal of Psychopharmacology (Oxford, England)*, *29*(3), 270–279. https://doi.org/10.1177/0269881114568039

Johnson, M. W., Garcia-Romeu, A., Cosimano, M. P., & Griffiths, R. R. (2014). Pilot Study of the 5-HT2AR Agonist Psilocybin in the Treatment of Tobacco Addiction. *Journal of Psychopharmacology (Oxford, England)*, *28*(11), 983–992. https://doi.org/10.1177/0269881114548296

Johnson, M. W., Griffiths, R. R., Hendricks, P. S., & Henningfield, J. E. (2018). The abuse potential of medical psilocybin according to the 8 factors of the Controlled Substances Act. *Neuropharmacology*, *142*, 143–166. https://doi.org/10.1016/j.neuropharm.2018.05.012

Johnson, M. W., Richards, W. A., & Griffiths, R. R. (2008). Human hallucinogen research: Guidelines for safety. *Journal of Psychopharmacology*, *22*(6), 603–620. https://doi.org/10.1177/0269881108093587

Kaelen, M., Barrett, F. S., Roseman, L., Lorenz, R., Family, N., Bolstridge, M., Curran, H. V., Feilding, A., Nutt, D. J., & Carhart-Harris, R. L. (2015). LSD enhances the emotional response to music. *Psychopharmacology*, *232*(19), 3607–3614. https://doi.org/10.1007/s00213-015-4014-y

Kaelen, M., Giribaldi, B., Raine, J., Evans, L., Timmermann, C., Rodriguez, N., Roseman, L., Feilding, A., Nutt, D., & Carhart-Harris, R. (2018). The hidden therapist: evidence for a central role of music in psychedelic therapy. *Psychopharmacology*, *235*(2), 505–519. https://doi.org/10.1007/s00213-017-4820-5

Kangaslampi, S., Hausen, A., & Rauteenmaa, T. (2020). Mystical experiences in retrospective reports of first times using a psychedelic in Finland. *Journal of Psychoactive Drugs*, *52*(4), 309–318. https://doi.org/10.1080/02791072.2020.1767321

Kaplan, J., Mandel, L. R., Stillman, R., Walker, R. W., VandenHeuvel, W. J. A., Gillin, J. C., & Wyatt, R. J. (1974). Blood and urine levels of N,N-dimethyltryptamine following administration of psychoactive dosages to human subjects. *Psychopharmacologia*, *38*(3), 239–245. https://doi.org/10.1007/BF00421376

Katz, M. M., Waskow, I. E., & Olsson, J. (1968). Characterizing the psychological state produced by LSD. *Journal of Abnormal Psychology*, *73*(1), 1–14. https://doi.org/10.1037/h0020114

Kettner, H., Glowacki, D., Wall, J., Carhart-Harris, R., Roseman, L., & Hardy, J. (2025). Observational cohort study of a group-based VR program to improve mental health and well-being in people with life-threatening illnesses. *Frontiers in Virtual Reality*, *5*, 1466362.

Khan, M., Carter, G. T., Aggarwal, S. K., & Holland, J. (2021). Psychedelics for Brain Injury: A Mini-Review. *Frontiers in Neurology*, *12*. https://doi.org/10.3389/fneur.2021.685085

Klaiber, A., Schmid, Y., Becker, A. M., Straumann, I., Erne, L., Jelusic, A., Thomann, J., Luethi, D., & Liechti, M. E. (2024). Acute dose-dependent effects of mescaline in a double-blind placebo-controlled study in healthy subjects. *Translational Psychiatry*, *14*(1), 1–8. https://doi.org/10.1038/s41398-024-03116-2

Klock, J. C., Boerner, U., & Becker, C. E. (1974). Coma, Hyperthermia and Bleeding Associated with Massive LSD Overdose. *Western Journal of Medicine*, *120*(3), 183–188.

Ko, K., Knight, G., Rucker, J. J., & Cleare, A. J. (2022). Psychedelics, Mystical Experience, and Therapeutic Efficacy: A Systematic Review. *Frontiers in Psychiatry*, *13*. https://doi.org/10.3389/fpsyt.2022.917199

Ko, K., Kopra, E. I., Cleare, A. J., & Rucker, J. J. (2023). Psychedelic therapy for depressive symptoms: A systematic review and meta-analysis. *Journal of Affective Disorders*, *322*, 194–204. https://doi.org/10.1016/j.jad.2022.09.168

Konstantinidou, H., Chartonas, D., Rogalski, D., & Lee, T. (2023). Will this tablet make me happy again? The contribution of relational prescribing in providing a pragmatic and psychodynamic framework for prescribers. *BJPsych Advances*, *29*(4), 265–273. https://doi.org/10.1192/bja.2022.43

Kopra, E. I., Ferris, J. A., Rucker, J. J., McClure, B., Young, A. H., Copeland, C. S., & Winstock, A. R. (2022). Adverse experiences resulting in emergency medical treatment seeking following the use of lysergic acid diethylamide (LSD). *Journal of Psychopharmacology*, *36*(8), 956–964.

Kopra, E. I., Ferris, J. A., Winstock, A. R., Young, A. H., & Rucker, J. J. (2022). Adverse experiences resulting in emergency medical treatment seeking following the use of magic mushrooms. *Journal of Psychopharmacology*, *36*(8), 965–973.

Kopra, E. I., Penttinen, J., Rucker, J. J., & Copeland, C. S. (2025). Psychedelic-related deaths in England, Wales and Northern Ireland (1997–2022). *Progress in Neuro-Psychopharmacology and Biological Psychiatry*, *136*, 111177. https://doi.org/10.1016/j.pnpbp.2024.111177

Kramer, E. N., Reddy, K., & Shapiro, B. (2023). A suicide attempt following psilocybin ingestion in a patient with no prior psychiatric history. *Psychiatry Research Case Reports*, *2*(1), 100118.

Krebs, T. S., & Johansen, P.-Ø. (2012). Lysergic acid diethylamide (LSD) for alcoholism: Meta-analysis of randomized controlled trials. *Journal of Psychopharmacology (Oxford, England)*, *26*(7), 994–1002. https://doi.org/10.1177/0269881112439253

Krebs, T. S., & Johansen, P.-Ø. (2013a). Over 30 million psychedelic users in the United States. *F1000Research*, *2*, 98. https://doi.org/10.12688/f1000research.2-98.v1

Krebs, T. S., & Johansen, P.-Ø. (2013b). Psychedelics and Mental Health: A Population Study. *PLOS ONE*, *8*(8), e63972. https://doi.org/10.1371/journal.pone.0063972

Krupnick, J. L., Sotsky, S. M., Simmens, S., Moyer, J., Elkin, I., Watkins, J., & Pilkonis, P. A. (1996). The role of the therapeutic alliance in psychotherapy and pharmacotherapy outcome: Findings in the National Institute of Mental Health Treatment of Depression Collaborative Research Program. *Journal of Consulting and Clinical Psychology*, *64*(3), 532–539. https://doi.org/10.1037/0022-006X.64.3.532

Le Nedelec, M., Glue, P., Winter, H., Goulton, C., Broughton, L., & Medlicott, N. (2018). Acute low-dose ketamine produces a rapid and robust increase in plasma BDNF without altering brain BDNF concentrations. *Drug delivery and translational research*, *8*(3), 780–786. https://doi.org/10.1007/s13346-017-0476-2

Lea, T., Amada, N., Jungaberle, H., Schecke, H., & Klein, M. (2020). Microdosing psychedelics: Motivations, subjective effects and harm reduction. *International Journal of Drug Policy*, *75*, 102600. https://doi.org/10.1016/j.drugpo.2019.11.008

Leary, T., Litwin, G. H., & Metzner, R. (1963). Reactions to psilocybin administered in a supportive environment. *The Journal of Nervous and Mental Disease*, *137*(6), 561.

Lee, M. A., & Shlain, B. (1985). *Acid dreams: The CIA, LSD, and the sixties rebellion* (Vol. 1001). Grove Press.

Leuner, H. (1967). Present state of psycholytic therapy and its possibilities. In H. Abramson (Ed.), *The use of LSD in psychotherapy and alcoholism* (pp. 101–116).

Levin, A. W., Lancelotta, R., Sepeda, N. D., Gukasyan, N., Nayak, S., Wagener, T. L., Barrett, F. S., Griffiths, R. R., & Davis, A. K. (2024). The therapeutic alliance between study participants and intervention facilitators is associated with acute effects and clinical outcomes in a psilocybin-assisted therapy trial for major depressive disorder. *PloS One*, *19*(3), e0300501. https://doi.org/10.1371/journal.pone.0300501

Lewis, V., Bonniwell, E. M., Lanham, J. K., Ghaffari, A., Sheshbaradaran, H., Cao, A. B., Calkins, M. M., Bautista-Carro, M. A., Arsenault, E., Telfer, A., Taghavi-Abkuh, F. F., Malcolm, N. J., El Sayegh, F., Abizaid, A., Schmid, Y., Morton, K., Halberstadt, A. L., Aguilar-Valles, A., & McCorvy, J. D. (2023). A non-hallucinogenic LSD analog with therapeutic potential for mood disorders. *Cell reports*, *42*(3), 112203. https://doi.org/10.1016/j.celrep.2023.112203

Lewis-Healey, E., Pallavicini, C., Cavanna, F., D'Amelio, T., Fuente, L. A. D. L., Copa, D., Müller, S., Bruno, N., Tagliazucchi, E., & Bekinschtein, T. (2024). *Time-resolved neural and experience dynamics of medium- and high-dose DMT* (p. 2024.12.19.629418). bioRxiv. https://doi.org/10.1101/2024.12.19.629418

Liardi, A., Rosas, F. E., Carhart-Harris, R. L., Blackburne, G., Bor, D., & Mediano, P. A. (2024). Null models for comparing information decomposition across complex systems. arXiv preprint arXiv:2410.11583.

Liechti, M. E. (2017). Modern Clinical Research on LSD. *Neuropsychopharmacology: Official Publication of the American College of Neuropsychopharmacology*, *42*(11), 2114–2127. https://doi.org/10.1038/npp.2017.86

Lienert, G. A. (1959). Changes in the factor structure of intelligence tests produced by d-lysergic acid diethylamide (LSD). In P. B. Bradley (Ed.), *Neuro-Psychopharmacology* (pp. 461–465). Elsevier.

Lii, T. R., Smith, A. E., Flohr, J. R., Okada, R. L., Nyongesa, C. A., Cianfichi, L. J., Hack, L. M., Schatzberg, A. F., & Heifets, B. D. (2023). Randomized Trial of Ketamine Masked by Surgical Anesthesia in Depressed Patients. *medRxiv: The Preprint Server for Health Sciences*, 2023.04.28.23289210. https://doi.org/10.1101/2023.04.28.23289210

Lindenblatt, H., Krämer, E., Holzmann-Erens, P., Gouzoulis-Mayfrank, E., & Kovar, K. A. (1998). Quantitation of psilocin in human plasma by high-performance liquid chromatography and electrochemical detection: comparison of liquid-liquid extraction with automated on-line solid-phase extraction. *Journal of chromatography. B, Biomedical sciences and applications*, *709*(2), 255–263. https://doi.org/10.1016/s0378-4347(98)00067-x

Lindquist M. A. (2024). Sliding windows analysis can undo the effects of preprocessing when applied to fMRI data. *bioRxiv : the preprint server for biology*, 2023.10.06.561221. https://doi.org/10.1101/2023.10.06.561221

Luan, L. X., Eckernäs, E., Ashton, M., Rosas, F. E., Uthaug, M. V., Bartha, A., Jagger, S., Gascon-Perai, K., Gomes, L., Nutt, D. J., Erritzøe, D., Carhart-Harris, R. L., & Timmermann, C. (2024). Psychological and physiological effects of extended DMT. *Journal of Psychopharmacology*, *38*(1), 56–67. https://doi.org/10.1177/02698811231196877

Luppi, A. I., Carhart-Harris, R. L., Roseman, L., Pappas, I., Menon, D. K., & Stamatakis, E. A. (2021). LSD alters dynamic integration and segregation in the human brain. *NeuroImage*, *227*, 117653. https://doi.org/10.1016/j.neuroimage.2020.117653

Luppi, A. I., Mediano, P. A. M., Rosas, F. E., Allanson, J., Pickard, J., Carhart-Harris, R. L., Williams, G. B., Craig, M. M., Finoia, P., Owen, A. M., Naci, L., Menon, D. K., Bor, D., & Stamatakis, E. A. (2024). A synergistic workspace for human consciousness revealed by Integrated Information Decomposition. *eLife*, *12*, RP88173. https://doi.org/10.7554/eLife.88173

Luppi, A. I., Mediano, P. A. M., Rosas, F. E., Harrison, D. J., Carhart-Harris, R. L., Bor, D., & Stamatakis, E. A. (2021). What it is like to be a bit: an integrated information decomposition account of emergent mental phenomena. *Neuroscience of consciousness*, *2021*(2), niab027. https://doi.org/10.1093/nc/niab027

Lynn, S. J., McDonald, C. W., Sleight, F. G., & Mattson, R. E. (2023). Cross-validation of the ego dissolution scale: Implications for studying psychedelics. *Frontiers in Neuroscience*, *17*. https://doi.org/10.3389/fnins.2023.1267611

Lyons, T., Spriggs, M., Kerkelä, L., Rosas, F. E., Roseman, L., Mediano, P. a. M., Timmermann, C., Oestreich, L., Pagni, B. A., Zeifman, R. J., Hampshire, A., Trender, W., Douglass, H. M., Girn, M., Godfrey, K., Kettner, H., Sharif, F., Espasiano, L., Gazzaley, A., … Carhart-Harris, R. L. (2024). H*uman brain changes after first psilocybin use* (p. 2024.10.11.617955). bioRxiv. https://doi.org/10.1101/2024.10.11.617955

MacCallum, C. A., Lo, L. A., Pistawka, C. A., & Deol, J. K. (2022). Therapeutic use of psilocybin: Practical considerations for dosing and administration. *Frontiers in Psychiatry*, *13*, 1040217. https://doi.org/10.3389/fpsyt.2022.1040217

Marini J. L. (1981). Serotonergic and dopaminergic effects of yawning in the cat. *Pharmacology, biochemistry, and behavior*, *15*(5), 711–715. https://doi.org/10.1016/0091-3057(81)90010-1

Marks, M., & Cohen, I. G. (2021). Psychedelic therapy: A roadmap for wider acceptance and utilization. *Nature Medicine*, *27*(10), 1669–1671. https://doi.org/10.1038/s41591-021-01530-3

Malitz, S., Esecover, H., Wilkens, B., & Hoch, P. H. (1960). Some observations on psilocybin, a new hallucinogen, in volunteer subjects. *Comprehensive Psychiatry*, *1*, 8–17. https://doi.org/10.1016/s0010-440x(60)80045-4

Marrocu, A., Kettner, H., Weiss, B., Zeifman, R. J., Erritzoe, D., & Carhart-Harris, R. L. (2024). Psychiatric risks for worsened mental health after psychedelic use. *Journal of Psychopharmacology (Oxford, England)*, *38*(3), 225–235. https://doi.org/10.1177/02698811241232548

Martinotti, G., Santacroce, R., Pettorruso, M., Montemitro, C., Spano, M. C., Lorusso, M., di Giannantonio, M., & Lerner, A. G. (2018). Hallucinogen Persisting Perception Disorder: Etiology, Clinical Features, and Therapeutic Perspectives. *Brain Sciences*, *8*(3), 47. https://doi.org/10.3390/brainsci8030047

McAlpine, R. G., Blackburne, G., & Kamboj, S. K. (2024). Development and psychometric validation of a novel scale for measuring 'psychedelic preparedness'. *Scientific reports*, *14*(1), 3280. https://doi.org/10.1038/s41598-024-53829-z

McClelland, H., Cleare, S., & O'Connor, R. C. (2023). Suicide Risk in Personality Disorders: A Systematic Review. *Current Psychiatry Reports*, *25*(9), 405–417. https://doi.org/10.1007/s11920-023-01440-w

McGuire, A. L., Cohen, I. G., Sisti, D., Baggott, M., Celidwen, Y., Devenot, N., Gracias, S., Grob, C., Harvey, I., Kious,

B., Marks, M., Mithoefer, M., Nielson, E., Öngür, D., Pallas, A., Peterson, A., Schenberg, E. E., Summergrad, P., Waters, B., … Yaden, D. B. (2024). Developing an Ethics and Policy Framework for Psychedelic Clinical Care: A Consensus Statement. *JAMA Network Open*, *7*(6), e2414650. https://doi.org/10.1001/jamanetworkopen.2024.14650

McKay, K. M., Imel, Z. E., & Wampold, B. E. (2006). Psychiatrist effects in the psychopharmacological treatment of depression. *Journal of Affective Disorders*, *92*(2–3), 287–290. https://doi.org/10.1016/j.jad.2006.01.020

McLane, H., Hutchinson, C., Wikler, D., Howell, T., & Knighton, E. (2021, December 22). *Respecting autonomy in altered states: Navigating ethical quandaries in psychedelic therapy*. Journal of Medical Ethics Blog. https://blogs.bmj.com/medical-ethics/2021/12/22/respecting-autonomy-in-altered-states-navigating-ethical-quandaries-in-psychedelic-therapy/

McNamee, S., Devenot, N., & Buisson, M. (2023). Studying Harms Is Key to Improving Psychedelic-Assisted Therapy—Participants Call for Changes to Research Landscape. *JAMA Psychiatry*, *80*(5), 411–412. https://doi.org/10.1001/jamapsychiatry.2023.0099

McWilliams, N. (2004). *Psychoanalytic Psychotherapy: A Practitioner's Guide*. The Guilford Press.

Mediano, P. A. M., Rosas, F. E., Luppi, A. I., Carhart-Harris, R. L., Bor, D., Seth, A. K., & Barrett, A. B. (2021). *Towards an extended taxonomy of information dynamics via Integrated Information Decomposition* (arXiv:2109.13186). arXiv. https://doi.org/10.48550/arXiv.2109.13186

Mediano, P. A. M., Rosas, F. E., Timmermann, C., Roseman, L., Nutt, D. J., Feilding, A., Kaelen, M., Kringelbach, M. L., Barrett, A. B., Seth, A. K., Muthukumaraswamy, S., Bor, D., & Carhart-Harris, R. L. (2024). Effects of External Stimulation on Psychedelic State Neurodynamics. *ACS chemical neuroscience*, *15*(3), 462–471. https://doi.org/10.1021/acschemneuro.3c00289

Meltzer, H., Simonovic, M., Fang, V., & Goode, D. (1981). Neuroendocrine effects of psychotomimetic drugs. *McLean Hospital Journal*, *6*(2), 115–137.

Mertens, L. J., Wall, M. B., Roseman, L., Demetriou, L., Nutt, D. J., & Carhart-Harris, R. L. (2020). Therapeutic mechanisms of psilocybin: Changes in amygdala and prefrontal functional connectivity during emotional processing after psilocybin for treatment-resistant depression. *Journal of psychopharmacology (Oxford, England)*, *34*(2), 167–180. https://doi.org/10.1177/0269881119895520

Meshkat, S., Malik, T., Zeifman, R., Swainson, J., Zhang, Y., Burback, L., ... & Bhat, V. (2025). Psychedelics and Suicide-Related Outcomes: A Systematic Review. Journal of Clinical Medicine, 14(5), 1416.

Michael, P., Luke, D., & Robinson, O. (2021). An Encounter With the Other: A Thematic and Content Analysis of DMT Experiences From a Naturalistic Field Study. *Frontiers in Psychology*, *12*. https://doi.org/10.3389/fpsyg.2021.720717

Michaels, T. I., Purdon, J., Collins, A., & Williams, M. T. (2018). Inclusion of people of color in psychedelic-assisted psychotherapy: A review of the literature. *BMC Psychiatry*, *18*(1), 245. https://doi.org/10.1186/s12888-018-1824-6

Mintz, K. T., Gammer, B., Khan, A. J., Shaub, G., Levine, S., & Sisti, D. (2022). Physical Disability and Psychedelic Therapies: An Agenda for Inclusive Research and Practice. *Frontiers in psychiatry*, *13*, 914458. https://doi.org/10.3389/fpsyt.2022.914458

Mitchell, J. M., Ot'alora G, M., van der Kolk, B., Shannon, S., Bogenschutz, M., Gelfand, Y., Paleos, C., Nicholas, C. R., Quevedo, S., Balliett, B., Hamilton, S., Mithoefer, M., Kleiman, S., Parker-Guilbert, K., Tzarfaty, K., Harrison, C., de Boer, A., Doblin, R., Yazar-Klosinski, B., & MAPP2 Study Collaborator Group (2023). MDMA-assisted therapy for moderate to severe PTSD: a randomized, placebo-controlled phase 3 trial. *Nature medicine*, *29*(10), 2473–2480. https://doi.org/10.1038/s41591-023-02565-4

Mortaheb, S., Fort, L. D., Mason, N. L., Mallaroni, P., Ramaekers, J. G., & Demertzi, A. (2024). Dynamic Functional Hyperconnectivity After Psilocybin Intake Is Primarily Associated With Oceanic Boundlessness. Biological psychiatry. *Cognitive neuroscience and neuroimaging*, *9*(7), 681–692. https://doi.org/10.1016/j.bpsc.2024.04.001

Mosig, Y. D. (2006). Conceptions of the self in Western and Eastern psychology. *Journal of Theoretical and Philosophical Psychology*, *26*(1–2), 39–50. https://doi.org/10.1037/h0091266

Munn, H. (1973). *The Mushrooms of Language in Hallucinogens and Shamanism* (pp. 86–122). Oxford University Press.

Murphy, R., Kettner, H., Zeifman, R., Giribaldi, B., Kartner, L., Martell, J., Read, T., Murphy-Beiner, A., Baker-Jones, M., Nutt, D., Erritzoe, D., Watts, R., & Carhart-Harris, R. (2022). Therapeutic Alliance and Rapport Modulate Responses to Psilocybin Assisted Therapy for Depression. *Frontiers in pharmacology*, *12*, 788155. https://doi.org/10.3389/fphar.2021.788155

Murphy, R. J., Muthukumaraswamy, S., & de Wit, H. (2024). Microdosing Psychedelics: Current Evidence From Controlled Studies. *Biological Psychiatry: Cognitive Neuroscience and Neuroimaging*, *9*(5), 500–511. https://doi.org/10.1016/j.bpsc.2024.01.002

Murphy, R. J., Sumner, R., Evans, W., Ponton, R., Ram, S., Godfrey, K., Forsyth, A., Cavadino, A., Krishnamurthy Naga, V., Smith, T., Hoeh, N. R., Menkes, D. B., & Muthukumaraswamy, S. (2023). Acute Mood-Elevating Properties of Microdosed Lysergic Acid Diethylamide in Healthy Volunteers: A Home-Administered Randomized Controlled Trial. *Biological psychiatry*, *94*(6), 511–521. https://doi.org/10.1016/j.biopsych.2023.03.013

Murray, C.H., Frohlich, J., Haggarty, C.J., Tare, I., Lee, R. & de Wit, H. (2024) Neural complexity is increased after low doses of LSD, but not moderate to high doses of oral THC or methamphetamine. *Neuropsychopharmacology*. 49 (7), 1120–1128. https://doi.org/10.1038/s41386-024-01809-2.

Muthukumaraswamy, S.D. (2014) The use of magnetoencephalography in the study of psychopharmacology (pharmaco-MEG). *Journal of Psychopharmacology*. 28 (9), 815–829. https://doi.org/10.1177/0269881114536790.

Muthukumaraswamy, S., Baggott, M., Schenberg, E. E., Repantis, D., Wolff, M., Forsyth, A., & Noorani, T. (2025). *Psychedelic Assisted Therapy as a Complex Intervention: Implications for clinical trial design*. OSF. https://doi.org/10.31234/osf.io/caup9_v1

Muthukumaraswamy, S.D., Carhart-Harris, R.L., Moran, R.J., Brookes, M.J., Williams, T.M., Errtizoe, D., Sessa, B., Papadopoulos, A., Bolstridge, M., Singh, K.D., Feilding, A., Friston, K.J. & Nutt, D.J. (2013) Broadband cortical desynchronization underlies the human psychedelic state. *The Journal of Neuroscience: The Official Journal of the Society for Neuroscience.* 33 (38), 15171–15183. https://doi.org/10.1523/JNEUROSCI.2063-13.2013.

Myran, D. T., Pugliese, M., Xiao, J., Kaster, T. S., Husain, M. I., Anderson, K. K., Fabiano, N., Wong, S., Fiedorowicz, J. G., Webber, C., Tanuseputro, P., & Solmi, M. (2024). Emergency Department Visits Involving Hallucinogen Use and Risk of Schizophrenia Spectrum Disorder. *JAMA Psychiatry*. https://doi.org/10.1001/jamapsychiatry.2024.3532

Nardou, R., Sawyer, E., Song, Y.J., Wilkinson, M., Padovan-Hernandez, Y., de Deus, J.L., Wright, N., Lama, C., Faltin, S., Goff, L.A., Stein-O'Brien, G.L. & Dölen, G. (2023) Psychedelics reopen the social reward learning critical period. *Nature.* 618 (7966), 790–798. https://doi.org/10.1038/s41586-023-06204-3.

Nayak, S. M., Bradley, M. K., Kleykamp, B. A., Strain, E. C., Dworkin, R. H., & Johnson, M. W. (2023). Control Conditions in Randomized Trials of Psychedelics: An ACTTION Systematic Review. *The Journal of Clinical Psychiatry*, *84*(3), 22r14518. https://doi.org/10.4088/JCP.22r14518

Neumann, J., Dhein, S., Kirchhefer, U., Hofmann, B., & Gergs, U. (2024). Effects of hallucinogenic drugs on the human heart. *Frontiers in pharmacology*, *15*, 1334218. https://doi.org/10.3389/fphar.2024.1334218

Nicholas, C. R., Banks, M. I., Lennertz, R. C., Wenthur, C. J., Krause, B. M., Riedner, B. A., ... & Raison, C. L. (2024). Co-administration of midazolam and psilocybin: differential effects on subjective quality versus memory of the psychedelic experience. *Translational Psychiatry*, *14*(1), 372.

Nichols, D. E. (2016). Psychedelics. *Pharmacological Reviews*, *68*(2), 264–355. https://doi.org/10.1124/pr.115.011478

Nichols, D. E., & Grob, C. S. (2018). Is LSD toxic?. *Forensic science international*, *284*, 141–145. https://doi.org/10.1016/j.forsciint.2018.01.006

Nichols, D. E., & Walter, H. (2020). The History of Psychedelics in Psychiatry. *Pharmacopsychiatry*, *54*, 151–166. https://doi.org/10.1055/a-1310-3990

Nikolic, M., Mediano, P., Froese, T., Reydellet, D., & Palenicek, T. (2024). *Psilocybin alters brain activity related to sensory and cognitive processing in a time-dependent manner* (p. 2024.09.09.24313316). medRxiv. https://doi.org/10.1101/2024.09.09.24313316

Noorani, T. (2020, July 21). *The Pollan Effect: Psychedelic Research between World and Word*. Hot Spots, Fieldsights. https://culanth.org/fieldsights/the-pollan-effect-psychedelic-research-between-world-and-word

Noorani, T., Bedi, G., & Muthukumaraswamy, S. (2023). Dark loops: Contagion effects, consistency and chemosocial matrices in psychedelic-assisted therapy trials. *Psychological Medicine*, *53*(13), 5892–5901. https://doi.org/10.1017/S0033291723001289

Noorani, T., Garcia-Romeu, A., Swift, T. C., Griffiths, R. R., & Johnson, M. W. (2018). Psychedelic therapy for smoking cessation: Qualitative analysis of participant accounts. *Journal of psychopharmacology (Oxford, England)*, *32*(7), 756–769. https://doi.org/10.1177/0269881118780612

Norcross, J. C., & Lambert, M. J. (2018). Psychotherapy relationships that work III. *Psychotherapy*, *55*(4), 303–315. https://doi.org/10.1037/pst0000193

Nour, M. M., Evans, L., & Carhart-Harris, R. L. (2017). Psychedelics, Personality and Political Perspectives. *Journal of Psychoactive Drugs*, *49*(3), 182–191. https://doi.org/10.1080/02791072.2017.1312643

Nour, M. M., Evans, L., Nutt, D., & Carhart-Harris, R. L. (2016). Ego-Dissolution and Psychedelics: Validation of the Ego-Dissolution Inventory (EDI). *Frontiers in Human Neuroscience*, *10*. https://doi.org/10.3389/fnhum.2016.00269

Nutt, D., Spriggs, M., & Erritzoe, D. (2023). Psychedelics therapeutics: What we know, what we think, and what we need to research. *Neuropharmacology*, *223*, 109257. https://doi.org/10.1016/j.neuropharm.2022.109257

O'Callaghan, C., Hubik, D. J., Dwyer, J., Williams, M., & Ross, M. (2020). Experience of Music Used With Psychedelic Therapy: A Rapid Review and Implications. *Journal of Music Therapy*, *57*(3), 282–314. https://doi.org/10.1093/jmt/thaa006

Olson D. E. (2020). The Subjective Effects of Psychedelics May Not Be Necessary for Their Enduring Therapeutic Effects. *ACS pharmacology & translational science*, *4*(2), 563–567. https://doi.org/10.1021/acsptsci.0c00192

Oram, M. (2014). Efficacy and enlightenment: LSD psychotherapy and the drug amendments of 1962. Journal of the History of Medicine and Allied Sciences, 69(2), 221-250.

Orłowski, P., Hobot, J., Ruban, A., Szczypiński, J., & Bola, M. (2024). The relation between naturalistic use of psychedelics and perception of emotional stimuli: An event-related potential study comparing non-users and experienced users of classic psychedelics. *Journal of psychopharmacology (Oxford, England)*, *38*(1), 68–79. https://doi.org/10.1177/02698811231216322

Orłowski, P., Hobot, J., Ruban, A., Szczypiński, J., & Bola, M. (2024). Naturalistic use of psychedelics does not modulate processing of self-related stimuli (but it might modulate attentional mechanisms): An event-related potentials study. *Psychophysiology*, *61*(8), e14583. https://doi.org/10.1111/psyp.14583

Orsolini, L., Papanti, G. D., De Berardis, D., Guirguis, A., Corkery, J. M., & Schifano, F. (2017). The "Endless Trip" among the NPS Users: Psychopathology and Psychopharmacology in the Hallucinogen-Persisting Perception Disorder. A Systematic Review. *Frontiers in Psychiatry*, *8*, 240. https://doi.org/10.3389/fpsyt.2017.00240.

Ort, A., Smallridge, J.W., Sarasso, S., Casarotto, S., Rotz, R. von, Casanova, A., Seifritz, E., Preller, K.H., Tononi, G. & Vollenweider, F.X. (2023) TMS-EEG and resting-state EEG applied to altered states of consciousness: oscillations, complexity, and phenomenology. *iScience*. 26 (5). https://doi.org/10.1016/j.isci.2023.106589.

Ortiz Bernal, A. M., Raison, C. L., Lancelotta, R. L., & Davis, A. K. (2022). Reactivations after 5-methoxy-N,N-dimethyltryptamine use in naturalistic settings: An initial exploratory analysis of the phenomenon's predictors and its emotional valence. *Frontiers in psychiatry*, *13*, 1049643. https://doi.org/10.3389/fpsyt.2022.1049643

Ott, J. (1999). Pharmahuasca: Human Pharmacology of Oral DMT Plus Harmine. *Journal of Psychoactive Drugs*, *31*(2), 171–177. https://doi.org/10.1080/02791072.1999.10471741

Pahnke, W. N. (1963). *Drugs and mysticism: An analysis of the relationship between psychedelic drugs and the mystical consciousness: A thesis*. Harvard University.

Pahnke, W. N. (1966). Drugs and mysticism. *International Journal of Parapsychology*, *8*(2), 295–314.

Pahnke, W. N. (1969). Psychedelic drugs and mystical experience. *International Psychiatry Clinics*, *5*(4), 149–162.

Palitsky, R., Kaplan, D. M., Perna, J., Bosshardt, Z., Maples-Keller, J. L., Levin-Aspenson, H. F., Zarrabi, A. J., Peacock, C., Mletzko, T., Rothbaum, B. O., Raison, C. L., Grant, G. H., & Dunlop, B. W. (2024). A framework for assessment of adverse events occurring in psychedelic-assisted therapies. *Journal of Psychopharmacology*, *38*(8), 690–700. https://doi.org/10.1177/02698811241265756

Pallavicini, C., Cavanna, F., Zamberlan, F., de la Fuente, L. A., Ilksoy, Y., Perl, Y. S., Arias, M., Romero, C., Carhart-Harris, R., Timmermann, C., & Tagliazucchi, E. (2021). Neural and subjective effects of inhaled N,N-dimethyltryptamine in natural settings. *Journal of Psychopharmacology (Oxford, England)*, *35*(4), 406–420. https://doi.org/10.1177/0269881120981384

Pallavicini, C., Vilas, M.G., Villarreal, M., Zamberlan, F., Muthukumaraswamy, S., Nutt, D., Carhart-Harris, R. & Tagliazucchi, E. (2019) Spectral signatures of serotonergic psychedelics and glutamatergic dissociatives. *NeuroImage*. *200*, 281–291. https://doi.org/10.1016/j.neuroimage.2019.06.053.

Passie, T. (1997). *Psycholytic and psychedelic therapy research 1931-1995: A complete international bibliography.* Laurentius Publishers (Hannover).

Passie, T., Guss, J., & Krähenmann, R. (2022). Lower-dose psycholytic therapy – A neglected approach. *Frontiers in Psychiatry*, *13*. https://doi.org/10.3389/fpsyt.2022.1020505

Passie, T., Halpern, J. H., Stichtenoth, D. O., Emrich, H. M., & Hintzen, A. (2008). The pharmacology of lysergic acid diethylamide: A review. *CNS Neuroscience & Therapeutics*, *14*(4), 295–314. https://doi.org/10.1111/j.1755-5949.2008.00059.x

Paterson, N. E., Darby, W. C., & Sandhu, P. S. (2015). N,N-Dimethyltryptamine-Induced Psychosis. Clinical Neuropharmacology, 38(4), 141–143. https://doi.org/10.1097/WNF.0000000000000078

Penn, A. D., Phelps, J., Rosa, W. E., & Watson, J. (2021). Psychedelic-Assisted Psychotherapy Practices and Human Caring Science: Toward a Care-Informed Model of Treatment. Journal of *Humanistic Psychology*, *64*(4), 592–617. https://doi.org/10.1177/0022167821101101

Percival, R. S. (2015). Descartes' Model of Mind. In R. L. Cautin & S. O. Lilienfeld (Eds.), *The Encyclopedia of Clinical Psychology* (1st ed., pp. 1–8). Wiley. https://doi.org/10.1002/9781118625392.wbecp494

Petitmengin, C. (2006). Describing one's subjective experience in the second person: An interview method for the science of consciousness. *Phenomenology and the Cognitive Sciences*, *5*(3), 229–269. https://doi.org/10.1007/s11097-006-9022-2

Petitmengin, C., Remillieux, A., & Valenzuela-Moguillansky, C. (2019). Discovering the structures of lived experience. *Phenomenology and the Cognitive Sciences*, *18*(4), 691–730. https://doi.org/10.1007/s11097-018-9597-4

Petranker, R., Kim, J., & Anderson, T. (2022). Microdosing as a Response to the Meaning Crisis: A Qualitative Analysis. *Journal of Humanistic Psychology*, 00221678221075076. https://doi.org/10.1177/00221678221075076

Petri, G., Expert, P., Turkheimer, F., Carhart-Harris, R., Nutt, D., Hellyer, P. J., & Vaccarino, F. (2014). Homological scaffolds of brain functional networks. *Journal of The Royal Society Interface*, *11*(101), 20140873. https://doi.org/10.1098/rsif.2014.0873

Pic-Taylor, A., da Motta, L. G., de Morais, J. A., Junior, W. M., Santos, A.deF., Campos, L. A., Mortari, M. R., von Zuben, M. V., & Caldas, E. D. (2015). Behavioural and neurotoxic effects of ayahuasca infusion (Banisteriopsis caapi and Psychotria viridis) in female Wistar rat. *Behavioural processes*, *118*, 102–110. https://doi.org/10.1016/j.beproc.2015.05.004

Pompili, M., Girardi, P., Ruberto, A., & Tatarelli, R. (2005). Suicide in borderline personality disorder: A meta-analysis. *Nordic Journal of Psychiatry*, *59*(5), 319–324. https://doi.org/10.1080/08039480500320025

Preller, K. H., Burt, J. B., Ji, J. L., Schleifer, C. H., Adkinson, B. D., Stämpfli, P., Seifritz, E., Repovs, G., Krystal, J. H., Murray, J. D., Vollenweider, F. X., & Anticevic, A. (2018). Changes in global and thalamic brain connectivity in LSD-induced altered states of consciousness are attributable to the 5-HT2A receptor. *eLife*, *7*, e35082. https://doi.org/10.7554/eLife.35082

Preller, K. H., Duerler, P., Burt, J. B., Ji, J. L., Adkinson, B., Stämpfli, P., Seifritz, E., Repovš, G., Krystal, J. H., Murray, J. D., Anticevic, A., & Vollenweider, F. X. (2020). Psilocybin Induces Time-Dependent Changes in Global Functional Connectivity. *Biological Psychiatry*, *88*(2), 197–207. https://doi.org/10.1016/j.biopsych.2019.12.027

Preller, K. H., Herdener, M., Pokorny, T., Planzer, A., Kraehenmann, R., Stämpfli, P., Liechti, M. E., Seifritz, E., & Vollenweider, F. X. (2017). The Fabric of Meaning and Subjective Effects in LSD-Induced States Depend on Serotonin 2A Receptor Activation. *Current Biology*, *27*(3), 451–457. https://doi.org/10.1016/j.cub.2016.12.030

Preller, K. H., & Vollenweider, F. X. (2016). Phenomenology, Structure, and Dynamic of Psychedelic States. *Current topics in behavioral neurosciences*, *36*, 221–256. https://doi.org/10.1007/7854_2016_459

Preuss, C. V., Kalava, A., & King, K. C. (2023). Prescription of Controlled Substances: Benefits and Risks. In *StatPearls*. StatPearls Publishing. http://www.ncbi.nlm.nih.gov/books/NBK537318/

Price, D. D., Finniss, D. G., & Benedetti, F. (2008). A Comprehensive Review of the Placebo Effect: Recent Advances and Current Thought. *Annual Review of Psychology*, *59*, 565–590. https://doi.org/10.1146/annurev.psych.59.113006.095941

Pronovost-Morgan, C., Greenway, K., & Roseman, L. (2025). An international Delphi consensus for reporting of Setting in Psychedelic Clinical Trials. *Nature Medicine (in press)*. https://doi.org/10.21203/rs.3.rs-5428217/v1

Pronovost-Morgan, C., Hartogsohn, I., & Ramaekers, J. G. (2023). Harnessing placebo: Lessons from psychedelic science. *Journal of psychopharmacology (Oxford, England)*, *37*(9), 866–875. https://doi.org/10.1177/02698811231182602

Raison, C. L., Sanacora, G., Woolley, J., Heinzerling, K., Dunlop, B. W., Brown, R. T., Kakar, R., Hassman, M., Trivedi, R. P., Robison, R., Gukasyan, N., Nayak, S. M., Hu, X., O'Donnell, K. C., Kelmendi, B., Sloshower, J., Penn, A. D., Bradley, E., Kelly, D. F., … Griffiths, R. R. (2023). Single-Dose Psilocybin Treatment for Major Depressive Disorder: A Randomized Clinical Trial. *JAMA*, *330*(9), 843–853. https://doi.org/10.1001/jama.2023.14530

Raval, N. R., Johansen, A., Donovan, L. L., Ros, N. F., Ozenne, B., Hansen, H. D., & Knudsen, G. M. (2021). A Single Dose of Psilocybin Increases Synaptic Density and Decreases 5-HT2A Receptor Density in the Pig Brain. *International journal of molecular sciences*, *22*(2), 835. https://doi.org/10.3390/ijms22020835

Reckweg, J. T., van Leeuwen, C. J., Henquet, C., van Amelsvoort, T., Theunissen, E. L., Mason, N. L., Paci, R., Terwey, T. H., & Ramaekers, J. G. (2023). A phase 1/2 trial to assess safety and efficacy of a vaporized 5-methoxy-N,N-dimethyltryptamine formulation (GH001) in patients with treatment-resistant depression. Frontiers in Psychiatry, 14, 1133414. https://doi.org/10.3389/FPSYT.2023.1133414/BIBTEX

Reckweg, J., Mason, N. L., van Leeuwen, C., Toennes, S. W., Terwey, T. H., & Ramaekers, J. G. (2021). A Phase 1, Dose-Ranging Study to Assess Safety and Psychoactive Effects of a Vaporized 5-Methoxy-N, N-Dimethyltryptamine Formulation (GH001) in Healthy Volunteers. *Frontiers in pharmacology*, *12*, 760671. https://doi.org/10.3389/fphar.2021.760671

Reckweg, J. T., Uthaug, M. V., Szabo, A., Davis, A. K., Lancelotta, R., Mason, N. L., & Ramaekers, J. G. (2022). The clinical pharmacology and potential therapeutic applications of 5-methoxy-N,N-dimethyltryptamine (5-MeO-DMT). *Journal of Neurochemistry*, *162*(1), 128–146. https://doi.org/10.1111/jnc.15587

Resnik, D. B., & McCann, D. J. (2015). Deception by Research Participants. *The New England Journal of Medicine*, *373*(13), 1192–1193. https://doi.org/10.1056/NEJMp1506985

Riba, J., Anderer, P., Jané, F., Saletu, B. & Barbanoj, M.J. (2004) Effects of the South American Psychoactive Beverage Ayahuasca on Regional Brain Electrical Activity in Humans: A Functional Neuroimaging Study Using Low-Resolution Electromagnetic Tomography. *Neuropsychobiology*. 50 (1). https://doi.org/10.1159/000077946.

Riba, J., McIlhenny, E. H., Bouso, J. C., & Barker, S. A. (2015). Metabolism and urinary disposition of N,N-dimethyltryptamine after oral and smoked administration: A comparative study. *Drug Testing and Analysis*, *7*(5), 401–406. https://doi.org/10.1002/dta.1685

Riba, J., Rodríguez-Fornells, A., Strassman, R. J., & Barbanoj, M. J. (2001). Psychometric assessment of the Hallucinogen Rating Scale. *Drug and Alcohol Dependence*, *62*(3), 215–223. https://doi.org/10.1016/s0376-8716(00)00175-7

Rocha, J. M., Rossi, G. N., de Lima Osório, F., Bouso, J. C., de Oliveira Silveira, G., Yonamine, M., Campos, A. C., Bertozi, G., Cecílio Hallak, J. E., & dos Santos, R. G. (2021). Effects of Ayahuasca on the Recognition of Facial Expressions of Emotions in Naive Healthy Volunteers: A Pilot, Proof-of-Concept, Randomized Controlled Trial. *Journal of Clinical Psychopharmacology*, *41*(3), 267. https://doi.org/10.1097/JCP.0000000000001396

Rogers, C. R. (1957). The necessary and sufficient conditions of therapeutic personality change. *Journal of Consulting Psychology*, *21*(2), 95–103. https://doi.org/10.1037/h0045357

Romeo, B., Kervadec, E., Fauvel, B., Strika-Bruneau, L., Amirouche, A., Verroust, V., Piolino, P., & Benyamina, A. (2024). Safety and risk assessment of psychedelic psychotherapy: A meta-analysis and systematic review. *Psychiatry Research*, *335*, 115880. https://doi.org/10.1016/j.psychres.2024.115880

Roseman L. (2024). A reflection on paradigmatic tensions within the FDA advisory committee for MDMA-assisted therapy. *Journal of psychopharmacology (Oxford, England)*, 2698811241309611. Advance online publication. https://doi.org/10.1177/02698811241309611

Roseman, L., Demetriou, L., Wall, M. B., Nutt, D. J., & Carhart-Harris, R. L. (2018). Increased amygdala responses to emotional faces after psilocybin for treatment-resistant depression. *Neuropharmacology*, *142*, 263–269. https://doi.org/10.1016/j.neuropharm.2017.12.041

Roseman, L., Erritzoe, D., Nutt, D., Carhart-Harris, R., & Timmermann, C. (2024). Interrupting the psychedelic experience through contextual manipulation to study experience efficacy. *JAMA Network Open*, *7*(7), e2422181-e2422181.

Roseman, L., Haijen, E., Idialu-Ikato, K., Kaelen, M., Watts, R., & Carhart-Harris, R. (2019). Emotional breakthrough and psychedelics: Validation of the Emotional Breakthrough Inventory. *Journal of Psychopharmacology*, *33*(9), 1076–1087. https://doi.org/10.1177/0269881119855974

Roseman, L., Nutt, D. J., & Carhart-Harris, R. L. (2018). Quality of Acute Psychedelic Experience Predicts Therapeutic Efficacy of Psilocybin for Treatment-Resistant Depression. *Frontiers in pharmacology*, *8*, 974. https://doi.org/10.3389/fphar.2017.00974

Roseman, L., Sereno, M. I., Leech, R., Kaelen, M., Orban, C., McGonigle, J., Feilding, A., Nutt, D. J., & Carhart-Harris, R. L. (2016). LSD alters eyes-closed functional connectivity within the early visual cortex in a retinotopic fashion. *Human brain mapping*, *37*(8), 3031–3040. https://doi.org/10.1002/hbm.23224

Rosenberg, D. E., Isbell, H., Miner, E. J., & Logan, C. R. (1964). The effect of N,N-dimethyltryptamine in human subjects tolerant to lysergic acid diethylamide. *Psychopharmacologia*, *5*(3), 217–227. https://doi.org/10.1007/BF00413244

Ross, S., Bossis, A., Guss, J., Agin-Liebes, G., Malone, T., Cohen, B., Mennenga, S. E., Belser, A., Kalliontzi, K.,

Babb, J., Su, Z., Corby, P., & Schmidt, B. L. (2016). Rapid and sustained symptom reduction following psilocybin treatment for anxiety and depression in patients with life-threatening cancer: A randomized controlled trial. *Journal of Psychopharmacology (Oxford, England)*, *30*(12), 1165–1180. https://doi.org/10.1177/0269881116675512

Rucker, J. J. H., Iliff, J., & Nutt, D. J. (2018). Psychiatry & the psychedelic drugs. Past, present & future. *Neuropharmacology*, *142*, 200–218. https://doi.org/10.1016/j.neuropharm.2017.12.040

Rucker, J. J. H., Jafari, H., Mantingh, T., Bird, C., Modlin, N. L., Knight, G., Reinholdt, F., Day, C., Carter, B., & Young, A. (2021). Psilocybin-assisted therapy for the treatment of resistant major depressive disorder (PsiDeR): Protocol for a randomised, placebo-controlled feasibility trial. *BMJ Open*, *11*(12), e056091. https://doi.org/10.1136/bmjopen-2021-056091

Rucker, J. J. H., Roberts, C., Seynaeve, M., Young, A. H., Suttle, B., Yamamoto, T., Ermakova, A. O., Dunbar, F., & Wiegand, F. (2024). Phase 1, placebo-controlled, single ascending dose trial to evaluate the safety, pharmacokinetics and effect on altered states of consciousness of intranasal BPL-003 (5-methoxy-N,N-dimethyltryptamine benzoate) in healthy participants. *Journal of psychopharmacology (Oxford, England)*, *38*(8), 712–723. https://doi.org/10.1177/02698811241246857

Sabé, M., Sulstarova, A., Glangetas, A., De Pieri, M., Mallet, L., Curtis, L., Richard-Lepouriel, H., Penzenstadler, L., Seragnoli, F., Thorens, G., Zullino, D., Preller, K., Böge, K., Leucht, S., Correll, C. U., Solmi, M., Kaiser, S., & Kirschner, M. (2025). Reconsidering evidence for psychedelic-induced psychosis: an overview of reviews, a systematic review, and meta-analysis of human studies. *Molecular psychiatry*, *30*(3), 1223–1255. https://doi.org/10.1038/s41380-024-02800-5

Saeger, H. N., & Olson, D. E. (2022). Psychedelic-inspired approaches for treating neurodegenerative disorders. *Journal of Neurochemistry*, *162*(1), 109–127. https://doi.org/10.1111/jnc.15544

Sanders, J. W., Millière, R., Daily, Z. G., Carhart-Harris, R., & Timmermann, C. (2024). *DMT micro-phenomenology.* OSF. https://doi.org/10.31234/osf.io/2m9d4

Sanz, C., Pallavicini, C., Carrillo, F., Zamberlan, F., Sigman, M., Mota, N., Copelli, M., Ribeiro, S., Nutt, D., Carhart-Harris, R., & Tagliazucchi, E. (2021). The entropic tongue: Disorganization of natural language under LSD. *Consciousness and Cognition*, *87*. https://doi.org/10.1016/J.CONCOG.2020.103070

Schartner, M.M., Carhart-Harris, R.L., Barrett, A.B., Seth, A.K. & Muthukumaraswamy, S.D. (2017) Increased spontaneous MEG signal diversity for psychoactive doses of ketamine, LSD and psilocybin. *Scientific Reports*. 7 (1), 46421. https://doi.org/10.1038/srep46421.

Schenberg, E.E., Alexandre, J.F.M., Filev, R., Cravo, A.M., Sato, J.R., Muthukumaraswamy, S.D., Yonamine, M., Waguespack, M., Lomnicka, I., Barker, S.A. & Silveira, D.X. da (2015) Acute Biphasic Effects of Ayahuasca. *PLOS ONE*. 10 (9), e0137202. https://doi.org/10.1371/journal.pone.0137202.

Schenberg, E. E., Tófoli, L. F., Rezinovsky, D., & Silveira, D. X. D. (2017). Translation and cultural adaptation of the States of Consciousness Questionnaire (SOCQ) and statistical validation of the Mystical Experience Questionnaire (MEQ30) in Brazilian Portuguese. *Archives of Clinical Psychiatry (São Paulo)*, *44*, 1–5. https://doi.org/10.1590/0101-60830000000105

Schmid, Y., Enzler, F., Gasser, P., Grouzmann, E., Preller, K. H., Vollenweider, F. X., Brenneisen, R., Müller, F., Borgwardt, S., & Liechti, M. E. (2015). Acute Effects of Lysergic Acid Diethylamide in Healthy Subjects. *Biological psychiatry, 78*(8), 544–553. https://doi.org/10.1016/j.biopsych.2014.11.015

Shadani, S., Conn, K., Andrews, Z. B., & Foldi, C. J. (2024). Potential Differences in Psychedelic Actions Based on Biological Sex. *Endocrinology*, *165*(8), bqae083. https://doi.org/10.1210/endocr/bqae083

Shen, H.-W., Jiang, X.-L., C. Winter, J., & Yu, A.-M. (2010). Psychedelic 5-Methoxy-N,N-dimethyltryptamine: Metabolism, Pharmacokinetics, Drug Interactions, and Pharmacological Actions. *Current Drug Metabolism*, *11*(8), 659. https://doi.org/10.2174/138920010794233495

Shen, H. W., Wu, C., Jiang, X. L., & Yu, A. M. (2010). Effects of monoamine oxidase inhibitor and cytochrome P450 2D6 status on 5-methoxy-N,N-dimethyltryptamine metabolism and pharmacokinetics. *Biochemical Pharmacology*, *80*(1), 122–128. https://doi.org/10.1016/J.BCP.2010.02.020

Shinozuka, K., Tewarie, P. K. B., Luppi, A., Lynn, C., Roseman, L., Muthukumaraswamy, S., Nutt, D. J., Carhart-Harris, R., Deco, G., & Kringelbach, M. L. (2025). LSD flattens the hierarchy of directed information flow in fast whole-brain dynamics. *Imaging Neuroscience*, *3*, imag_a_00420. https://doi.org/10.1162/imag_a_00420

Shulgin A, Shulgin A (1997) TIHKAL: The Continuation. Berkeley, CA: Transform Press

Shulgin, A. T., Shulgin, L. A., & Jacob, P., 3rd (1986). A protocol for the evaluation of new psychoactive drugs in man. M*ethods and findings in experimental and clinical pharmacology*, *8*(5), 313–320.

Siegel, J.S., Subramanian, S., Perry, D., Kay, B.P., Gordon, E.M., et al. (2024) Psilocybin desynchronizes the human brain. *Nature*. 632 (8023), 131–138. https://doi.org/10.1038/s41586-024-07624-5.

Silva, F., Butlen-Ducuing, F., Guizzaro, L., & Balabanov, P. (2025). A review of psychedelics trials completed in depression, informed by European regulatory perspectives. *Neuroscience Applied*, *4*, 105516. https://doi.org/10.1016/j.nsa.2025.105516

Simonsson, O., Goldberg, S. B., & Hendricks, P. S. (2024). Into the wild frontier: Mapping the terrain of adverse events in psychedelic-assisted therapies. *Journal of psychopharmacology (Oxford, England)*, 2698811241292944. Advance online publication. https://doi.org/10.1177/02698811241292944

Simonsson, O., Goldberg, S. B., Osika, W., Stenfors, C. U. D., Chaturvedi, S., Swords, C. M., Narayanan, J., & Hendricks, P. S. (2025). Longitudinal associations of naturalistic psychedelic use with psychotic and manic symptoms. *Psychological Medicine*, *55*, e99. https://doi.org/10.1017/S0033291725000716

Simonsson, O., Hendricks, P. S., Chambers, R., Osika, W., & Goldberg, S. B. (2023). Prevalence and associations of challenging, difficult or distressing experiences using classic psychedelics. *Journal of Affective Disorders*, *326*, 105–110. https://doi.org/10.1016/j.jad.2023.01.073

Simonsson, O., Mosing, M. A., Osika, W., Ullén, F., Larsson, H., Lu, Y., & Wesseldijk, L. W. (2024). Adolescent Psychedelic Use and Psychotic or Manic Symptoms. *JAMA Psychiatry*, *81*(6), 579–585. https://doi.org/10.1001/jamapsychiatry.2024.0047

Singleton, S. P., Luppi, A. I., Carhart-Harris, R. L., Cruzat, J., Roseman, L., Nutt, D. J., Deco, G., Kringelbach, M. L., Stamatakis, E. A., & Kuceyeski, A. (2022). Receptor-informed network control theory links LSD and psilocybin to a flattening of the brain's control energy landscape. *Nature communications*, *13*(1), 5812. https://doi.org/10.1038/s41467-022-33578-1

Sklerov, J., Levine, B., Moore, K. A., King, T., & Fowler, D. (2005). A fatal intoxication following the ingestion of 5-methoxy-N,N-dimethyltryptamine in an ayahuasca preparation. *Journal of Analytical Toxicology*, *29*(8), 838–841. https://doi.org/10.1093/JAT/29.8.838

Skosnik, P.D., Sloshower, J., Safi-Aghdam, H., Pathania, S., Syed, S., Pittman, B. & D'Souza, D.C. (2023) Sub-acute effects of psilocybin on EEG correlates of neural plasticity in major depression: Relationship to symptoms. *Journal of Psychopharmacology*. 37 (7), 687–697. https://doi.org/10.1177/02698811231179800.

Smallwood, J., Turnbull, A., Wang, H., Ho, N. S. P., Poerio, G. L., Karapanagiotidis, T., Konu, D., Mckeown, B., Zhang, M., Murphy, C., Vatansever, D., Bzdok, D., Konishi, M., Leech, R., Seli, P., Schooler, J. W., Bernhardt, B., Margulies, D. S., & Jefferies, E. (2021). The neural correlates of ongoing conscious thought. *iScience*, *24*(3), 102132. https://doi.org/10.1016/j.isci.2021.102132

Soto-Angona, Ó., Fortea, A., Fortea, L., Martínez-Ramírez, M., Santamarina, E., López, F. J. G., Knudsen, G. M., & Ona, G. (2024). Do classic psychedelics increase the risk of seizures? A scoping review. *European Neuropsychopharmacology*, *85*, 35–42. https://doi.org/10.1016/j.euroneuro.2024.05.002

Spriggs, M. J., Giribaldi, B., Lyons, T., Rosas, F. E., Kärtner, L. S., Buchborn, T., Douglass, H. M., Roseman, L., Timmermann, C., Erritzoe, D., Nutt, D. J., & Carhart-Harris, R. L. (2023). Body mass index (BMI) does not predict responses to psilocybin. *Journal of Psychopharmacology (Oxford, England)*, *37*(1), 107–116. https://doi.org/10.1177/02698811221131994

Spriggs, M. J., Sumner, R.L., McMillan, R.L., Moran, R.J., Kirk, I.J. & Muthukumaraswamy, S.D. (2018) Indexing sensory plasticity: Evidence for distinct Predictive Coding and Hebbian learning mechanisms in the cerebral cortex. *NeuroImage*. 176, 290–300. https://doi.org/10.1016/j.neuroimage.2018.04.060.

Stauffer, C. S., Anderson, B. T., Ortigo, K. M., & Woolley, J. (2020). Psilocybin-Assisted Group Therapy and Attachment: Observed Reduction in Attachment Anxiety and Influences of Attachment Insecurity on the Psilocybin Experience. *ACS Pharmacology & Translational Science*, *4*(2), 526–532. https://doi.org/10.1021/acsptsci.0c00169

Stoliker, D., Novelli, L., Khajehnejad, M., Biabani, M., Barta, T., Greaves, M. D., Williams, M., Chopra, S., Bazin, O., Simonsson, O., Chambers, R., Barrett, F., Deco, G., Seth, A., Preller, K. H., Carhart-Harris, R., Sundram, S., Egan, G. F., & Razi, A. (2025). Psychedelics Align Brain Activity with Context. *bioRxiv*, 2025.03.09.642197. https://doi.org/10.1101/2025.03.09.642197

Stoliker, D., Novelli, L., Vollenweider, F. X., Egan, G. F., Preller, K. H., & Razi, A. (2024). Neural Mechanisms of Resting-State Networks and the Amygdala Underlying the Cognitive and Emotional Effects of Psilocybin. *Biological Psychiatry*, *96*(1), 57–66. https://doi.org/10.1016/j.biopsych.2024.01.002

Stone, A. L., Storr, C. L., & Anthony, J. C. (2006). Evidence for a hallucinogen dependence syndrome developing soon after onset of hallucinogen use during adolescence. *International Journal of Methods in Psychiatric Research*, *15*(3), 116–130. https://doi.org/10.1002/mpr.188

Strassman, R. J., & Qualls, C. R. (1994). Dose-response study of N,N-dimethyltryptamine in humans. I. Neuroendocrine, autonomic, and cardiovascular effects. *Archives of General Psychiatry*, *51*(2), 85–97. https://doi.org/10.1001/archpsyc.1994.03950020009001

Strassman, R. J., Qualls, C. R., Uhlenhuth, E. H., & Kellner, R. (1994). Dose-response study of N,N-dimethyltryptamine in humans. II. Subjective effects and preliminary results of a new rating scale. *Archives of General Psychiatry*, *51*(2), 98–108. https://doi.org/10.1001/archpsyc.1994.03950020022002

Strassman, R. J., Qualls, C. R., & Berg, L. M. (1996). Differential tolerance to biological and subjective effects of four closely spaced doses of N,N-dimethyltryptamine in humans. *Biological Psychiatry*, *39*(9), 784–795. https://doi.org/10.1016/0006-3223(95)00200-6

Studerus, E., Gamma, A., & Vollenweider, F. X. (2010). Psychometric Evaluation of the Altered States of Consciousness Rating Scale (OAV). *PLoS ONE*, *5*(8), e12412. https://doi.org/10.1371/journal.pone.0012412

Studerus, E., Kometer, M., Hasler, F., & Vollenweider, F. X. (2011). Acute, subacute and long-term subjective effects of psilocybin in healthy humans: A pooled analysis of experimental studies. *Journal of Psychopharmacology (Oxford, England)*, *25*(11), 1434–1452. https://doi.org/10.1177/0269881110382466

Sumner, R. L., & Lukasiewicz, K. (2023). Psychedelics and neural plasticity. *BMC neuroscience*, *24*(1), 35. https://doi.org/10.1186/s12868-023-00809-0

Sumner, R. L., McMillan, R., Spriggs, M. J., Campbell, D., Malpas, G., Maxwell, E., Deng, C., Hay, J., Ponton, R., Sundram, F., & Muthukumaraswamy, S. D. (2020). Ketamine improves short-term plasticity in depression by enhancing sensitivity to prediction errors. *European neuropsychopharmacology : the journal of the European College of Neuropsychopharmacology*, *38*, 73–85. https://doi.org/10.1016/j.euroneuro.2020.07.009

Szára, S. (1956). Dimethyltryptamin: Its metabolism in man; the relation to its psychotic effect to the serotonin metabolism. *Experientia*, *12*(11), 441–442. https://doi.org/10.1007/BF02157378

Szigeti, B., & Heifets, B. D. (2024). Expectancy Effects in Psychedelic Trials. *Biological Psychiatry: Cognitive Neuroscience and Neuroimaging*, *9*(5), 512–521. https://doi.org/10.1016/j.bpsc.2024.02.004

Szigeti, B., Kartner, L., Blemings, A., Rosas, F., Feilding, A., Nutt, D. J., Carhart-Harris, R. L., & Erritzoe, D. (2021). Self-blinding citizen science to explore psychedelic microdosing. *eLife*, *10*, e62878. https://doi.org/10.7554/eLife.62878

Szigeti, B., Weiss, B., Rosas, F. E., Erritzoe, D., Nutt, D., & Carhart-Harris, R. (2024). Assessing expectancy and suggestibility in a trial of escitalopram v. psilocybin for depression. *Psychological Medicine*, *54*(8), 1717–1724. doi:10.1017/S0033291723003653

Szmulewicz, A. G., Valerio, M. P., & Smith, J. M. (2015). Switch to mania after ayahuasca consumption in a man with bipolar disorder: a case report. *International Journal of Bipolar Disorders*, *3*(1). https://doi.org/10.1186/S40345-014-0020-Y

Tagliazucchi E. (2022). Language as a Window Into the Altered State of Consciousness Elicited by Psychedelic Drugs. *Frontiers in pharmacology*, *13*, 812227. https://doi.org/10.3389/fphar.2022.812227

Tagliazucchi, E., Roseman, L., Kaelen, M., Orban, C., Muthukumaraswamy, S. D., Murphy, K., ... & Carhart-Harris, R. (2016). Increased global functional connectivity correlates with LSD-induced ego dissolution. *Current biology*, *26*(8), 1043-1050.

Tanaka, E., Kamata, T., Katagi, M., Tsuchihashi, H., & Honda, K. (2006). A fatal poisoning with 5-methoxy-N,N-diisopropyltryptamine, Foxy. *Forensic Science International*, *163*(1–2), 152–154. https://doi.org/10.1016/J.FORSCIINT.2005.11.026

Thomas, C. W., Blanco-Duque, C., Bréant, B. J., Goodwin, G. M., Sharp, T., Bannerman, D. M., & Vyazovskiy, V. V. (2022). Psilocin acutely alters sleep-wake architecture and cortical brain activity in laboratory mice. *Translational psychiatry*, *12*(1), 77. https://doi.org/10.1038/s41398-022-01846-9

Timmermann, C., Bauer, P. R., Gosseries, O., Vanhaudenhuyse, A., Vollenweider, F., Laureys, S., Singer, T., Mind and Life Europe (MLE) ENCECON Research Group, Antonova, E., & Lutz, A. (2023). A neurophenomenological approach to non-ordinary states of consciousness: hypnosis, meditation, and psychedelics. *Trends in cognitive sciences*, *27*(2), 139–159. https://doi.org/10.1016/j.tics.2022.11.006

Timmermann, C., Kettner, H., Letheby, C., Roseman, L., Rosas, F. E., & Carhart-Harris, R. L. (2021). Psychedelics alter metaphysical beliefs. *Scientific reports*, *11*(1), 22166. https://doi.org/10.1038/s41598-021-01209-2

Timmermann, C., Roseman, L., Haridas, S., Rosas, F.E., Luan, L., Kettner, H., Martell, J., Erritzoe, D., Tagliazucchi, E., Pallavicini, C., Girn, M., Alamia, A., Leech, R., Nutt, D.J. & Carhart-Harris, R.L. (2023) Human brain effects of DMT assessed via EEG-fMRI. *Proceedings of the National Academy of Sciences*. 120 (13), e2218949120. https://doi.org/10.1073/pnas.2218949120.

Timmermann, C., Roseman, L., Schartner, M., Milliere, R., Williams, L. T. J., Erritzoe, D., Muthukumaraswamy, S.,

Ashton, M., Bendrioua, A., Kaur, O., Turton, S., Nour, M. M., Day, C. M., Leech, R., Nutt, D. J., & Carhart-Harris, R. L. (2019). Neural correlates of the DMT experience assessed with multivariate EEG. *Scientific Reports*, *9*(1), 16324. https://doi.org/10.1038/s41598-019-51974-4

Timmermann, C., Roseman, L., Williams, L., Erritzoe, D., Martial, C., Cassol, H., Laureys, S., Nutt, D., & Carhart-Harris, R. (2018). DMT Models the Near-Death Experience. *Frontiers in psychology*, *9*, 1424. https://doi.org/10.3389/fpsyg.2018.01424

Timmermann, C., Watts, R., & Dupuis, D. (2022). Towards psychedelic apprenticeship: Developing a gentle touch for the mediation and validation of psychedelic-induced insights and revelations. *Transcultural psychiatry*, *59*(5), 691–704. https://doi.org/10.1177/13634615221082796

Tupper, K. W. (2009). Ayahuasca healing beyond the Amazon: The globalization of a traditional indigenous entheogenic practice. *Global Networks*, *9*(1), 117–136. https://doi.org/10.1111/j.1471-0374.2009.00245.x

Umbricht, D., Koller, R., Vollenweider, F. X., & Schmid, L. (2002). Mismatch negativity predicts psychotic experiences induced by NMDA receptor antagonist in healthy volunteers. *Biological psychiatry*, *51*(5), 400–406. https://doi.org/10.1016/s0006-3223(01)01242-2

United States Drug Enforcement Administration. (2024) *Drug Scheduling.* Last accessed on 11th Dec 2024: https://www.dea.gov/drug-information/drug-scheduling#:~:text=Schedule%20I%20drugs%2C%20substances%2C%20or,)%2C%20methaqualone%2C%20and%20peyote.

U.S. Embassy in Peru. (2025, January 23). *Health alert: Do not use ayahuasca/kambo*. Retrieved March 22, 2025, from https://pe.usembassy.gov/health-alert-do-not-use-ayahuasca-kambo/

Uthaug, M. V., Lancelotta, R., Bernal, A. M. O., Davis, A. K., & Ramaekers, J. G. (2020). A comparison of reactivation experiences following vaporization and intramuscular injection (IM) of synthetic 5-methoxy-N,N-dimethyltryptamine (5-MeO-DMT) in a naturalistic setting. *Journal of Psychedelic Studies*, *4*(2), 104–113. https://doi.org/10.1556/2054.2020.00123

Uthaug, M. V., Lancelotta, R., Szabo, A., Davis, A. K., Riba, J., & Ramaekers, J. G. (2019). Prospective examination of synthetic 5-methoxy-N,N-dimethyltryptamine inhalation: effects on salivary IL-6, cortisol levels, affect, and non-judgment. *Psychopharmacology*, *237*(3), 773. https://doi.org/10.1007/S00213-019-05414-W

Uthaug, M. V., Mason, N. L., Toennes, S. W., Reckweg, J. T., de Sousa Fernandes Perna, E. B., Kuypers, K. P. C., van Oorsouw, K., Riba, J., & Ramaekers, J. G. (2021). A placebo-controlled study of the effects of ayahuasca, set and setting on mental health of participants in ayahuasca group retreats. *Psychopharmacology*, *238*(7), 1899–1910. https://doi.org/10.1007/s00213-021-05817-8

Valenzuela-Moguillansky, C., & Vásquez-Rosati, A. (2019). An Analysis Procedure for the Micro-Phenomenological Interview. *Constructivist Foundations*, *14*(2), Article 2.

van Amsterdam, J., Opperhuizen, A., & Brink, W. van den. (2011). Harm potential of magic mushroom use: A review. *Regulatory Toxicology and Pharmacology*, *59*(3), 423–429. https://doi.org/10.1016/J.YRTPH.2011.01.006

Varela, F. (1996). Neurophenomenology: A Methodological Remedy for the Hard Problem. *Journal of Consciousness Studies*, *3*(4), 330–349.

Vargas, M.V., Dunlap, L.E., Dong, C., Carter, S.J., Tombari, R.J., Jami, S.A., Cameron, L.P., Patel, S.D., Hennessey, J.J., Saeger, H.N., McCorvy, J.D., Gray, J.A., Tian, L. & Olson, D.E. (2023) Psychedelics promote neuroplasticity through the activation of intracellular 5-HT2A receptors. *Science*. 379 (6633), 700–706. https://doi.org/10.1126/science.adf0435.

Varker, T., Watson, L., Gibson, K., Forbes, D., & O'Donnell, M. L. (2021). Efficacy of Psychoactive Drugs for the Treatment of Posttraumatic Stress Disorder: A Systematic Review of MDMA, Ketamine, LSD and Psilocybin. *Journal of Psychoactive Drugs*, *53*(1), 85–95. https://doi.org/10.1080/02791072.2020.1817639

Varley T. F. (2024). Generalized decomposition of multivariate information. *PloS one*, *19*(2), e0297128. https://doi.org/10.1371/journal.pone.0297128

Varley, T. F., Havert, D., Fosque, L., Alipour, A., Weerawongphrom, N., Naganobori, H., O'Shea, L., Pope, M., & Beggs, J. (2024). The serotonergic psychedelic N,N-dipropyltryptamine alters information-processing dynamics in in vitro cortical neural circuits. *Network neuroscience (Cambridge, Mass.)*, *8*(4), 1421–1438. https://doi.org/10.1162/netn_a_00408

Viña, S. (2025). Educational moderation of gender disparities in psychedelic health outcomes. *Academia Mental Health and Well-Being*, *2*(1). https://doi.org/10.20935/MHealthWellB7580

Vogt, S. B., Ley, L., Erne, L., Straumann, I., Becker, A. M., Klaiber, A., Holze, F., Vandersmissen, A., Mueller, L., Duthaler, U., Rudin, D., Luethi, D., Varghese, N., Eckert, A., & Liechti, M. E. (2023). Acute effects of intravenous DMT in a randomized placebo-controlled study in healthy participants. *Translational Psychiatry*, *13*(1), 1–9. https://doi.org/10.1038/s41398-023-02477-4

Vollenweider, F. X., & Kometer, M. (2010). The neurobiology of psychedelic drugs: Implications for the treatment of mood disorders. *Nature Reviews Neuroscience*, *11*(9), 642–651. https://doi.org/10.1038/nrn2884

Wall, M. B., Lam, C., Ertl, N., Kaelen, M., Roseman, L., Nutt, D. J., & Carhart-Harris, R. L. (2023). Increased low-frequency brain responses to music after psilocybin therapy for depression. *Journal of affective disorders*, *333*, 321–330. https://doi.org/10.1016/j.jad.2023.04.081

Warren, J. M., Dham-Nayyar, P., & Alexander, J. (2013). Recreational use of naturally occurring dimethyltryptamine - contributing to psychosis?. *The Australian and New Zealand journal of psychiatry*, *47*(4), 398–399. https://doi.org/10.1177/0004867412462749

Watts, R., Day, C., Krzanowski, J., Nutt, D., & Carhart-Harris, R. (2017). Patients' Accounts of Increased "Connectedness" and "Acceptance" After Psilocybin for Treatment-Resistant Depression. *Journal of Humanistic Psychology*, *57*(5), 520–564. https://doi.org/10.1177/0022167817709585

Williams, M. T., Reed, S., & Aggarwal, R. (2020). Culturally informed research design issues in a study for MDMA-assisted psychotherapy for posttraumatic stress disorder. *Journal of Psychedelic Studies*, *4*(1), 40–50. https://doi.org/10.1556/2054.2019.016

Wolff, M., Mertens, L. J., Walter, M., Enge, S., & Evens, R. (2022). The Acceptance/Avoidance-Promoting Experiences Questionnaire (APEQ): A theory-based approach to psychedelic drugs' effects on psychological flexibility. *Journal of Psychopharmacology (Oxford, England)*, *36*(3), 387–408. https://doi.org/10.1177/02698811211073758

Wood, M. J., McAlpine, R. G., & Kamboj, S. K. (2024). Strategies for resolving challenging psychedelic experiences: Insights from a mixed-methods study. *Scientific Reports*, *14*(1), 28817. https://doi.org/10.1038/s41598-024-79931-w

World Inequality Database. (n.d.). *Income comparator*. Retrieved March 24, 2025, from https://wid.world/income-comparator/

Wsół A. (2023). Cardiovascular safety of psychedelic medicine: current status and future directions. *Pharmacological reports : PR*, *75*(6), 1362–1380. https://doi.org/10.1007/s43440-023-00539-4

Yaden, D. B., Goldy, S. P., Weiss, B., & Griffiths, R. R. (2024). Clinically relevant acute subjective effects of psychedelics beyond mystical experience. *Nature Reviews Psychology*, *3*(9), 606–621. https://doi.org/10.1038/s44159-024-00345-6

Yaden, D. B., & Griffiths, R. R. (2021). The Subjective Effects of Psychedelics Are Necessary for Their Enduring Therapeutic Effects. *ACS Pharmacology & Translational Science*, *4*(2), 568–572. https://doi.org/10.1021/acsptsci.0c00194

Yaden, D. B., Nayak, S. M., Gukasyan, N., Anderson, B. T., & Griffiths, R. R. (2022). The Potential of Psychedelics for End of Life and Palliative Care. *Current Topics in Behavioral Neurosciences*, *56*, 169–184. https://doi.org/10.1007/7854_2021_278

Yao, Y., Guo, D., Lu, T.-S., Liu, F.-L., Huang, S.-H., Diao, M.-Q., Li, S.-X., Zhang, X.-J., Kosten, T. R., Shi, J., Bao, Y.-P., Lu, L., & Han, Y. (2024). Efficacy and safety of psychedelics for the treatment of mental disorders: A systematic review and meta-analysis. *Psychiatry Research*, *335*, 115886. https://doi.org/10.1016/j.psychres.2024.115886

Yeo, B. T., Krienen, F. M., Sepulcre, J., Sabuncu, M. R., Lashkari, D., Hollinshead, M., Roffman, J. L., Smoller, J. W., Zöllei, L., Polimeni, J. R., Fischl, B., Liu, H., & Buckner, R. L. (2011). The organization of the human cerebral cortex estimated by intrinsic functional connectivity. *Journal of neurophysiology*, *106*(3), 1125–1165. https://doi.org/10.1152/jn.00338.2011

Yerubandi, A., Thomas, J. E., Bhuiya, N. M. M. A., Harrington, C., Villa Zapata, L., & Caballero, J. (2024). Acute Adverse Effects of Therapeutic Doses of Psilocybin: A Systematic Review and Meta-Analysis. *JAMA Network Open*, *7*(4), e245960. https://doi.org/10.1001/jamanetworkopen.2024.5960

Younger, J., Gandhi, V., Hubbard, E., & Mackey, S. (2012). Development of the Stanford Expectations of Treatment Scale (SETS): A tool for measuring patient outcome expectancy in clinical trials. *Clinical Trials (London, England)*, *9*(6), 767–776. https://doi.org/10.1177/1740774512465064

Yu, A. M., Idle, J. R., Herraiz, T., Küpfer, A., & Gonzalez, F. J. (2003). Screening for endogenous substrates reveals that CYP2D6 is a 5-methoxyindolethylamine O-demethylase. *Pharmacogenetics*, *13*(6), 307–319. https://doi.org/10.1097/01.FPC.0000054094.48725.B7

Zafar, R., Siegel, M., Harding, R., Barba, T., Agnorelli, C., Suseelan, S., Roseman, L., Wall, M., Nutt, D. J., & Erritzoe, D. (2023). Psychedelic therapy in the treatment of addiction: the past, present and future. *Frontiers in psychiatry*, *14*, 1183740. https://doi.org/10.3389/fpsyt.2023.1183740

Zeifman, R. J., Singhal, N., Breslow, L., & Weissman, C. R. (2021). On the Relationship between Classic Psychedelics

and Suicidality: A Systematic Review. *ACS Pharmacology & Translational Science*, *4*(2), 436–451. https://doi.org/10.1021/acsptsci.1c00024

Zeifman, R. J., & Wagner, A. C. (2020). Exploring the case for research on incorporating psychedelics within interventions for borderline personality disorder. *Journal of Contextual Behavioral Science*, *15*, 1–11. https://doi.org/10.1016/j.jcbs.2019.11.001

Zeifman, R. J., Yu, D., Singhal, N., Wang, G., Nayak, S. M., & Weissman, C. R. (2022). Decreases in Suicidality Following Psychedelic Therapy: A Meta-Analysis of Individual Patient Data Across Clinical Trials. *The Journal of Clinical Psychiatry*, *83*(2), 21r14057. https://doi.org/10.4088/JCP.21r14057

Zelen M. (1979). A new design for randomized clinical trials. *The New England journal of medicine*, *300*(22), 1242–1245. https://doi.org/10.1056/NEJM197905313002203

Zhang, T., Schoene, A. M., Ji, S., & Ananiadou, S. (2022). Natural language processing applied to mental illness detection: A narrative review. *Npj Digital Medicine, 5*(1), 1–13. https://doi.org/10.1038/s41746-022-00589-7